\documentclass[graphics,floatfix,footinbib,tightenlines,nobibnotes,aps,prb,twocolumn,longbibliography]{revtex4-2}
\usepackage{soul}
\usepackage{amsmath,amssymb,wasysym}
\usepackage{hyperref}
\hypersetup{
colorlinks = true,
linkcolor = [rgb]{1,0,0},
citecolor = [rgb]{0,0,1},
urlcolor = [rgb]{0., 0., 0.545098}}
\usepackage{color}
\usepackage{xcolor}
\usepackage{braket}

\usepackage[utf8]{inputenc}
\usepackage{bm}
\usepackage{verbatim}
\usepackage{graphicx}
\usepackage{amsmath}
\usepackage{color}
\usepackage{float}
\usepackage{bbm}
\usepackage{amssymb}
\usepackage{slashed}
\usepackage{wasysym,bm,bbm,dsfont,braket}

\begin{document}
\title{High-temperature superconductivity from kinetic energy}
\author{Hanbit Oh}
\thanks{These two authors contributed equally}
\thanks{Corresponding author: \href{mailto:hoh22@jh.edu}{hoh22@jh.edu}}
\author{Hui Yang}
\thanks{These two authors contributed equally}
\author{Ya-Hui Zhang}
\thanks{Corresponding author: \href{mailto:yzhan566@jhu.edu}{yzhan566@jhu.edu}}
\affiliation{William H. Miller III Department of Physics and Astronomy, Johns Hopkins University, Baltimore, Maryland, 21218, USA}

\date{\today}
\begin{abstract}
Superconductivity is usually assumed to arise from attractive interaction. In this work we show that strong pairing is possible soley from kinetic energy even without a net attraction. We demonstrate a high-temperature kinetic superconductor in a simple lattice model with nearest-neighbor hopping ($t$) projected onto a constrained Hilbert space, analogous to the $t$-$J$ model with $J=0$, where kinetic magnetism has been previously studied. Using density matrix renormalization group (DMRG) on cylinders up to width $L_y=8$, we find a superconducting ground state exhibiting a key difference from high-$T_c$ cuprates: both the pairing gap and phase stiffness \textit{increase} with doping ($x$). We find pairing gaps, determined from spin and single-electron charge gaps, exceeding $1.5t$. This model can be realized within the double Kondo lattice model, relevant to bilayer nickelates, in the limit of strong inter-layer spin coupling ($J_\perp/t \rightarrow +\infty$) and a balancing inter-layer repulsion ($V$). Importantly, the double Kondo model does not fundamentally restrict $J_\perp/t$, suggesting the potential for high critical temperatures ($T_c$) approaching $0.5t$. While this idealized limit predicts large pairing gaps, we show a smooth connection to the more realistic regime with $J_\perp \sim t$, albeit with a reduced pairing gap of approximately $0.1t$. Assuming $t \sim 10^3$ K in typical solid state systems, our model suggests the exciting possibility of achieving $T_c$  of hundreds of Kelvin. We propose searching for bilayer materials with reduced out-of-plane lattice constants to better approximate the conditions of our ideal model.
\end{abstract}
\maketitle

\section{Introduction}

Searching for a new mechanism of high-temperature superconductivity is one of the most important problems in condensed matter physics. In the familiar high Tc cuprate materials, the pairing exists only at small doping away from the Mott insulator~\cite{lee2006doping}. As a consequence, the critical temperature ($T_c$) is limited by the phase stiffness $\rho_s$ which is proportional to the doping level $x$~\cite{PhysRevLett.39.1201,Emery1995}. It is tempting to expect a larger $T_c$ if the optimal doping is larger. However, the pairing usually competes with the kinetic energy and thus the pairing gap decreases with the doping $x$. The opposite trend of the phase stiffness and the pairing gap over the doping leads to a small optimal doping level $x_p \approx 0.2$~\cite{lee2006doping}.  In this work, we ask whether it is possible to make both the pairing gap and the phase stiffness increase with the doping $x$ to achieve a larger optimal doping and thus higher $T_c$. Because the kinetic energy increases with $x$, this requires an unprecedented mechanism in which kinetic energy assists pairing. To our best knowledge, kinetic driven superconductor has never been unambiguously established even at the theoretical level.

In this work, we demonstrate the existence of kinetic superconductivity in an ideal model with only a nearest neighbor hopping projected into a constrained Hilbert space. The model can be reached in the strong coupling limit of the double Kondo lattice model~\cite{yang2024strong}. The double Kondo lattice model can be realized in a tetra-layer optical lattice simulator~\cite{yang2024strong} \textcolor{black}{within the ultracold atom
experiment platform~\cite{BOHRDT2021168651, Hirthe2023}} and was also proposed\cite{oh2023type,yang2024strong} by us to describe the recently discovered bilayer nickelate La$_3$Ni$_2$O$_7$ system~\cite{sun2023signatures}.  In the double Kondo model, we have mobile electrons in the two outer-layers, while the inner two layers host only spin moments coupled together by a large antiferromagnetic inter-layer spin-spin coupling $J_\perp$ (see Fig.~\ref{fig:1}).  Then the mobile electron in the outer layer couples to the local moments from the inner layer through a Kondo coupling $J_K$. The Kondo coupling can share the large $J_\perp$ to the outer two layers. As a result, the mobile electrons in the top and bottom layers have an effective strong antiferromagnetic inter-layer spin coupling $\tilde J_\perp$, despite that there is no inter-layer hopping $t_\perp$. 
\\
 \textcolor{black}{\indent Within above double-Kondo model setup, we present the ``\textsl{unambiguous demonstration}'' of superconductivity driven purely by kinetic energy without any net attractive interaction. While one might naively expect interlayer pairing from a mean-field decoupling of an effective interlayer exchange $\tilde{J}_\perp$, this ignores the inevitable presence of interlayer repulsion $V$ in real materials. In realistic scenarios where $V$ is comparable to or larger than $\tilde{J}_\perp$, thus the conventional mean-field theory predicts no superconductivity.
To understand how pairing still arises in this not-attractive interaction setting, it is crucial to separate the two roles of the $J_\perp$ term:
\begin{enumerate}
    \item[(I)] $J_\perp$  lowers the average energy of the doublon and holon states relative to the singlon by $\epsilon_0 \propto -J_\perp<0$, effectively inducing attraction. The resulting repulsive interaction now becomes $V_{\mathrm{eff}} = V + 2\epsilon_0$, which is lowered by $\epsilon_0<0$. 
    \item[(II)] $J_\perp$ introduces a spin splitting between the $S = 0$ and $S = 1$ states of the holon and doublon. For example for doublon states, the energy difference is $\Delta_d = \frac{1}{4}J_\perp$ in the large negative $J_K / J_\perp$ limit. This splitting penalizes spin-triplet configurations and frustrates single-particle hopping.
\end{enumerate}
Here, the doublon ($n_T = 2$), singlon ($n_T = 1$), and holon ($n_T = 0$) refer to the total conduction electrons per rung. Conventional mean-field pairing arises only from effect (I), and thus predicts superconductivity only for $V_{\mathrm{eff}} < 0$. In this work, we are interested in the the case $V > -2\epsilon_0$, where effect (I) is canceled and no pairing is expected from mean-field theory. In contrast to the common wisdom, we actually find strong pairing from a collaboration between the kinetic energy and the restricted Hilbert space from  the effect (II): Because the $\Delta_d$ splitting penalizes spin-triplet states, single-electron motion is frustrated. Remarkably, it turns out in the large $\Delta_d$ limit, forming a Cooper pair can lower the kinetic energy within the constrained low-energy subspace.  We illustrate this kinetic pairing mechanism by first taking the limit $\Delta_d \to \infty$, where the hopping term is projected into the low-energy Hilbert space, and then show that the pairing persists even for finite $\Delta_d$.  We will focus on the ideal limit of $V_{\mathrm{eff}}=0$, when the net attraction from $J_\perp$ is exactly canceled by $V$. But we emphasize that the superconductor exists in a wide range of $V_{\mathrm{eff}}$.  At exactly $V_{\mathrm{eff}}=0$, the kinetic pairing mechanism from the effect (II) plays the dominant role. Once we turn on a negative $V_{\mathrm{eff}}$, a net attraction obviously enhances pairing.  On the other hand, a finite net repulsion $V_{\mathrm{eff}}$ suppresses the pairing, but we will show that the superconductivity survives to a reasonably large value of positive $V_{\mathrm{eff}}$. 
}

\textcolor{black}{Our conceptual focus on the $V_{\mathrm{eff}}=0$ limit aims to unambiguously establish the mechanism of kinetic pairing. This is challenging because typical models combine kinetic (hopping) and interaction terms, making it difficult to isolate purely kinetic contributions. A similar difficulty arises in kinetic magnetism~\cite{PhysRev.147.392,haerter2005kinetic,PhysRevResearch.6.013307}, where distinguishing kinetic effects from interaction-driven magnetism (like Stoner) in models like the Hubbard model is hard. Nagaoka's seminal work~\cite{PhysRev.147.392} resolved this by demonstrating ferromagnetism in a model containing \textit{only} hopping within a constrained Hilbert space, definitively proving its kinetic origin. Analogously, we argue the clearest path to establishing kinetic pairing is to demonstrate superconductivity in a similar hopping-only model within a constrained space --- something not previously achieved, to our knowledge. Following this principle, we project the double Kondo model into a restricted Hilbert space by taking the strong coupling limit ($|J_K|, J_\perp \gg t$). This regime's low-energy physics is captured by the ESD model~\cite{yang2024strong}, comprising nearest-neighbor hopping ($t$) and an on-site interaction $\epsilon \propto V_{\mathrm{eff}}$ within a space of empty (E), single (S), and double (D) occupancy states (requiring $J_\perp > 0$). By setting $\epsilon=0$, analogous to Nagaoka's approach, we isolate kinetic effects. This allows for an unambiguous identification of kinetic pairing mechanisms, distinct from interaction-driven ones. Furthermore, tuning $\epsilon$ permits studying the interplay with net interactions, from attractive to repulsive. Moreover, Several interesting theoretical proposals have been made to account for high critical temperatures of bilayer nickelate.
}

\textcolor{black}{Using both analytical and numerical approaches, we demonstrate that the ground state of our model is superconducting over a wide doping range $x$. In the dilute doping limit, we provide an \textit{exact solution} using exact diagonalization, analogous in spirit to Nagaoka's ferromagnetism argument. For general doping, we perform density matrix renormalization group (DMRG) simulations. At low hole doping levels such as $x = \frac{1}{8}$, we observe robust power-law decay of pair-pair correlations with a small exponent, even for systems with width $L_y = 8$. We find that both the pairing gap and phase stiffness scale with $t x$ near $x = 0$ and $t(1 - x)$ near $x = 1$. The pairing gap can exceed $1.5t$, and the estimated critical temperature reaches up to $T_c \sim 0.5t$.
Unlike previous DMRG studies on the conventional $t-J$ model~\cite{PhysRevB.92.195139,PhysRevB.99.235117,PhysRevB.53.251,PhysRevB.62.R14633,PhysRevB.102.115136,PhysRevB.55.R14701,PhysRevB.60.R753,Huang2018,PhysRevB.98.140505,PhysRevB.95.155116,PhysRevResearch.2.033073,PhysRevLett.127.097002,PhysRevLett.127.097003,doi:10.1073/pnas.2109978118,PhysRevX.4.031040,Jiang2021,PhysRevLett.125.157002,PhysRevLett.119.067002,https://doi.org/10.1002/qute.202000126,PhysRevLett.88.117001,PhysRevB.64.100506,PhysRevB.79.220504,PhysRevLett.113.046402,PhysRevLett.122.167001,PhysRevLett.80.1272,grusdt2018meson,chen2018two,zhu2015quasiparticle,zhu2015charge,sun2019localization,zhu2016exact,chen2024phase,wang2015variational}, we find no evidence of stripe order or other translation-symmetry-breaking phases in our model. This absence of competing orders provides clear support for a kinetically-driven superconducting ground state and suggests a route to very high $T_c$, potentially scaling with the Fermi energy.
}

We offer intuitions on why the electrons can pair up to minimize the kinetic energy. 
Along with the exact-diagonalization method of the two-electron problem, we provide analytical treatments in a special limit with a layer-opposite Zeeman field. In these special cases, we can really demonstrate pairing in the two-dimensional (2D) limit.  For the general case, we do not have an exact analytical solution, but we show that a generalized slave boson theory is in qualitative agreement with the DMRG results.  In our model, we have two different carriers, the doublon $\ket{d}$ and the holon $\ket{h}$ states with charges $+1$ and $-1$ respectively. The $x=0$ and $x=1$ limits correspond to a product of $\ket{d}$ state and a product of $\ket{h}$ state respectively. The normal states close to $x=0$ and $x=1$ are two different Fermi liquids (FL), one of which violates the conventional Luttinger theorem of the Fermi volume by half of the Brillouin zone per flavor~\cite{PhysRevLett.84.3370,yang2024strong}. Interestingly the model can be adjusted to have a particle-hole symmetry relating these two FLs at $x=0.5$. As a consequence a symmetric FL is simply forbidden at $x=0.5$ and the superconductor can not be viewed as a descendant of a Fermi liquid. This challenges the conventional theoretical frameworks of superconductivity. We offer a variational wavefunction based on our generalized slave-boson theory at the doping $x=0.5$ to demonstrate the possibility of superconductivity without the need of a Fermi liquid normal state.

Although the model is achieved at the limit $J_\perp/t\gg1$, we show that the superconductor can survive to the more realistic regime with $J_\perp \sim t$. We also demonstrate that the superconductor is robust when increasing $V$ so that there is a net repulsive interaction. Therefore, we believe that it is promising to realize the mechanism in a real material, such as in a bilayer nickelate system similar to the recently studied La$_3$Ni$_2$O$_7$.  Because the theoretical upper bound of $T_c$ in this class of model is at the order of $0.5 t$, it may not be unimaginable to achieve a $T_c$ at least at the order of $0.1 t$ by enhancing $J_\perp$ in real systems, for example through reducing the z-axis lattice constant. Given that $t$ is typically thousands Kelvin in solid state systems, the model may offer a plausible route toward room-temperature superconductivity.

The rest of the paper is organized in the following way. In Sec.\ref{sec:2}, we introduce the double Kondo lattice model and the ESD model in the strong coupling limit. In Sec.\ref{sec:3}, we show DMRG evidence for superconductivity in the ESD model with only the hopping term. Sec.\ref{sec:4} provides an exact solution to the two-electron and two-hole problems of the ESD model in the 2D limit. 

\textcolor{black}{In Sec.\ref{sec:6}, we formulate the fermion-boson theory for the dilute limit, and the slave boson mean field theory for general doping showing that the mean-field theory captures the phase diagram in qualitative agreement with our DMRG results.} In Sec.\ref{sec:7}, we demonstrate that the superconductor survives to the realistic regime of $J_\perp/t$ for two full models with large negative $J_K$ and large positive $ J_K$. We emphasize that the bilayer nickelate is described by a bilayer type II t-J model instead of a simple bilayer one-orbital model. In Sec.\ref{sec:8}, we provide some insights on the pairing mechanism and propose a route to experimentally search for a bilayer nickelate system with a reduced z-axis lattice constant to closely approach our ideal model. We summarize our results and conclude the paper in Sec.\ref{sec:9}.

\section{Model}\label{sec:2}
\subsection{The double Kondo model}

\begin{figure}[ht]
\centering
\includegraphics[width=\linewidth]{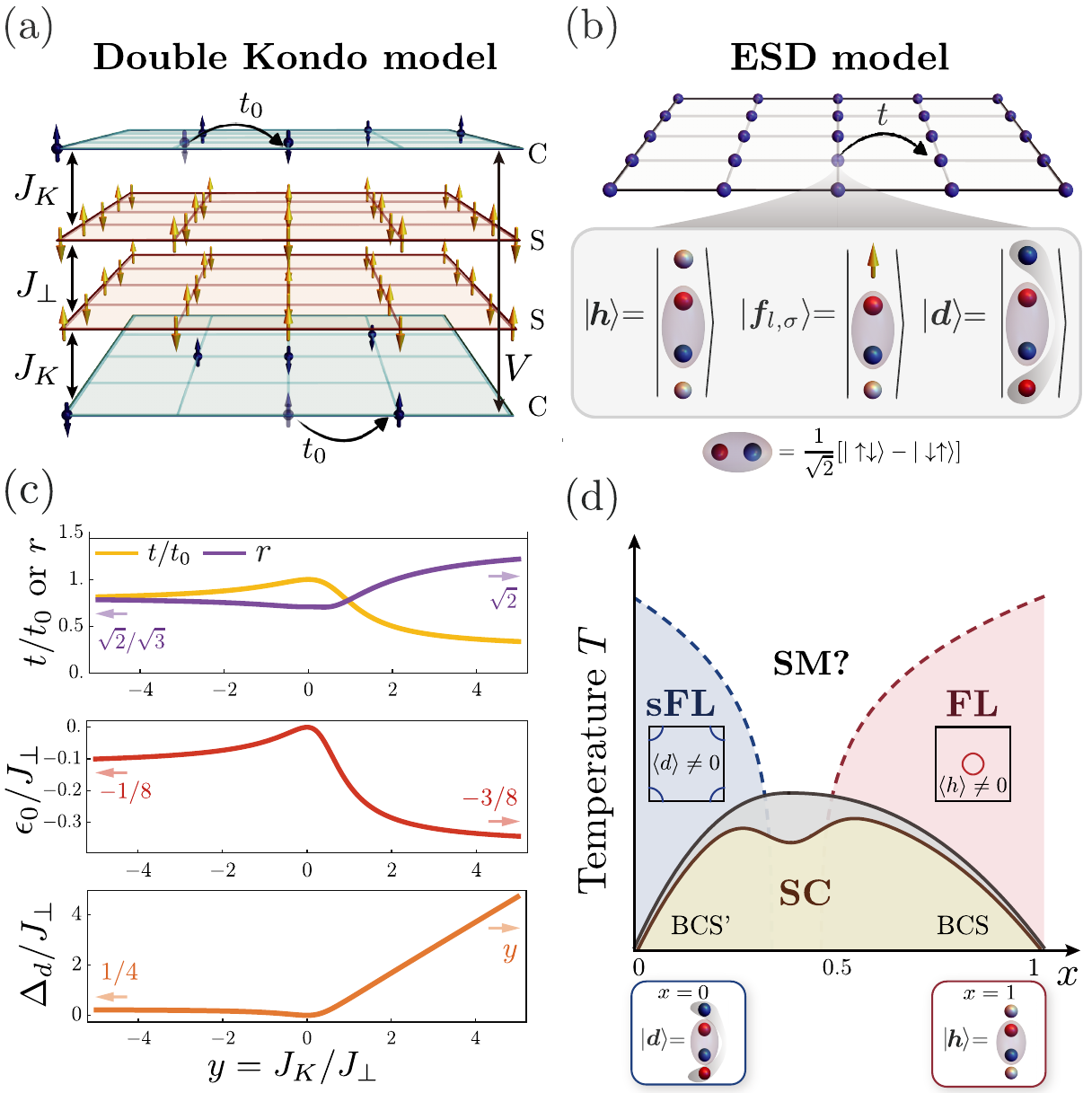}
\vspace{-15pt}
\caption{\textbf{(a,b) Illustration of the double Kondo model and the ESD model.}
(a) The double Kondo model consists of two copies of the conventional spin-$1/2$ Kondo model, with conduction electrons (C-layer) and localized moments (S-layer) interacting through the Kondo coupling $J_K$. We use the labels $l = t, b$ to distinguish the top and bottom layers. The two S-layers are coupled by an on-site inter-layer antiferromagnetic coupling $J_\perp$. \textcolor{black}{We add repulsive interaction $V$ between C-layers.} 
(b) In the strong-coupling limit ($|J_K|, J_\perp \gg t$), the double Kondo model reduces to a simpler model termed the ESD model. The ESD model retains only six states on each rung, which combines the two layers: one empty (E) state $|h\rangle$, four singlon (S) states $|f_{l\sigma}\rangle$, and one doublon (D) state $|d\rangle$. The inset illustrates these six states in the specific limit $|J_\perp| \gg |J_K|$; however, in general, they possess complex internal structures that depend on the ratio $J_K/J_\perp$.
\textbf{(c) Dependence of the parameters in the ESD model.}
The figure shows the dependence of $t/t_0$, $r$, $\epsilon_0$, and $\Delta_d$ on $J_K/J_\perp$. In the limit $J_K \to -\infty$, $r \to \sqrt{2}/\sqrt{3}$, while $r \to \sqrt{2}$ in the $J_K \to +\infty$ limit. Under the particle-hole transformation, $r$ transforms to $1/r$, with $x$ transforming to $1 - x$, where $x$ is the hole doping per site per layer.
\textbf{(d) Schematic phase diagram of the ESD model.}
The solid black and brown lines indicate the pairing and coherence temperature scales, respectively. In our ESD model, a superconducting pairing dome extends from two distinct Fermi liquids: the conventional Fermi liquid (FL) near $x = 1$ and a second Fermi liquid (sFL) near $x = 0$. The inset illustrates the hole pocket around $\vec{k} = (\pi,\pi)$ and the electron pocket around $\vec{k} = (0,0)$ for the sFL and FL, respectively. The nature of the normal state near $x = 0.5$ remains an open question.
}
\label{fig:1}
\end{figure}

We consider a double Kondo model, consisting of two separate spin-1/2 Kondo models in two layers coupled by an inter-layer spin-spin coupling, $J_\perp$. Each Kondo model involves itinerant electrons coupled to localized spin moments through the Kondo coupling, $J_K$ (see Fig.~\ref{fig:1}(a)).

The Hamiltonian of the double Kondo model is given by
\begin{align}
\hat{H}_{\mathrm{DK}} = &-t_0 \sum_{l=t,b} \sum_{\langle i,j \rangle} P c^\dagger_{i;l;\sigma} c_{j;l;\sigma} P + V \sum_{i} n_{i;t} n_{i;b} \nonumber \\
&+ J_{K} \sum_{i} \sum_{l=t,b} \vec{s}_{c;i;l} \cdot \vec{S}_{i;l} + J_{\perp} \sum_{i} \vec{S}_{i;t} \cdot \vec{S}_{i;b},
\label{eq:Ham_double_kondo}
\end{align}
where $c_{i;l;\sigma}$ is the creation operator for a conduction electron at site $i$ in layer $l$ with spin $\sigma$, and $\vec{S}_{i;l}$ denotes the localized spin moment at site $i$ in layer $l$. Here, $l=t$ ($l=b$) labels the top (bottom) layer.

The spin operator of the itinerant electron is defined as $\vec{s}_{c;i;l} = \frac{1}{2} \sum_{\sigma,\sigma'} c^\dagger_{i;l;\sigma} \vec{\sigma}_{\sigma \sigma'} c_{i;l;\sigma'}$, where $\vec{\sigma}$ represents the vector of Pauli matrices and $\sigma = \uparrow, \downarrow$. We consider the regime of strong $J_\perp$, where the two Kondo models are strongly coupled, despite the absence of direct inter-layer hopping ($t_\perp = 0$). We also neglect intra-layer spin-spin couplings, as they do not qualitatively alter the physics in the large-$J_K$, large-$J_\perp$ limit considered here. The projection operator $P$ is introduced to prevent double occupancy of conduction electrons at each site, effectively modeling a large Hubbard $U$. However, in the limit $J_K, J_\perp \gg t_0$, the Hubbard U is not essential. The strong $J_K$ and $J_\perp$ terms sufficiently suppress double occupancy within each layer, so the same physics can be obtained in a model without the projection. The model has a $(U(1)_t \times U(1)_b \times SU(2)_S)/\mathbb{Z}_2$ symmetry, meaning that the top and bottom layer has separate charge conservation, while sharing the same spin rotation symmetry.

\subsection{The ESD  model in the strong coupling limit}
\begin{figure}[h]
    \centering
\includegraphics[width=1\linewidth]{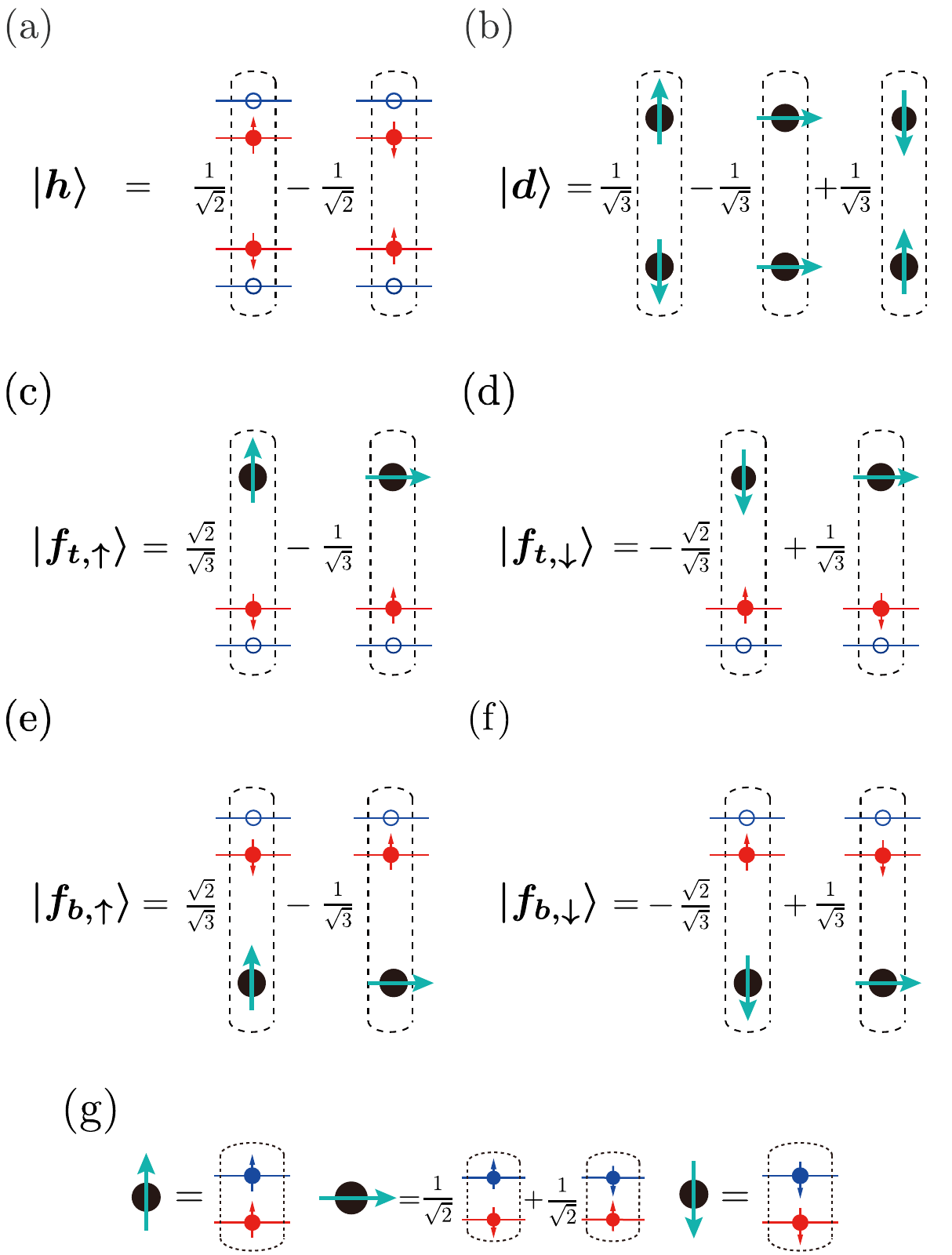}
\caption{\textbf{Illustration of the six states in the ESD-$t$-$J$ model in the $J_K/J_\perp \to -\infty$ limit.}
The black/red circles denote an electron in the C/S layer of the double Kondo model, and empty circles represent empty states. In the context of bilayer nickelates, the black/red circles correspond to the $d_{x^2-y^2}$/$d_{z^2}$ orbitals. A black circle with a green arrow denotes a spin-one moment arising from occupation of both orbitals. (a-f) The wave functions of the $|h\rangle$, $|d\rangle$, and $|f_{l,\sigma}\rangle$ states are shown. We emphasize that the $d_{z^2}$ orbital (the S layer in the double Kondo model) remains active. For example, $|f_{t,\uparrow}\rangle$ is not simply a spin-up electron in the $d_{x^2-y^2}$ orbital combined with a rung singlet of $d_{z^2}$ orbitals; it also includes a component of a polaron state. This polaron state comprises a spin-down electron in the $d_{x^2-y^2}$ orbital combined with an $S = 1$ excitation that breaks the rung singlet of the $d_{z^2}$ spin moments. (g) The spin-triplet states formed by two electrons in the C and S layers (the two $e_g$ orbitals of each nickel atom) are illustrated. Each $S = 1$ state is labeled by its $S_z$ value ($+1$, $0$, and $-1$). 
}
    \label{fig:2}
\end{figure}

In the strong coupling limit ($|J_K|, J_\perp,V \gg \textcolor{black}{t_0}$), the double Kondo model reduces to a simpler model, the so-called \textit{ESD model} (see Fig.~\ref{fig:1}(b)) \cite{yang2024strong}. In this limit, we first analytically solve the $J_K$ and $J_\perp$ terms on each rung (labeled by $i$) and then treat the hopping term $t$ as a first-order perturbation within the resulting degenerate subspace. Each site $i$ here represents a super-site encompassing all four layers shown in Fig.~\ref{fig:1}(a). As detailed in Appendix~\ref{sec:s1}, we need only to consider six states, classified by the total number of electrons ($n_T$) and the total spin ($S_T$):

\begin{itemize}
    \item Empty state $|h\rangle$: $n_T = 0$, $S_T = 0$.
    \item Singlon states $|l, \sigma\rangle$: $n_T = 1$, $S_T = 1/2$.
    \item Doublon state $|d\rangle$: $n_T = 2$, $S_T = 0$.
\end{itemize}
There is one empty state, four {singlon} states (specified by the layer index $l = t, b$ and spin index $\sigma=\uparrow,\downarrow$), and one doublon state. The acronym ESD stands for Empty (E), {Singlon} (S), and Doublon (D). The Hilbert space of our ESD model thus comprises these six states. The microscopic internal structures of these states, which depend on $J_K/J_\perp$, are derived in Appendix~\ref{sec:s1}. We emphasize that these internal structures can be complex and vary with $J_K/J_\perp$. In the large, negative $J_K$ limit relevant to bilayer nickelates, the wave functions of these six states are given in Fig.~\ref{fig:2}. Within our low-energy ESD model, these detailed internal structures are not important; we can simply label the states by their quantum numbers $n_T$ and $S_T$.

Projecting the double Kondo model Hamiltonian, Eq.~(\ref{eq:Ham_double_kondo}), onto the restricted Hilbert space leads to the ESD model:
\begin{align}
\hat{H}_{\mathrm{ESD}} = -t \sum_{l,\textcolor{black}{\sigma}\langle i,j \rangle} P c^\dagger_{i;l;\sigma} c_{j;l;\sigma} P + \textcolor{black}{V_{\mathrm{eff}}} \sum_{i} (n_{h;i} + n_{d;i}),
\label{eq:Ham_esd}
\end{align}
with the density operators $n_{h;i} = |h\rangle_i \langle h|_i$ and $n_{d;i} = |d\rangle_i \langle d|_i$. Here, $\textcolor{black}{V_{\mathrm{eff}}}$ characterizes the net interaction, with $\textcolor{black}{V_{\mathrm{eff}}} < 0$ corresponding to an attractive interaction and $\textcolor{black}{V_{\mathrm{eff}}} > 0$ corresponding to a repulsive interaction. We emphasize that the hopping $t$ is rescaled compared to $t_0$ in the double Kondo model, depending on $J_K/J_\perp$. \textcolor{black}{ We assume mirror reflection symmetry that exchanges the two layer remains unbroken, and therefore the two layers have equal densities.}

Within the ESD Hilbert space, the creation (annihilation) operator is given by
\begin{align}
c_{i,l,\uparrow} &= |h\rangle_i \langle l,\uparrow|_i + r |\bar{l},\downarrow\rangle_i \langle d|_i, \label{eq:electron1} \\
c_{i,l,\downarrow} &= |h\rangle_i \langle l,\downarrow|_i - r |\bar{l},\uparrow\rangle_i \langle d|_i. \label{eq:electron2}
\end{align}
where $r$ is a parameter that depends on $J_K/J_\perp$ in the original double Kondo model.

\begin{figure}
    \centering
\includegraphics[width=\linewidth]{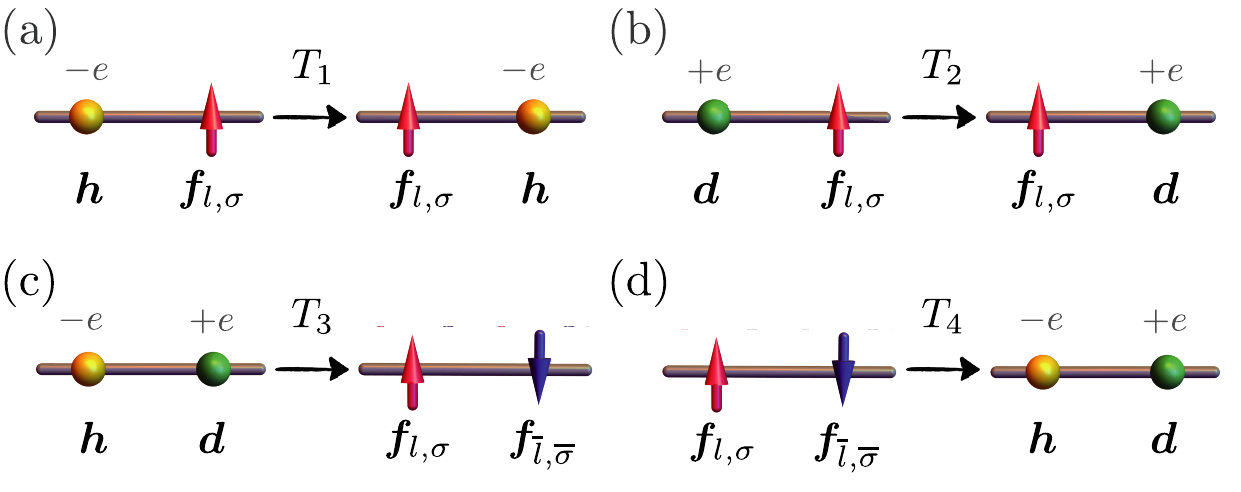}
    \caption{\textbf{Schematic illustrations of four distinct hopping processes between two adjacent sites.}
The hopping processes can be categorized into four types. $T_1$ and $T_2$ represent exchange processes between $|h\rangle$ and $|f\rangle$ (or $|d\rangle$ and $|f\rangle$), while $T_3$ and $T_4$ describe processes in which a pair of $|h\rangle$ and $|d\rangle$ states on a nearest-neighbor bond are converted into a pair of $|f_{l;\sigma}\rangle$ and $|f_{\bar{l},\bar{\sigma}}\rangle$ states, and vice versa. The $T_3$ and $T_4$ processes are unique to the ESD model and contribute to kinetic pairing.
}
    \label{fig:5}
\end{figure}

Moreover, we have the constraint $n_{d;i} + n_{f;i} + n_{h;i} = 1$ at each site $i$. On average, we have $\frac{1}{2} n_f + n_d = x$, with the hole doping per site defined as $n_T = 2(1 - x)$. The factor of 2 arises from the two layers, and the density per layer is $1 - x$. From $n_T = n_f + 2n_d$, one can derive $n_T = 1 + n_d - n_h$. Thus, the $d$ and $h$ states can be interpreted as carrying charges $+1$ and $-1$, respectively, while the singlon state is neutral and represents a spin moment. In contrast to the conventional $t$-$J$ model, which has carriers with either $+1$ or $-1$ charge, here we have both. In Fig.~\ref{fig:5}, we illustrate the hopping term. The hopping can be divided into two categories: (1) exchange of a singlon state ($f$) and a nearby $h$ or $d$ state; (2) annihilation of a pair of $d$ and $h$ states to create a pair of $f$ states, and the reverse process. Category (1) also exists in the conventional $t$-$J$ model, but category (2) is unique to our ESD model. As we will see later, this leads to interesting new physics not present in the familiar $t$-$J$ model.

The model has three parameters: $t$, $r$, and $\textcolor{black}{V_{\mathrm{eff}}}$. The parameters $t$ and $r$ are determined by the internal structure of the six states, while $\textcolor{black}{V_{\mathrm{eff}}}$ represents the net interaction arising from both the $J_\perp$ and $V$ terms. Specifically, $\textcolor{black}{V_{\mathrm{eff}}} = \epsilon_0 + V/2$, where
 \begin{eqnarray}
  \epsilon_{0}&=&
   \frac{J_\perp}{4} \left[-1 - 
   \sqrt{1 -2\frac{J_K}{J_\perp}+ 4\left(\frac{J_K}{J_\perp}\right)^2}\right. \nonumber
   \\
   &&\quad 
   \left. +2 \sqrt{1 -\frac{J_K}{J_\perp}+\left( \frac{J_K}{J_\perp}\right)^2}\right].
\end{eqnarray}
Figure~\ref{fig:1}(c) shows the dependence of $t/t_0$, $r$, and $\epsilon_0/J_\perp$ on $J_K/J_\perp$. For simplicity, we set $t = 1$ throughout this study. We primarily focus on the case where $\textcolor{black}{V_{\mathrm{eff}}} = 0$, implying that the $V$ term exactly cancels the attraction from the $J_\perp$ term. In this case, our ESD model contains only one term: nearest-neighbor hopping.
We also note that, because we consider finite $\textcolor{black}{V_{\mathrm{eff}}}$, the model is fermionic, and the singlon states play a crucial role. Therefore, it does not reduce to a simple hard-core boson theory justified only in the limit $\textcolor{black}{V_{\mathrm{eff}}} \to -\infty$. \textcolor{black}{The $T_c$ analysis in the hard-core bosonic theory corresponding to $\textcolor{black}{V_{\mathrm{eff}}} \rightarrow -\infty$ in our model is well studied in the previous literature \cite{Schlömer2024,Carrasquilla_2013}.}

\textbf{Particle-hole symmetry at $r=1$} 
We can define a particle-hole (PH) transformation, which transforms the six states in the following way:
$\ket{h}_i\rightarrow\ket{d}_i$,  $\ket{d}_i\rightarrow-\ket{h}_i$, $\ket{l,\uparrow}_i\rightarrow(-1)^i\ket{\overline{l},\downarrow}_i$ and $\ket{l,\downarrow}_i\rightarrow-(-1)^i\ket{\overline{l},\uparrow}$. where $(-1)^i$ means $(-1)^{x+y}$ and $\bar l$ is the opposite layer of $l$.

Under this particle-hole transformation, the creation operators transform as
\begin{align}
c_{i,l,\uparrow} &\rightarrow (-1)^i (|d\rangle_i \langle \bar{l},\downarrow|_i + r|l,\uparrow\rangle_i \langle h|_i), \label{eq:electron3} \\
c_{i,l,\downarrow} &\rightarrow (-1)^i (-|d\rangle_i \langle \bar{l},\uparrow|_i + r|l,\downarrow\rangle_i \langle h|_i). \label{eq:electron4}
\end{align}
One can see that the PH transformation maps $r$ to $1/r$, in addition to $x \rightarrow 1 - x$. At the special parameter value $r = 1$, we have $c_{i,l,\sigma} \rightarrow (-1)^i c_{i,l,\sigma}^\dagger$. On a square lattice, the ESD model with $r = 1$ exhibits a particle-hole (PH) symmetry, relating $x$ to $1-x$.

{
From the ESD model at $V_\mathrm{eff}=0$, we can gain an intuitive picture of the pairing mechanism driven by kinetic energy. 
The essential ingredient here is the Hilbert-space restriction imposed by large $J_\perp$, where spin–triplet doublons is strongly penalized by $\Delta_d\gg 0$ from spin–singlet doublons ($d$ state).  
As a result, single-electron hopping is frustrated [Fig.~\ref{fig:8}(a)], since it inevitably involves entering the spin-triplet sector. 
By contrast, a spin-singlet Cooper pair can propagate lowering the kinetic energy [Fig.~\ref{fig:8}(b)].
Thus, the ground state becomes superconducting state in stead of a Fermi liquid to lower the kinetic energy.
For completeness, we note that $J_\perp$ also induces an effective attraction that lowers the doublon energy, $\epsilon_{0}<0$, in Fig.~\ref{fig:1}(c). 
But here, we set $V_{\mathrm{eff}}=0$, by introducing a repulsion $V$ that exactly cancels this attraction, thus safely isolate the Hilbert-space–restriction effect alone.  
In the sections that follow, we confirm this picture using comprehensive analysis.
}
\begin{figure}[b]
    \centering
\includegraphics[width=\linewidth]{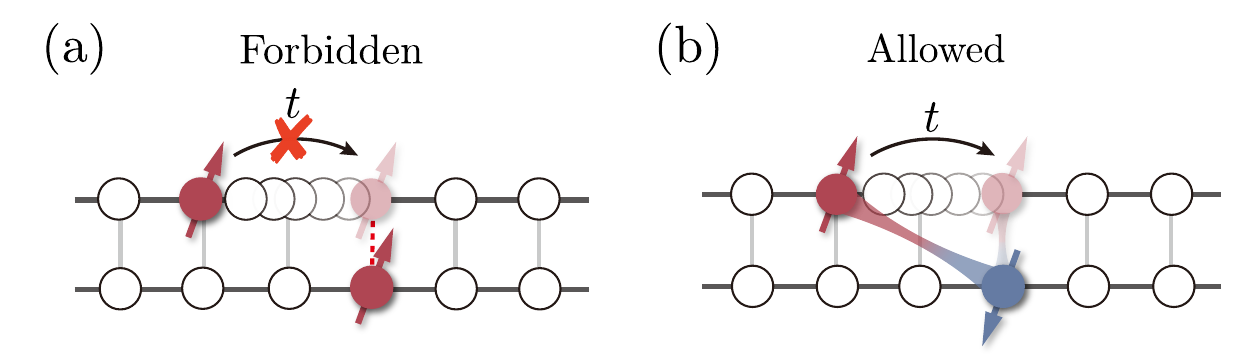}
\caption{{\textbf{Pairing mechanism from kinetic energy in ESD model.} 
(a) When a single-electron hops, it always access the spin-triplet sector (red-dashed line), which should be excluded by our Hilbert-space restriction. Thus, this process is forbidden, and cannot lower kinetic energy.
(b) In contrast, if the two electrons form a spin-singlet pair, the object can move freely without accessing the triplet sector, thereby can lowering the kinetic energy.
This illustrates how kinetic energy alone can drive pairing within a restricted Hilbert space.
}}
    \label{fig:8}
\end{figure}

\begin{figure}[tb]
    \centering
\includegraphics[width=1.05\linewidth]{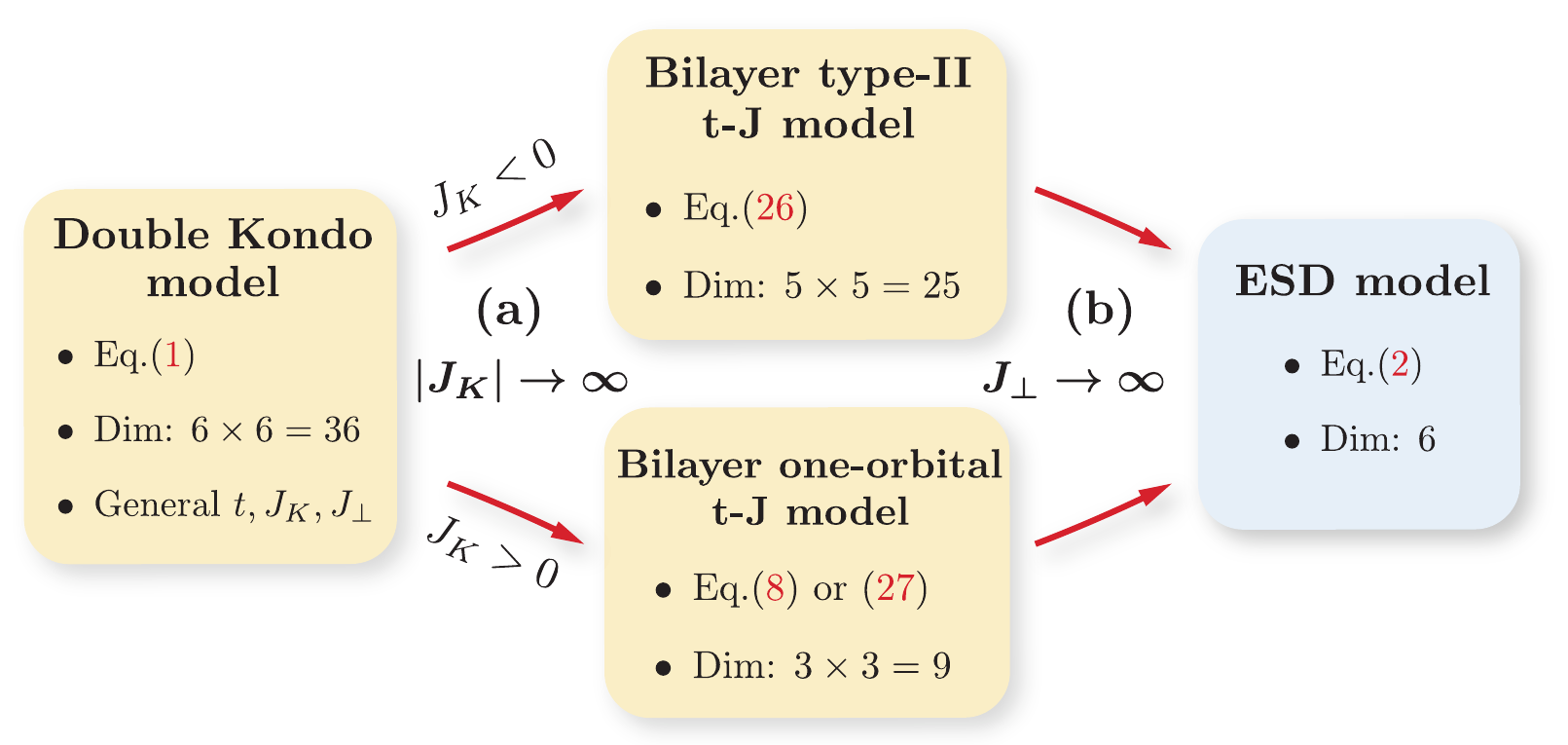}\vspace{-10pt}
    \caption{\textcolor{black}{\textbf{Summary of various theoretical models discussed in this work.} The ESD model can be viewed as a descendant of the double Kondo model. 
(a) By taking the limit $|J_{K}| \rightarrow \infty$, the double Kondo model maps onto either the bilayer type-II $t$-$J$ model ($J_K<0$) or the bilayer one-orbital $t$-$J$ model ($J_K>0$).
(b) Both of these models further reduce to the ESD model in the strong coupling limit $|J_K|, J_\perp \gg t$. 
The arrows in the schematic indicate the flow toward more effective low-energy models, by a progressive reduction in the Hilbert space dimension per rung.}}
    \label{fig:chart}
\end{figure}

\textcolor{black}{\subsection{The bilayer one-orbital t-J model}
We can also realize the ESD model in a simpler model, the bilayer one-orbital $t$–$J$ model. This model is defined on a bilayer square lattice as
\begin{align}
    \hat{H}=&-t \sum_{l,\sigma,\langle ij \rangle}Pc^\dagger_{i;l;\sigma}c_{j;l;\sigma}P+\mathrm{h.c.}
    \nonumber\\
    &+V\sum_{i} n_{i;t}n_{i;b}
    +J_\perp \sum_i \vec s_{i;t}\cdot \vec s_{i;b},
    \label{eq:Ham_one_orbit_t_J}
\end{align}
where $P$ is the projection operator enforcing no double occupancy. Here, $l = t, b$ denotes the top and bottom layers, while $i, j$ label lattice sites. The model also possesses a $(U(1)_t \times U(1)_b \times SU(2)_S)/\mathbb{Z}_2$ symmetry, reflecting separate charge conservation in each layer and global spin rotation symmetry. 
This model has been previously discussed in the context of bilayer optical lattices such as in mixed-dimensional setup \cite{Hirthe2023}, where the interlayer interaction $V$ was neglected—effectively leading to a net attractive interaction. In contrast, our work explicitly includes a repulsive interlayer interaction $V$, which can be comparable in strength to $J_\perp$. This allows us to explore pairing mechanisms even in regimes where the net interaction is not attractive.  We stress that the normal state of the bilayer one-orbital t-J model is remarkable as demonstrated later. We remark that though the bilayer one-orbital t-J model is naturally realized in the bilayer system, but at the same time, it can be derived from the double Kondo model in the $J_K\rightarrow \infty$ limit. 
\\
\\
In the strong interlayer interaction limit ($J_\perp \gg t$), the low-energy Hilbert space is again restricted to six states per each rung site: the holon ($h$), doublon ($d$), and four singlon states $f_{l,\sigma}$. In this regime, the effective model reduces to the ESD model given in Eq.(\ref{eq:Ham_esd}). Through straightforward algebra, we find that the parameter $\textcolor{black}{V_{\mathrm{eff}}} = \frac{1}{2}(V - \frac{J_\perp}{4})$, and the coefficient $r = \frac{1}{\sqrt{2}}$ should be used in Eqs.(\ref{eq:electron1}–\ref{eq:electron2}). In Fig. \ref{fig:chart}, we summarize the various theoretical models discussed in this work together.  Because both the double Kondo model and this bilayer one-orbital model flow to the ESD model in the large $J_\perp$ limit, they should share the same physics in the strong coupling regime.
}

\section{Numerical evidence of superconductivity from DMRG}\label{sec:3}
To investigate the possibility of superconductivity in the ESD model, we performed density matrix renormalization group (DMRG) simulations \cite{PhysRevLett.69.2863} using the TeNPy package \cite{tenpy}. DMRG calculations are typically restricted to quasi-one-dimensional systems, and we approach the two-dimensional limit by increasing $L_y$. We achieved $L_y = 6$ for finite-system DMRG and $L_y = 8$ for infinite-system DMRG at small hole doping. Throughout this paper, we primarily employ open boundary conditions along the $x$-direction unless otherwise specified. In the $y$-direction, we use open boundary conditions (OBC) for $L_y = 2$ and periodic boundary conditions (PBC) for $L_y > 2$. We also exploit the three $U(1)$ quantum numbers $N_t, N_b, S_z$ corresponding to the total charges in the top, and bottom layer and the total spin $S_z$.

In the following, we present evidence for a Luther-Emery liquid ground state \cite{lutheremergy}, a precursor to superconductivity in quasi-one dimension, across a broad doping range $x$ for the ESD model at $\textcolor{black}{V_{\mathrm{eff}}} = 0$. This evidence includes a finite spin gap and charge-1e gap, a vanishing charge-2e gap, power-law pair-pair correlations, and finite phase stiffness. We note that calculations around $x \sim 0.5$ are significantly more challenging than those for $x$ close to $0$ or $1$. Consequently, for $L_y \geq 4$, technical limitations prevent us from definitively concluding that the ground state is superconducting in the region around $x = 0.5$, although it is clear that the single-electron charge gap remains finite and large there. The key question is whether Cooper pairs condense. Conversely, for small doping such as $x = 1/8$, we are confident that the ground state is superconducting, even in the two-dimensional limit.

\begin{figure}[tb]
    \centering
\includegraphics[width=\linewidth]{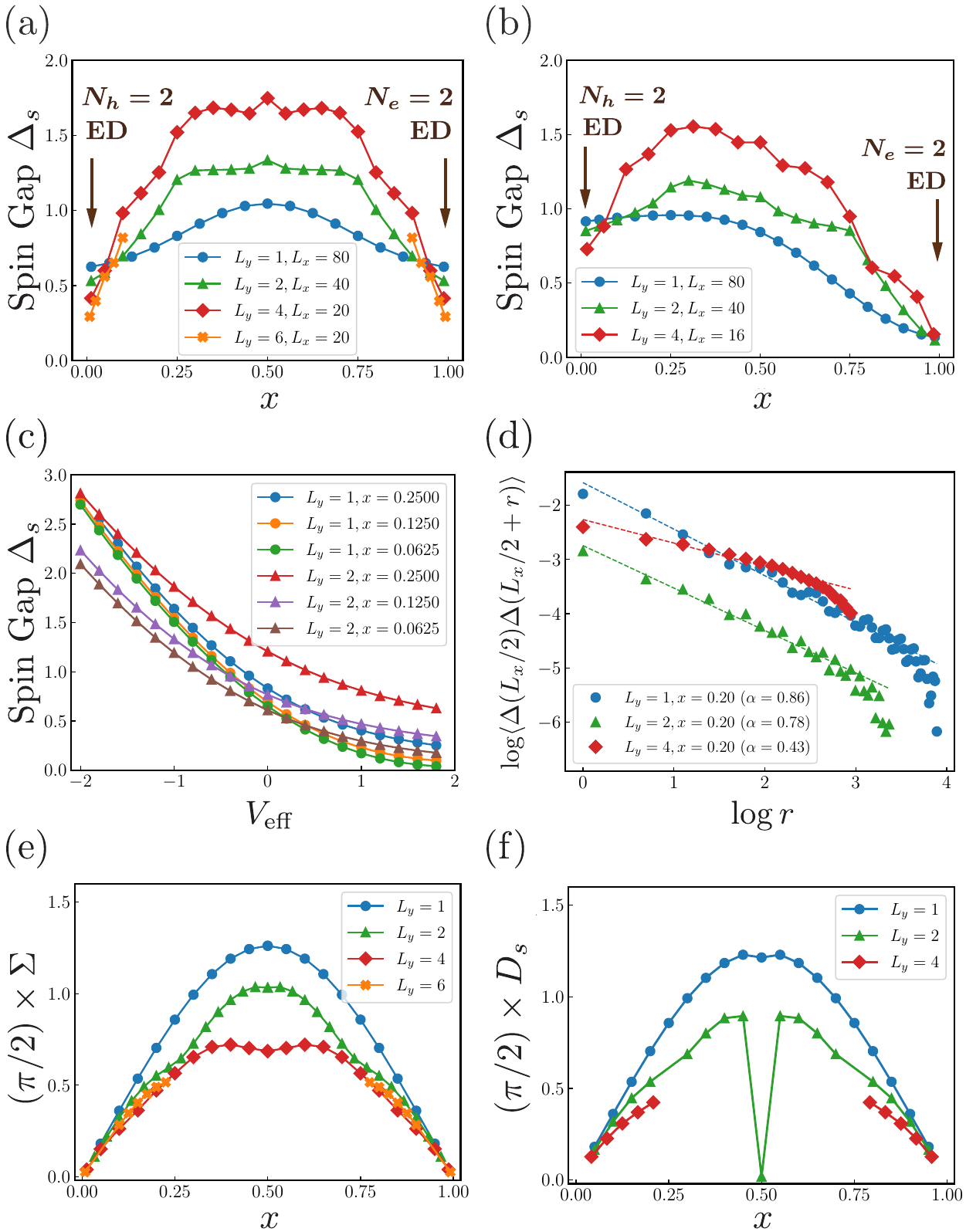}
\vspace{-18pt}
\caption{\textbf{Finite-system DMRG results for the ESD model, Eq.~(\ref{eq:Ham_esd}).} All energy scales are given in units of $t = 1$.
(a) Doping dependence of the spin gap at $r = 1$ and $\textcolor{black}{V_{\mathrm{eff}}} = 0$ for different system sizes ($L_y = 1, 2, 4, 6$) with bond dimension $\chi = 3000$. A dome-like behavior of the spin gap as a function of $x$ is consistently observed, although the quantitative values vary with system size $L_y$. In the limits $x \to 0$ and $x \to 1$ (indicated by arrows), the DMRG results agree with the \textbf{exact solution} values (see Table~\ref{table:1}). Due to numerical limitations, converged results for $L_y = 6$ are obtained only for $x < 0.1$ and $x > 0.9$.
(b) Doping dependence of the spin gap at $r = \sqrt{2}/\sqrt{3}$ and $\textcolor{black}{V_{\mathrm{eff}}} = 0$, corresponding to the type-II $t$-$J$ model. The bond dimension used was $\chi = 3000$. The dome is shifted to lower doping compared with the results at $r = 1$.
(c) Dependence of the spin gap on $\textcolor{black}{V_{\mathrm{eff}}}$ at $r = 1$. Different markers denote different system sizes: $L_y = 1$, $L_x = 80$ (circles) and $L_y = 2$, $L_x = 40$ (triangles).
(d) Pair-pair correlation function at $x = 0.2$ for various $L_y = 1, 2, 4$ with $\chi = 3000$, using $r = 1$ and $\textcolor{black}{V_{\mathrm{eff}}} = 0$. All pair correlation functions exhibit power-law decay with an exponent $\alpha$ that decreases with increasing $L_y$.
(e,f) Doping dependence of the total weight $\Sigma$ and the superfluid weight $D_s$ at $r = 1$ and $\textcolor{black}{V_{\mathrm{eff}}} = 0$. We use $L_x = 40$ ($L_y = 1$), $L_x = 30$ ($L_y = 2$), and $L_x = 20$ ($L_y = 4, 6$) with $\chi = 3000$. Open boundary conditions (OBC) are used in the $x$-direction for $\Sigma$ calculations, while periodic boundary conditions (PBC) are used for $D_s$ calculations.
    }
    \label{fig:3}
    \vspace{-20pt}
\end{figure}

\textbf{Finite spin gap}
Figure~\ref{fig:3}(a) shows the doping dependence of the spin gap, defined as the energy difference $\Delta_s = E[S_z=1] - E[S_z=0]$, at $r = 1$ and $\textcolor{black}{V_{\mathrm{eff}}} = 0$. For all system sizes considered ($L_y = 1, 2, 4$), we observe a universal dome-shaped behavior in the spin gap, centered at $x = 0.5$. The spin gap increases with system size $L_y$, reaching nearly $1.8t$ around $x = 0.5$ for $L_y = 4$. Due to numerical limitations, simulations for $L_y = 6$ are restricted to the doping ranges $0 < x < 0.1$ and $0.9 < x < 1$. Nevertheless, we still observe an increase in the spin gap with $x$, in stark contrast to the conventional $t$-$J$ model. Appendix Fig.~\ref{fig:s1} shows the not fully converged spin gap data for $L_y = 6$, which indicates that the spin gap can still reach $1.7t$ for $x = 0.25$.

The spin gap in Fig.~\ref{fig:3}(a) exhibits particle-hole symmetry at $r = 1$. When $r$ is varied, the qualitative physics remains consistent, although the center of the dome shifts. This shift is driven by the competition between the $h$ and $d$ states: increasing $r$ favors the $d$ state and shifts the dome center to higher doping, while decreasing $r$ favors the $h$ state, shifting the optimal doping to lower values than $x = 0.5$.
Figure~\ref{fig:3}(b) shows results for $r = \sqrt{2}/\sqrt{3}$, which corresponds to the type-II $t$-$J$ model used for bilayer nickelates. We find that the maximal gap now occurs around $x = 0.31$. Figure~\ref{fig:3}(c) shows that the spin gap persists with a finite net repulsive interaction ($\textcolor{black}{V_{\mathrm{eff}}} > 0$).

\textbf{Pair-pair correlations} A finite spin gap implies a gap for single-electron excitations, but it remains to be shown whether Cooper pairs are gapless. We calculate the charge-1e gap, $\Delta_{1e} = E[N+1] + E[N-1] - 2E[N]$, and the charge-2e gap, $\Delta_{2e} = E[N+2] + E[N-2] - 2E[N]$ (see Appendix~\ref{sec:s2}). Here, $N$ is the total number of electrons. For the charge-2e gap calculation, we always fix $N_t = N_b$ and the total $S_z = 0$, ensuring no imbalance between the two layers, and the gap thus corresponds to an inter-layer spin-singlet Cooper pair. As shown in Fig.~\ref{fig:s3}, for $L_y = 4$ and doping $x = 0.1, 0.2, 0.3$, the charge-1e gap $\Delta_{1e}$ is finite and comparable in magnitude to the spin gap. In contrast, the charge-2e gap $\Delta_{2e}$ vanishes in the $L_x \rightarrow \infty$ limit, indicating gapless Cooper pairs in this regime.
Furthermore, Fig.~\ref{fig:3}(d) illustrates the power-law decay of the pair correlation function, $\langle \Delta^\dagger(i) \Delta(j) \rangle$, where the $s$-wave pairing operator is defined as $\Delta(i) = \epsilon_{\sigma\sigma'} c_{i;t,\sigma} c_{i;b,\sigma'} = -2 |h\rangle_i \langle d|_i$. This power-law decay signifies that the system resides in a Luther-Emery liquid phase \cite{lutheremergy}. The exponent $\alpha$ decreases with increasing $L_y$ (see Fig.~\ref{fig:3}(d)), suggesting a trend towards true long-range superconducting order as the system approaches the two-dimensional limit.

\begin{figure}[tb]
    \centering
\includegraphics[width=\linewidth]{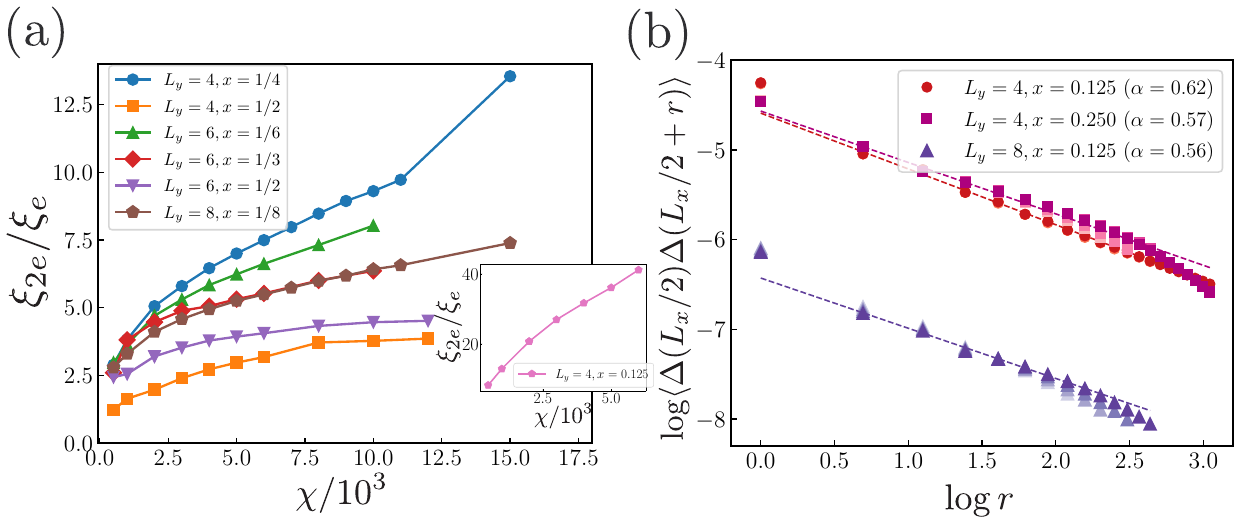}
\caption{\textbf{Infinite-system DMRG results for the ESD model, Eq.~(\ref{eq:Ham_esd}).}
(a) Dependence of the ratio $\xi_{2e}/\xi_{1e}$ on bond dimension. Correlation lengths for different operators are obtained using the transfer matrix technique in the appropriate symmetry sectors ($\delta N_t$, $\delta N_b$, $\delta S^z$). Here $N_t, N_b$ label the total number of electrons in the top and bottom layer, while $S_z$ labels the z component of the total spin.  The charge-2e correlation length corresponds to the pairing operator $\Delta$ in the ($1$, $1$, $0$) symmetry sector, and the charge-1e correlation length corresponds to the electron operator in the ($1$, $0$, $\frac{1}{2}$) symmetry sector.
(b) Pair-pair correlation function from iDMRG simulations with bond dimension \textcolor{black}{with different bond dimension $\chi$}. The red circles and red squares correspond to results for $L_x = 2$, $L_y = 4$, $x = 1/8$ and $L_x = 2$, $L_y = 2$, $x = 2/8$, respectively. 
\textcolor{black}{The darkness (brighter to dark) denotes the different bond dimension, $\chi=5\times 10^3,6\times 10^3,8\times 10^3,10^4$.}
The purple triangles correspond to results for $L_x = 1$, $L_y = 8$, $x = 1/8$. \textcolor{black}{The darkness (brighter to dark) denotes the different bond dimension, $\chi=5\times 10^3,7\times 10^3,10^4,15\times 10^3$.}
    }
    \label{fig:4}

\end{figure}

\textbf{Phase stiffness and total weight}
We have shown that the pairing gap can exceed $1.5t$. The superconducting critical temperature $T_c$ also depends on the phase stiffness. We therefore calculate the superfluid stiffness $D_s$ \cite{khodel1990superfluidity,PhysRevLett.39.1201} and the total weight $\Sigma$ from the $f$-sum rule \cite{Resta_2018}.
Following the formulation in Ref.~\cite{PhysRevB.102.165137}, $D_s$ is given by the second derivative of the ground-state energy $\mathcal{E}_0$ with respect to $A_x$, imposing periodic boundary conditions in the $x$-direction. Meanwhile, $\Sigma$ is given by the expectation value of the second derivative of the Hamiltonian:
\begin{equation}
D_s = D = \frac{1}{2V} \left. \frac{\partial^2 \mathcal{E}_0}{\partial A_x^2} \right|_{\vec{A}=0}, \quad
\Sigma = \frac{1}{2V} \left\langle \left. \frac{\partial^2 \hat{H}}{\partial A_x^2} \right|_{\vec{A}=0} \right\rangle_0,
\label{E0}
\end{equation}
where $\langle \hat{\mathcal{O}} \rangle_0 \equiv \langle G(0) | \hat{\mathcal{O}} | G(0) \rangle$ denotes the expectation value in the ground state $|G(0)\rangle$ of the Hamiltonian in the absence of $A_x$. The ESD model coupled to $A_x$ is
\begin{equation*}
\hat{H}_{\mathrm{ESD}}(A_x) = -t \sum_{l,\sigma} \sum_{\langle i,j \rangle_x} P [c_{i;l,\sigma}^\dagger e^{-iA_x} c_{j;l,\sigma} + \text{h.c.}] P,
\end{equation*}
where $\langle i,j \rangle_x$ denotes nearest neighbors along the $x$-direction. We emphasize that we calculate the Drude weight $D$, which is equivalent to the phase stiffness ($D_s = D$) for gapped states \cite{PhysRevB.102.165137}.

Figures~\ref{fig:3}(e) and \ref{fig:3}(f) show the doping dependence of $\Sigma$ and $D_s$. Due to numerical difficulties in simulating periodic boundary conditions, we calculate $\Sigma$ with open boundary conditions. However, we use periodic boundary conditions for the $D_s$ calculations. Consequently, we can achieve larger system sizes for $\Sigma$ than for $D_s$. Specifically, we obtain converged $\Sigma$ results for all $x$ for $L_y = 1, 2, 4$ and for $0 < x < 0.25$ and $0.75 < x < 1$ for $L_y = 6$. For $D_s$, we obtain results for all $x$ for $L_y = 1, 2$ and for $0 < x < 0.25$ and $0.75 < x < 1$ for $L_y = 4$.

There are several remarks. First, the total weight exhibits a dome-like structure similar to the spin gap, although the $L_y = 4$ data show a dip at $x = 0.5$. For $L_y = 1, 2$, we find that the phase stiffness $D_s$ closely tracks the total weight ($D_s/\Sigma \approx 1$), indicating that nearly all of the total weight resides in the Drude weight. The only exception is at $x = 0.5$ for $L_y = 2$, where the Drude weight vanishes. This point corresponds to an insulating state (see Appendix Fig.~\ref{fig:s2}), \textcolor{black}{where the similar tendency is observed in the bilayer one-orbital model without $V$ \cite{Schlömer2024}}. At $x = 0.5$, the Cooper pair filling is $1/2$ per site. For $L_y = 2$, this corresponds to one Cooper pair per unit cell, consistent with an insulating state. We ignore this special point. At generic fillings for $L_y = 2$, $D/\Sigma \approx 1$, and the system is in a Luther-Emery liquid phase. For $L_y = 4$, we again find $D/\Sigma \approx 1$ for $x < 0.21$, leading us to conclude that the total weight is nearly equal to the Drude weight at small doping. For $L_y = 6$, the total weight changes only slightly compared to $L_y = 4$, suggesting that the two-dimensional limit is approached. We then use our results for the total weight to estimate the phase stiffness $D_s$ in the two-dimensional limit for $x < 0.21$. We find that $\frac{\pi}{2} D_s$ increases with doping $x$, reaching $0.68t$ at $x = 0.25$.

In 2D, the Berezinskii–Kosterlitz–Thouless (BKT) transition temperature is given by $T_c = \frac{\pi}{2} D_s(T_c^-)$ \cite{Kosterlitz_1973}. Our results indicate that the pairing gap exceeds $\frac{\pi}{2} D_s$, suggesting that $T_c$ is limited by the phase stiffness. Although we cannot calculate $D_s$ exactly at $T_c$, our results suggest that $T_c/t$ can reach values on the order of $0.5$. For $x > 0.25$, we cannot obtain the phase stiffness for $L_y \geq 4$ due to the computational cost of periodic boundary condition calculations. It remains an open question whether $D_s/\Sigma$ deviates significantly from $1$ in this regime. Therefore, it is unclear whether the phase stiffness decreases with doping above $x = 0.25$ and exhibits a dip at $x = 0.5$.

\textbf{Infinite DMRG results}
As a complementary approach, we also present infinite DMRG (iDMRG) results in Fig.~\ref{fig:4}. Figure~\ref{fig:4}(a) shows the ratio of the pairing correlation length, $\xi_{2e}$, to the single-electron correlation length, $\xi_e$. We find that $\xi_{2e}/\xi_e$ generally increases with bond dimension $\chi$. In a Fermi liquid or Luttinger liquid phase, we expect $\xi_{2e}/\xi_e \sim 1/2$ because the Cooper pair is a composite operator. In our model, this ratio is always larger than $2$, indicating that single-electron excitations are gapped and that the low-energy physics is dominated by Cooper pairs rather than single electrons. We note that for $x = 0.5$, the correlation length $\xi_{2e}$ grows rather slowly with bond dimension, preventing us from definitively concluding whether the Cooper pairs are gapless or the system is in an insulating phase. The correlation lengths of various operators are shown in Appendix Fig.~\ref{fig:s5}.
Figure~\ref{fig:4}(b) shows that the pair correlation function exhibits power-law behavior at a fixed doping concentration, with increasing bond dimension ($\chi = 5000$, $7000$, $10000$, $15000$). For $x = 1/8$, we observe power-law pair-pair correlations even for $L_y = 8$, supporting our conclusion that the ground state is superconducting, even in the two-dimensional limit.

\textbf{Discussion for $x$ close to 0.5}
For $x$ near $0.5$, significantly larger bond dimensions are required to determine whether the pair-pair correlation exhibits power-law decay in both finite and infinite DMRG. We note that the finite and large spin gap, and consequently the finite and large charge-1e gap, imply gapped single-electron excitations. The remaining question is whether Cooper pairs condense. 
For an incommensurate filling close to $x=0.5$, an insulating phase may not be very likely and we are not aware of any other possible phases of the Cooper pair except a superconductor. Indeed, for $L_y = 4$, we find that the charge-2e gap vanishes with $1/L_x$ scaling at $x = 0.4$ from finite-system DMRG (see Appendix Fig.~\ref{fig:s3}). We therefore conjecture that the ground state is superconducting at incommensurate fillings near $0.5$, although confirming this requires significantly larger bond dimensions. For $x = 0.5$ exactly, we have confirmed that the $L_y = 2$ case is in an insulating phase with gapped Cooper pairs. For $L_y = 4$, the ratio $\xi_{2e}/\xi_e$ from iDMRG (see Fig.~\ref{fig:4}(a)) shows a tendency towards saturation with increasing bond dimension, suggesting a possible insulating state. However, a trivial, symmetric insulator is impossible in the two-dimensional limit. As we find no evidence of translational symmetry breaking (see Appendix Fig.~\ref{fig:s4}), we believe that $x = 0.5$ corresponds to a superconducting phase rather than a Cooper-pair insulator in the two-dimensional limit. Indeed, for $L_y = 6$, $\xi_{2e}/\xi_e$ grows more rapidly with bond dimension than for $L_y = 4$, consistent with this scenario. We leave a full resolution of this issue for fillings near $x = 0.5$ to future work.

\section{Exact solution of two-particle problems}\label{sec:4}
\begin{figure}[tb]
    \centering
\includegraphics[width=\linewidth]{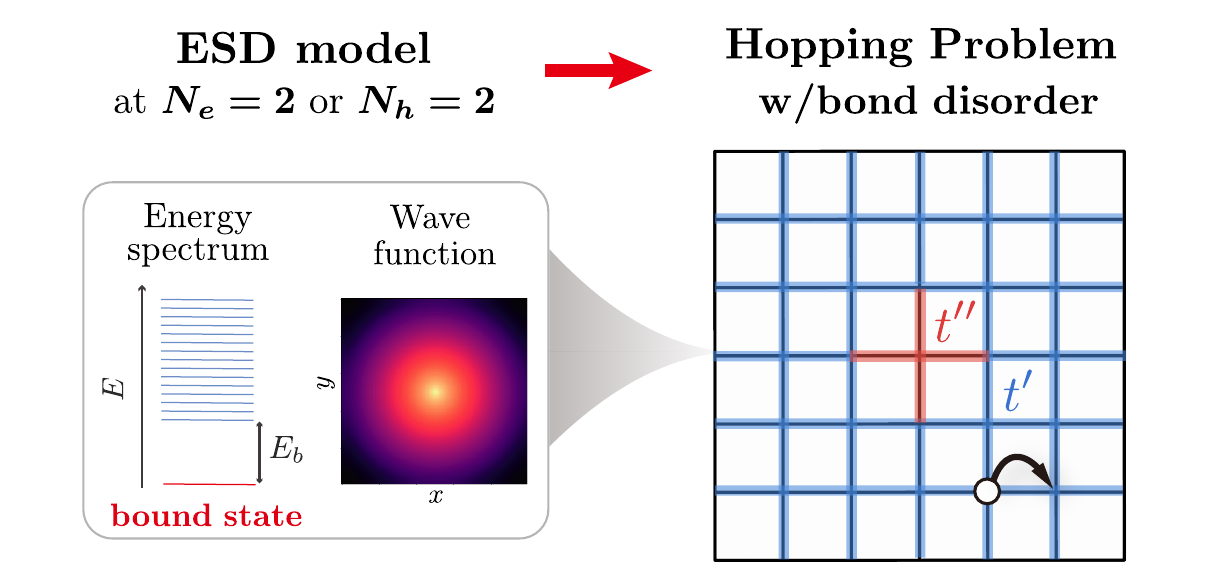}
\caption{\textcolor{black}{\textbf{Mapping from the ESD model with two holes into a single-electron hopping problem with a non-uniform hopping impurity at the origin.}
The two-body problem is derived as a tight-binding model with a bond disorder $t''$ at $\bm{R} = \bm{0}$ and a single impurity potential $w$. The hopping on other bonds is $t'$. For $N_h=2$, we have 
$(t',t'',w)=(2r^2t,2\sqrt{2}rt,2V_{\mathrm{eff}} )$. For $N_e=2$, we have 
$(t',t'',w)=(2t,2\sqrt{2}rt,2V_{\mathrm{eff}} )$.
When $t'' > t'$, the electrons tend to localize around $\bm{R} = \bm{0}$, corresponding to a spin-singlet Cooper-pair bound state. This bound state is identified by a positive binding energy ($E_b > 0$) in the energy spectrum, and its corresponding wave function is localized at the origin.
}
    }
    \label{fig:6_1}
\end{figure}

In this section, we provide an exact solution in the two-hole (or two-electron) problem in the $x=0$ (or $x=1$) limit.
\textcolor{black}{In the end of the derivation, it turns out that the ESD model with two holes or two electrons  is mapped to a single-electron hopping problem with a non-uniform hopping located at the origin (See Fig.~\ref{fig:6_1}). Then, the pairing state can be understood as a localized state in  the single hopping problem with the impurity. Moreover, the origin of the impurity is derived from the unconventional hopping process in the ESD model in Fig.~\ref{fig:5}. We can also obtain the exact wavefunction of the Cooper pair. The Cooper pair bound state survives to the repulsive regime with $V_{\mathrm{eff}}>0$, where it has a large size and cannot be understood in the simple Bose-Einstein-Condensation (BEC) picture.}
\\

We first consider the $N_h = 2$ case, which can be easily generalized to the $N_e= 2$ case in later. Let us consider two holes relative to the fully occupied vacuum near $x = 0$, which corresponds to a product state of doublons, $\ket{G} = \prod_i \ket{d}_i$. Relative to this ground state, we define a single-hole state as $\tilde{c}_{i,l,\sigma}^\dagger \ket{G}$, where the hole creation operator is given by $\tilde{c}_{i,l,\sigma}^\dagger = \ket{\overline{l},\overline{\sigma}}_i \bra{d}_i$. We can also define an creation operator of Cooper pair of holes, $\Delta_i^\dagger = \ket{h}_i \bra{d}_i $.

We consider a two hole states with spins $\sigma$ and $\sigma'$, reside in different layers separated by a distance $\bm{R}$.
With periodic boundary conditions, the center-of-mass momentum $\bm{Q}$ is a good quantum number. The two-body states can then be written as plane waves:
\begin{equation}
|\Psi^{(2h)}_{\sigma,\sigma'};\bm{Q},\bm{R} \neq \bm{0}\rangle = \frac{1}{\sqrt{N_s}} \sum_{\bm{i}} e^{i \bm{Q} \cdot \bm{i}} 
\tilde{c}_{\bm{i},t,\sigma}^\dagger
\tilde{c}_{\bm{i}+\bm{R},b,\sigma}^\dagger
\ket{G},
\end{equation}
where $N_s = L_x L_y$ is the total number of sites. 
The superscript $(2h)$ distinguishes two-hole states from two-electron states $(2e)$.

We here focus on the zero-momentum sector ($\bm{Q} = \bm{0}$) and the spin-singlet sector. A two-body state at the different site distinct by $\bm{R}\neq 0$ is given by, 
\begin{equation}
|\Psi^{(2h)};\bm{R}\rangle = \frac{1}{\sqrt{2N_s}} \sum_{\bm{i}} \left[ 
\tilde{c}_{\bm{i},t,\uparrow}^\dagger
\tilde{c}_{\bm{i}+\bm{R},b,\downarrow}^\dagger
-\tilde{c}_{\bm{i},t,\downarrow}^\dagger
\tilde{c}_{\bm{i}+\bm{R},b,\uparrow}^\dagger
\right]\ket{G}.
\end{equation}
A two-body state at the same site $\bm{R}=0$ corresponds to,
\begin{equation}
|\Psi^{(2h)};\bm{R}=\bm{0}\rangle = \frac{1}{\sqrt{N_s}} \sum_{\bm{i}} 
\Delta^{\dagger}_{i}
\ket{G}.
\end{equation}
These states form the complete Hilbert space for the two-body problem. We then project the ESD model onto this subspace and diagonalize it.
As mentioned previously, there are four distinct hopping processes (see Fig.~\ref{fig:5}). Among these, the $T_3$ and $T_4$ processes can hybridize $|\Psi^{(2h)};\bm{R} = \bm{0}\rangle$ and $|\Psi^{(2h)};\bm{R} \neq \bm{0}\rangle$, which plays a crucial role in pairing.

\begin{figure}[tb]
    \centering
\includegraphics[width=\linewidth]{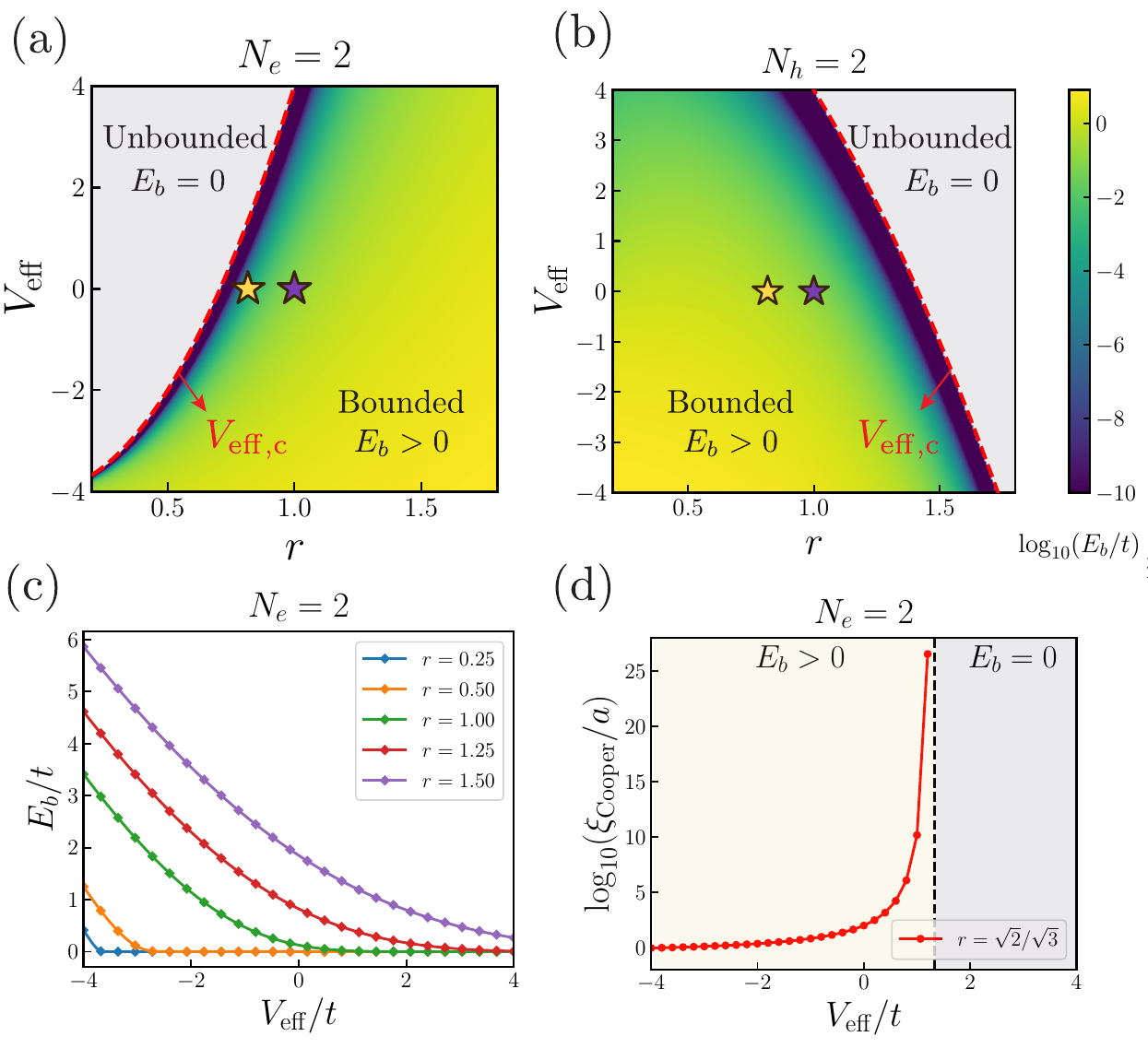}
\caption{
\textbf{Exact solution results for the $N_e=2$ and $N_h=2$ cases.}
(a,b) Phase diagrams in the $(r,V_{\mathrm{eff}})$ plane in the
thermodynamic limit, $L_x,L_y\to\infty$, for the $N_e=2$ and
$N_h=2$ cases. The color bar represents $\log_{10}(E_b/t)$.
The critical value $V_{\mathrm{eff},c}$ (dashed line) is determined
by Eq.~\ref{eq:Veffc_exact}.
{
The closed-form expressions are
$V_{\mathrm{eff},c}^{(2e)}=4t(2r^2-1)$ for the $N_e=2$ case and
$V_{\mathrm{eff},c}^{(2h)}=4t(2-r^2)$ for the $N_h=2$ case.} For $V_{\mathrm{eff}}<V_{\mathrm{eff},c}$, the binding energy is
finite and the two particles form a bound state.
For $V_{\mathrm{eff}}>V_{\mathrm{eff},c}$, the binding energy
vanishes in the thermodynamic limit and the ground state consists
of two freely moving particles.
We denote the points
$r=1$, $V_{\mathrm{eff}}=0$ and
$r=\sqrt{2}/\sqrt{3}$, $V_{\mathrm{eff}}=0$
with purple and yellow stars, respectively.
For $r=1$, the binding energy is positive in both the $x\to0$ and
$x\to1$ limits. For $r=\sqrt{2}/\sqrt{3}$, the binding energy is
positive only in the $x\to0$ limit.
(c) Dependence of the binding energy $E_b$ on $V_{\mathrm{eff}}$
for several values of $r$. Increasing $r$ and decreasing
$V_{\mathrm{eff}}$ favors pairing, as indicated by a larger $E_b$.
(d) Bound-state size as a function of $V_{\mathrm{eff}}$ at
$r=\sqrt{2}/\sqrt{3}$. The characteristic decay length is obtained
analytically from
$\xi_{\mathrm{Cooper}}=\sqrt{t'/E_b}$.
{The exact threshold (dashed line) is
$V_{\mathrm{eff},c}/t=1.333333$.}
}
\label{fig:6_2}
\end{figure}

Since our Hilbert space basis is labeled by the coordinate $\bm{R}$, we effectively have a tight-binding model for a single particle with non-uniform hopping (see Fig.~\ref{fig:6_1}). The explicit matrix form of $\hat{H}^{(2h)}$ is given by
\begin{equation}
\hat{H}^{(2h)} = -t' \sum_{\substack{\bm{i},\bm{\xi} \\ \bm{i},\bm{i}+\bm{\xi}\neq \bm{0}}} [c_{\bm{i}}^\dagger c_{\bm{i}+\bm{\xi}} + \text{h.c.}] - t'' \sum_{\bm{\xi}} [c_{\bm{0}}^\dagger c_{\bm{\xi}} + \text{h.c.}] + w c_{\bm{0}}^\dagger c_{\bm{0}}.
\label{eq:H_dis}
\end{equation}
Here, $\bm{\xi} = \hat{x}, \hat{y}$ represents the nearest-neighbor vectors on the square lattice. In the above equation, $c_{\bm{R}}^\dagger$ creates the state $|\Psi^{(2h)}; \bm{R}\rangle$. The effective hopping is $t' = 2r^2t$ if neither $\bm{i}$ nor $\bm{i} + \bm{\xi}$ is at the origin. However, if one of them is at the origin, the hopping amplitude is $t'' = 2\sqrt{2}rt$, which can be interpreted as a form of disorder that breaks translational symmetry in $\bm{R}$-space. We reiterate that the $t''$ term originates from the $T_3$ and $T_4$ processes in our ESD model (see Fig.~\ref{fig:5}).
Furthermore, there is an on-site potential $w = 2\textcolor{black}{V_{\mathrm{eff}}}$ when $\textcolor{black}{V_{\mathrm{eff}}}$ is nonzero. Depending on the disorder strength, controlled by $t''$ and $w$ (or equivalently by $r$ and $\textcolor{black}{V_{\mathrm{eff}}}$ in the original ESD model), the ground state can be either a bound state or delocalized.

{
For finite size, the energy spectrum and the wave function can be numerically obtained by
performing a diagonalization of an $N_s\times N_s$ matrix. We identify a
two-hole bound state by examining the binding energy,
$E_b=E_e-E_g$, where $E_g$ represents the ground-state energy of
$\hat H^{(2h)}$, while $E_e$ is the lowest energy of the continuum, which
converges to $E_e\to-4t'$ in the thermodynamic limit.
Here, $E_b>0$ indicates the existence of a two-hole Cooper-pair bound state.
In the thermodynamic limit, the binding energy can be obtained
exactly for the two-dimensional square lattice. In particular, $E_b>0$ is determined
by
\begin{equation}
\left[
\frac{t''^{\,2}}{t'^{\,2}}-1
-\frac{w}{4t'+E_b}
\right]
\frac{2}{\pi}
K\!\left[
\left(\frac{4t'}{4t'+E_b}\right)^2
\right]
=
\frac{t''^{\,2}}{t'^{\,2}},
\label{eq:Eb_exact}
\end{equation}
where $K(m)$ is the complete elliptic integral of the first kind. The corresponding binding threshold is obtained in closed form as
\begin{equation}
    V_{\rm eff,c}
    =
    2t'
    \left(
    \frac{t''^{\,2}}{t'^{\,2}}-1
    \right),
\label{eq:Veffc_exact}
\end{equation}
where a bound state exists for $V_{\rm eff}<V_{\rm eff,c}$ and
$E_b\rightarrow0$ as $V_{\rm eff}\rightarrow V_{\rm eff,c}^{-}$.
}

{The corresponding bound-state wave function can be written as
\begin{equation}
    |\Psi\rangle_{\rm bound}
    =
    \sum_{\bm R}
    f(\bm R)
    |\Psi^{(2h)};\bm R\rangle .
\end{equation}
The wave function $f(\bm R)$ can also be obtained exactly in the
thermodynamic limit from the lattice Green's function (see Appendix
Sec.~C). In the $E_b\ll t'$ limit, its long-distance behavior is
$f(\bm R)
    \sim
    K_0(R/\xi)
    \sim
    \frac{e^{-R/\xi}}{\sqrt{R}},
$
with $\xi=\sqrt{\frac{t'}{E_b}}$,
where $K_0$ is the modified Bessel function of the second kind.
We plot $f(\bm R)$ for $r=1$ and $V_{\rm eff}=0$ in the inset of
Fig.~7.
For comparison, in the one-dimensional limit $L_y=1$, both the
binding energy and the bound-state wave function reduce to elementary
closed-form expressions (see Appendix Sec.~C).
}
 
 We can generalize the two-hole problem to the two-electron problem, $N_e=2$ corresponding to $x\rightarrow 1$ limit. 
The only change in the two-electron Hamiltonian, $H^{(2e)}$, is now we have $t' = 2t $ while $t'' = 2\sqrt{2}rt$ and $w=2\textcolor{black}{V_{\mathrm{eff}}}$ are not changed (See Appendix Sec.\ref{sec:s3}). 
Hence, one can establish the relation, $E_{b}[r;N_e=2]=r^2E_{b}[1/r;N_h=2]$ at $\textcolor{black}{V_{\mathrm{eff}}}=0$ for general $r$. This gives $E_{b}[r=1;N_e=2]=E_{b}[r=1;N_h=2]$.

The main results are presented in Fig.\ref{fig:6_2}. 
The $r$, $\textcolor{black}{V_{\mathrm{eff}}}$ dependence of the binding energy is given in Fig.\ref{fig:6_2} (b) at $L_x, L_y \rightarrow \infty$ for $N_e=2$ case. 
The binding energy decreases with $\textcolor{black}{V_{\mathrm{eff}}}$, but can survive in the repulsive regime with $\textcolor{black}{V_{\mathrm{eff}}}>0$.
In Fig.~\ref{fig:6_2} (c-d), the phase diagram for both $N_e=2$ and $N_h=2$ cases are illustrated in $(r,\textcolor{black}{V_{\mathrm{eff}}})$ space. 
This proves that there exists broad phase space with a Cooper pair bound state in 2D even when the net interaction is repulsive with $\textcolor{black}{V_{\mathrm{eff}}} > 0$.  
For specific parameter $r=1,\sqrt{2}/\sqrt{3}$ and $\textcolor{black}{V_{\mathrm{eff}}}=0$, we tabulate the binding energy $E_b[N_e=2],E_b[N_h=2]$ for various $L_y$ in Table \ref{table:1}.
The binding energy reaches a finite value i.e. $E_b=0.127 t$ at $r=1$ in the two-dimensional limit. One can find the results from ED at finite $L_y$ agree well with the spin gap calculation from DMRG in Fig.\ref{fig:3} (a,b), with $\Delta_{s}=E_{b}$ in the $x\rightarrow 0$ and $x\rightarrow 1$ limit.
Our ED results then prove the finite pairing gap in the $x\rightarrow 0$ limit. Although the value is small, it likely grows with the doping $x$ quite rapidly according to our DMRG results at finite $L_y$. At the $x \rightarrow 0$ limit, the pairing gap decreases with $L_y$, but the gap actually increases with $L_y$ for larger $x$. Therefore we believe the pairing gap at larger $x$ is significantly larger than the two-hole problem studied here. From our DMRG calculation, it easily reaches more than $1.5 t$.

\textcolor{black}{If we use a large and negative $V_{\mathrm{eff}}$, we expect simple BEC picture, similar to the previous works of bipolaron superconductor with large net attractive interaction \cite{PhysRevLett.100.140405, PhysRevX.13.011010,PhysRevB.109.L220502}.  However, we emphasize that this simple BEC limit does not hold anymore in the regime of $V_{\mathrm{eff}}\geq 0$, which is our major focus in this work.  In Fig.~\ref{fig:6_2} (d), we calculating the Cooper pair size at $r=1$ with various $V_{\mathrm{eff}}$. The large size of the Cooper pair clearly demonstrates that superconductivity exists with net repulsion and can not be understood in the naive tight-bound Cooper pair picture.  Also, according to our DMRG calculation, the pairing gap actually increases with the doping $x$, which is also contrary to simple BEC picture. Therefore our mechanism of superconductivity is clearly beyond the previous theories assuming a net attraction. Our mechanism is likely more relevant to real materials where a large repulsion is usually inevitable. 
}

\begin{table}[tb]
    \centering
    {
\begin{tabular}{c|cccc|c}
\hline
\hline
$r=1$
& $L_y=1$
& $L_y=2$
& $L_y=4$
& $L_y=200$
& $L_x=L_y\rightarrow\infty$
\\
\hline
$E_b[N_e=2]$
& $0.619t$
& $0.507t$
& $0.333t$
& $0.127t$
& $0.126509t$
\\
\hline
\hline
$r=\sqrt{2}/\sqrt{3}$
& $L_y=1$
& $L_y=2$
& $L_y=4$
& $L_y=200$
& $L_x=L_y\rightarrow\infty$
\\
\hline
$E_b[N_e=2]$
& $0.131t$
& $0.091t$
& $0.048t$
& $0.001t$
& $0.000223t$
\\
$E_b[N_h=2]$
& $0.911t$
& $0.835t$
& $0.648t$
& $0.489t$
& $0.489523t$
\\
\hline
\hline
\end{tabular}
}
\caption{
\textbf{Dependence of the binding energy on $L_y$ at $r=1$ and
$r=\sqrt{2}/\sqrt{3}$ for $V_{\mathrm{eff}}=0$, obtained from the exact
solution of the two-particle problem.}
For finite $L_y$, the binding energies are numerically obtained with
$L_x=200$. The last column shows the analytical result in the
thermodynamic limit, obtained from Eq.~\ref{eq:Eb_exact}.
Due to particle-hole symmetry, the binding energies at $r=1$ are
identical for the $N_e=2$ and $N_h=2$ cases.
These exact two-particle results are consistent with the fDMRG
calculations shown in Fig.~\ref{fig:3}(a,b). The small deviations can
be attributed to finite-size effects in the fDMRG simulations.
}
    \label{table:1}
\end{table}

\textcolor{black}{
\section{Analytical treatment at finite density: small Fermi surface and doping tuned Feshbach resonance}\label{sec:6}
\subsection{Fermion-boson theory $x=0$}
Following the previous section, we consider the $x=0$ limit, where the ground state is a product state of doublons, $\ket{G} = \prod_i \ket{d}_i$.  But here we move to the case with a finite density of holes. Then the problem is not exactly solvable, but here we will provide a mean field theory which can at least qualitatively capture the doping dependence. Especially we will see that the pairing gap increases with the hole doping $x$. 
Similar to the previous treatment for the two-hole problem, we define a single-hole state, $\ket{\mathrm{hole}}_{i,l,\sigma} = \tilde{c}_{i,l,\sigma}^\dagger \ket{G}$, and on-site Cooper pair of holes state. Here, we used the hole creation operator, $\tilde{c}_{i,l,\sigma}^\dagger = \ket{\overline{l},\overline{\sigma}}_i \bra{d}_i$ and creation operator of Cooper pair of holes, $\Delta_i^\dagger = \ket{h}_i \bra{d}_i$.
At small hole doping $x$, the system has a dilute density of holes which interact with the on-site Cooper pair. The original ESD model can be mapped to a fermion-boson model:
\begin{eqnarray}
    H&=& -
   g \sum_{\langle i,j \rangle}\left[(\Delta_i^\dagger + \Delta_j^\dagger) \left[\tilde{c}_{i,b,\downarrow} \tilde{c}_{j,t,\uparrow} - \tilde{c}_{i,b,\uparrow} \tilde{c}_{j,t,\downarrow} + (t \leftrightarrow b) \right]\right.
    \nonumber  \\
    && +\mathrm{h.c.}]
   +\frac{t'}{2}\sum_{\langle i,j \rangle,l,\sigma}\left[\tilde{c}^\dagger_{i,l,\sigma} \tilde{c}_{j,l,\sigma} 
    +\tilde{c}_{i,l,\sigma}^\dagger
    \tilde{c}_{j,l,\sigma}
    \Delta_{j}^\dagger
    \Delta_{i}
\right]+\mathrm{h.c.}
    \nonumber \\
    &&
    -\mu \sum_i \tilde{c}_i^\dagger \tilde{c}_i 
    + (V_{\mathrm{eff}}-2\mu) \sum_i \Delta_i^\dagger \Delta_i,
    \label{eq:H_fermion-boson}
\end{eqnarray}
with a constraint $n_{i;\tilde c}+n_{i;\Delta}\leq 1$ at each site $i$. $n_{i;\tilde c}=\sum_{l\sigma}\ket{l\sigma}_i\bra{l\sigma}_i$ is the density of the hole state at site $i$ and $n_{i;\Delta}=\ket{h}_i\bra{h_i}$ is the density of the Cooper pair state at site $i$. In the above $g= \frac{t''}{2\sqrt{2}}$ is a fermion-boson coupling, corresponding to the non-uniform hopping term $t''=2\sqrt{2}rt$ in the two-hole problem (See Eq.~\ref{eq:H_dis} and Fig.~\ref{fig:6_1}). $t'=2r^2 t$ is a hopping term for the single hole.
Note that creation operator of $\ket{\Psi^{(2h)}}$, $c_{\bm{0}}^{\dagger}$ in Eq.~(\ref{eq:H_dis}) correspond to the $\Delta_{i}^{\dagger}$ in Eq.~(\ref{eq:H_fermion-boson}).
Now the non-uniform disorder term can be interpreted as the a Yukawa type coupling between the fermion and boson, $g$.
Here, $\mu$ is introduced to fix the doping $n_{\tilde c}+2n_{\Delta}=2x$ on average. In the same time we have the hard core constraint $n_{i;\tilde c}+n_{i;\Delta} \leq 1$. If this constraint is hold exactly,  at each site we must have $(n_{i;\tilde c},n_{i;\Delta})=(0,0),(1,0),(0,1)$. Here $(n_{i;\tilde c},n_{i;\Delta})=(0,0)$ labels the vacuum state, the $\ket{d}_i$ state in our original notation.  Therefore, if the constraint $n_{i;\tilde c}+n_{i;\Delta} \leq 1$ is treated exactly, the above model is an exact mapping of the original ESD model and there is no approximation so far. One can check that in the two-hole problem it reduces to the model in the previous section. 
\\
\indent
In the following we make an approximation: we ignore the hard-core constraint $n_{i;\tilde c}+n_{i;\Delta} \leq 1$ and now view $\tilde c_{i;l\sigma}$ and $\Delta_i$ as the usual  fermionic and bosonic operators applied on top  of the vacuum $\ket{G}$. This approximation should be justified in the dilute limit. After this approximation, we have a fermion-boson model which was actually widely used in the cold atom context \cite{GURARIE20072}. Thus one naturally expect a BEC to BCS crossover by tuning $V_{\mathrm{eff}}$ from negative to positive. Besides, the on-site energy of the boson $\Delta_i$ actually also depends on the chemical potential $\mu$ and thus we also expect a doping tuned Feshbach resonance in this model. As we will show,  this explains the unusual dome structure of  the pairing gap versus doping $x$ observed in our DMRG simulations.
\\
\indent
Next we discuss  the normal and superconducting states within this framework respectively.
First, in the high-temperature normal state, we can ignore the boson $\Delta_i$. The dilute density of holes $\tilde{c}_{i,l,\sigma}$ (with density $n_{\tilde c} = 2x$) form a small hole pocket centered at $\vec{k} = (\pi,\pi)$, with a Fermi surface area $A_{FS} = -\frac{x}{2}$ per spin and per layer. 
This is actually a quite unusual metallic state, contrasting sharply with the naive weak-coupling expectation of $A_{FS} = \frac{1-x}{2}$ if we use simple free fermion model in the $J_K=0$ limit of the double Kondo model.\\
\indent Regarding superconductivity, pairing in this unconventional normal state is mediated by virtual excitations of the boson $\Delta_i$ even if $V_{\mathrm{eff}}>0$. Let's assume a net repulsive interaction $V_{\mathrm{eff}}>0$, the virtual boson state can induce an effective four-fermion attraction for the hole $\tilde c$ which is estimated as $\tilde{V} \sim -\frac{t^2}{V_{\mathrm{eff}}-\mu}$. 
Because the denominator decreases as doping $x$ increases (increasing $\mu$), the pairing strength, and hence $T_c$, grows as $x$ increases towards $0.5$. 
This behavior is reminiscent of the Feshbach resonance mechanism in ultracold atomic systems~\cite{GURARIE20072,PhysRevLett.96.060401}, where a virtual molecular state becomes resonant with the Fermi surface. There have been some papers to understand the pairing of bilayer nickelate in analogy to this atomic Feshbach resonances \cite{lange2023feshbachresonancestronglyrepulsive,PhysRevB.110.L081113}. But in our theory, the resonance can be tuned by increasing the doping. 
Consequently, the system undergoes a BCS$'$-BEC crossover as doping approaches $x \approx 0.5$.  Note that the BEC side actually corresponds to the overdoped regime, which is quite remarkable and contrary to the naive expectation that the underdoped regime is more strongly coupled.
Here, the notation BCS$'$ is used to emphasize that the pairing instability emerges from an unconventional normal state with a small Fermi surface, distinguishing it from the standard BCS theory.\\}
\textcolor{black}{
\subsection{Fermion-boson theory at $x=1$} Near $x=1$, we can use the similar logic, yet the roles of the doublon $d$ and holon $h$ are reversed compared to the low-doping regime  $x=0$. Thus, the ground state is $\ket{G}=\prod_{i}\ket{h}_i$ and the fermion-boson model now is written with the single electron operator $c^{\dagger}_{i,l,\sigma}= \ket{l,\sigma}_{i} \bra{h}_i$ and one-site Cooper pair $\Delta_{i}^\dagger=\ket{d}_{i}\bra{h}_i$. 
In this case, the normal state has electron pockets centered at $\vec{k} = (0, 0)$, with Fermi surface area $A_{\mathrm{FS}}=\frac{1-x}{2}$ which agrees with the free fermion model with $J_K=0$ in the double Kondo model.
The pairing instability in this regime can also be described within the effective fermion-boson framework. As in the low-doping regime, the system exhibits a BCS–BEC crossover as $x$ decreases toward 0.5, driven by a doping controlled Feshbach resonance mechanism.
\\
\subsection{Application of Oshikawa-Luttinger theorem}
In the previous subsection, we find two different Fermi-liquid states with different Fermi-surface volume $A_{FS}=-x/2$ at small doping and $A_{FS}=(1-x)/2$ in large doping. At a fixed $x$, there is a difference of $1/2$ of Brillouin zone.  The result is surprising because it implies two different Fermi liquids without any symmetry breaking.
Here, we demonstrate that the application of the non-perturbative Luttinger theorem by Oshikawa \cite{PhysRevLett.84.3370} in the bilayer model actually allows a $Z_2$ index for the symmetric Fermi liquid. This is different from the familiar single layer model where only a $Z_1$ index is discovered. \\
\indent We consider a system where the total number of electrons per site (summed over both layers and spins) in the conduction C-layer is $n_T = 2(1 - x)$. Thus, the density per spin and per layer, or per flavor, is $\nu = (1 - x)/2$, where $x$ is the hole concentration in the C-layer.
The starting point of our model—the double Kondo model—does not include interlayer hopping $t_\perp$, but does have a finite interlayer exchange interaction $J_\perp$ (See Eq.(1)). This setup leads to a global $(U(1)_t \times U(1)_b \times SU(2))/\mathbb{Z}_2$ symmetry, along with a mirror reflection symmetry $\mathcal{M}$ between the layers.
We obtain the Luttinger relation  by employing Oshikawa's flux-threading argument\cite{PhysRevLett.84.3370} with $U(1)_t$, $U(1)_b$ sectors:
     \begin{align}
        A_{\mathrm{FS},t,\uparrow} + A_{\mathrm{FS},t,\downarrow} &= 2\nu \quad (\mathrm{mod}\ 1), \\
        A_{\mathrm{FS},b,\uparrow} + A_{\mathrm{FS},b,\downarrow} &= 2\nu \quad (\mathrm{mod}\ 1).
    \end{align}
    Due to the spin-rotation and mirror-reflection symmetries, we have additional following constraints on the Fermi surface areas:
    \begin{align}
        A_{\mathrm{FS},t,\uparrow} = A_{\mathrm{FS},t,\downarrow} &= A_{\mathrm{FS},b,\uparrow} = A_{\mathrm{FS},b,\downarrow} = A_{\mathrm{FS}}.
    \end{align}
The above three equalities lead to two distinct solutions: (I) $A_{\mathrm{FS}} = \nu \quad (\mathrm{mod}\ 1)$, and (II) $A_{\mathrm{FS}} = \nu - \frac{1}{2} \quad (\mathrm{mod}\ 1)$. Note that we can not thread a U(1) flux for each flavor separately: we do not have four U(1) symmetries in this model. Thus we can only fix the Fermi surface volume per flavor up to $1/2$ of the Brillouin zone (BZ). This leads to the $Z_2$ index.  We refer to solution (I) as the ``\textsl{conventional Fermi liquid}'', as it is adiabatically connected to the $J_K = 0$ limit of the double Kondo model, where the itinerant electron is connected to a free fermion model. In contrast, we refer to solution (II) as the ``\textsl{second Fermi liquid}'' to distinguish it from the conventional Fermi liquid.  Note  the distinction between the FL and sFL phase is topological and there must be either a phase transition or an intermediate phase between them.\\
\indent Our ESD model captures the transition between sFL to FL in the strong coupling regime ($|J_K|,J_\perp \gg t $), as illustrated in Fig.\ref{fig:r3}.  In the strong-coupling limit ($|J_K|,J_\perp \gg t $), a transition (indicated by the black arrow) occurs between the second Fermi liquid (sFL) at low doping and the conventional Fermi liquid (FL) at higher doping. The sFL phase features a hole pocket centered at $\vec{k} = (\pi, \pi)$ with Fermi surface area $A_{FS} = \nu - \frac{1}{2} = -\frac{x}{2}$, whereas the FL phase has an electron pocket at $\vec{k} = \Gamma$. In contrast, in the weak-coupling regime , only the conventional FL solution which has an electron pocket exists in the entire hole doping concentration, which is smoothly connected to each other. 
\begin{figure}[tb]
    \centering
\includegraphics[width=0.75\linewidth]{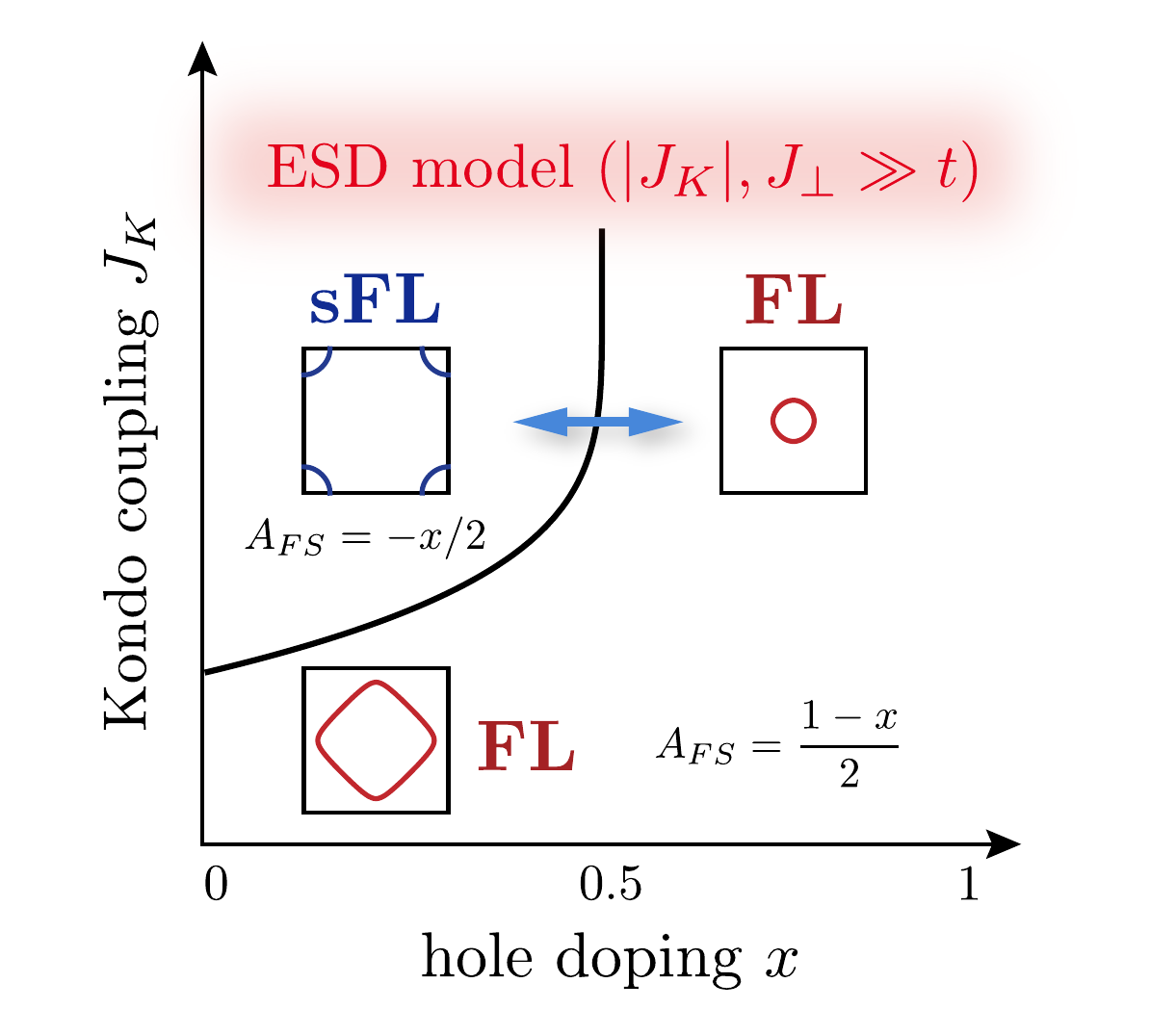}
\vspace{-10pt}
    \caption{\textcolor{black}{\textbf{Schematic phase diagram of the the normal state, in the $x$–$J_K$ plane at fixed $J_\perp$}. In the strong-coupling limit ($|J_K|,J_\perp \gg t$), a transition (indicated by the arrow) occurs between the second Fermi liquid (sFL) at low doping and the conventional Fermi liquid (FL) at higher doping. The sFL phase features a hole pocket centered at $\vec{k} = (\pi, \pi)$ with Fermi surface area $A_{FS} = \nu - \frac{1}{2} = -\frac{x}{2}$, whereas the FL phase has an electron pocket at $\vec{k} = \Gamma$. In contrast, in the weak-coupling regime, only the conventional FL solution which has an electron pocket exists in the entire hole doping concentration, which is smoothly connected to each other. Our ESD model is valid to use in $J_\perp\gg t$ limit and can demonstrate the transition between sFL to FL as increasing hole doping concentration.
    }}
    \label{fig:r3}
\end{figure}
}

\textcolor{black}{
\subsection{Bridging $x=0$ and $x=1$ limit}
We have discussed the nature of the normal states and their pairing mechanisms in the regimes near $x \sim 0$ and $x \sim 1$. To bridge these two scenarios and gain a comprehensive understanding of the full phase diagram—including the intermediate doping region around $x \sim 0.5$—we further perform a generalized slave-boson mean-field analysis.  We note that this slave boson theory is different from the usual slave boson theory for single layer t-J model.  Our treatment reduces to the fermion-boson model in the dilute limit where the approximation should be well justified. We use a generalized slave boson representation,}
\begin{align}
c_{i,l,\uparrow}&=h^\dagger_{i}
      f_{i,l,\uparrow}+r
f_{i,\overline{l},\downarrow}^\dagger d_{i},\\
c_{i,l,\downarrow}&= h^\dagger_{i} 
f_{i,l,\downarrow}-r
f_{i,\overline{l},\uparrow}^\dagger 
d_{i}, 
\end{align}
with $\overline{l}=-l$. Here, we have two slave boson operators, $h$, $d$ annihilating holon and doublon states, respectively. Meanwhile, $f_{l,\sigma}$ is the Abrikosov fermion which annihilates the singlon states. As in the conventional slave-boson theory, we have a U(1) gauge symmetry: $h_i \rightarrow e^{i\theta_i} h_i$, $d_i \rightarrow e^{i\theta_i} d_i$, and  $f_{i;l;\sigma} \rightarrow e^{i\theta_i} f_{i;l;\sigma} $. In this work, our ansatz always higgses the internal U(1) gauge field, so there is no fractionalization.

Substituting the parton construction, the ESD Hamiltonian becomes,
\begin{eqnarray}
    \hat{H}_{\mathrm{ESD}}
    &=&
-t\sum_{l,\langle i,j \rangle,\sigma}
\left[
    f^\dagger_{i,l,\sigma}
    f_{j,l,\sigma}
  h_{j}^\dagger
h_{i}
-r^2 
f_{j,\overline{l},\sigma}^\dagger
f_{i,\overline{l},\sigma}
d^\dagger_{i}
d_{j}\right]
\nonumber
\\
&&-t\sum_{l,\langle i,j \rangle}
\left[
r[f^\dagger_{i,l,\uparrow} f^\dagger_{j,\overline{l},\downarrow}-f^\dagger_{i,l,\downarrow} f^\dagger_{j,\overline{l},\uparrow}] h_i d_j+h.c. 
\right]
\nonumber
    \\
&&+\textcolor{black}{V_{\mathrm{eff}}} \sum_{i} (n_{d;i}+n_{h;i}). 
\label{eq:full}
\end{eqnarray}
with a constraint $n_{i;h}+n_{i;f}+n_{i;d}=1$ at each site $i$.

Then, it is clear to decompose $T_1,T_2,T_3,T_4$ terms illustrated in Fig.\ref{fig:5}. For example, the first line corresponds $T_1$,$T_2$, and second line corresponds $T_3$,$T_4$, respectively. 
We again emphasize the $T_3$, $T_4$ are relevant to the pairing, since only $T_3$, $T_4$ can offer the pairing order parameter $\Delta_f$ within the mean-field decoupling. 

From mean field decoupling, we reach the following mean-field Hamiltonian:
\begin{eqnarray}
    \hat{H}_{MF} 
    &=&\hat{H}_{f}+\hat{H}_{hd}, \label{esd_mft}
\end{eqnarray}
with 
\begin{eqnarray}
  \hat{H}_{f} &=&-t_{f} \sum_{l,\sigma}\sum_{\langle i,j \rangle}  f^{\dagger}_{i;l;\sigma}f_{j;l;\sigma}
+f^{\dagger}_{j;l;\sigma}f_{i;l;\sigma}+\sum_{i}\delta_{f}n_{f;i} 
  \nonumber \\
    &+&\Delta_{f} \sum_{\langle i,j\rangle }
f^{\dagger}_{i;t;\uparrow}f^{\dagger}_{j;b;\downarrow}
-f^{\dagger}_{i;t;\downarrow}f^{\dagger}_{i+x;b;\uparrow}
+(t\leftrightarrow b)
 +h.c. \notag\\
\hat{H}_{hd}&=& 
\sum_{i}
[\lambda_{h} h_{i}+\lambda_{d} d_{i} +h.c.]
+\delta_{h}n_{h;i}
+\delta_{d}n_{d;i} 
\label{eq:mean_field}
\end{eqnarray}
We have  $\delta_{f} =-\mu_{0}-\mu,  \delta_{h}=V_{\mathrm{eff}}-\mu_{0}-2\mu,
 \delta_{d}=V_{\mathrm{eff}}-\mu_{0}$. 
The two chemical potentials, $\mu_{0},\mu$ are introduced for the two constraints, $n_{h}+n_{f}+n_{d}=1$, and $n_{f}+2n_{h}=2x$. The explicit expressions of $t_f,\Delta_f,\lambda_h,\lambda_d$ are given in Appendix Sec.~\ref{sec:s4}.  We always have $\langle h \rangle \neq 0$ and $\langle d \rangle \neq 0$. 
By solving the mean-field Hamiltonian self-consistently, we achieve the mean-field solutions of the ESD model at $r=1$, $\textcolor{black}{V_{\mathrm{eff}}}=0$ in the two-dimensional limit in Fig.\ref{fig:8}.  One can see that the mean-field theory qualitatively agrees with the DMRG results on the doping dependence of $n_h,n_f,n_d$.  Especially, it captures correctly the value of $n_h=n_d\approx 0.25$ at $x=0.5$.  The mean field theory slightly underestimates $n_f$ at $x=0.5$. In either method, $n_f$ is significant around $x=0.5$, indicating that the $S=\frac{1}{2}$ singlon can not be simply ignored despite the large single electron gap.

\textcolor{black}{In the dilute limit, the slave boson theory can be reduced into the fermion-boson model, Eq.(\ref{eq:H_fermion-boson}).
For example, consider $x\sim 0$ limit, where the holon $h$ remains gapped, and the doublon $d$ condenses with $\langle d \rangle \neq 0$ and $n_d = 1 - 2x$. 
Then we can write the effective fermion-boson Hamiltonian by replacing the condensed doublon operator $d$ with its expectation value $\langle d \rangle$:
\begin{eqnarray}
H&=&-g\sum_{\langle i, j\rangle}\left[(h_i^\dag+h_j^\dag)
[f_{i,b,\downarrow}f_{j,t,\uparrow}-f_{i,b,\uparrow}f_{j,t,\downarrow}
+(t \leftrightarrow b )]\right.
\notag\\
&+&
\left. \mathrm{h.c.}\right]
+\frac{t'}{2}\langle d\rangle^2 
\sum_{\langle i, j\rangle, l,\sigma}
\left[f_{i,l,\sigma}^\dag f_{j,l,\sigma}+\mathrm{h.c.}\right]\notag\\
&+&\frac{t'}{2}
\sum_{\langle i, j\rangle, l,\sigma}
\left[f_{i,l,\sigma}^\dag f_{j,l,\sigma}
h_{j}^\dagger h_{i}
+\mathrm{h.c.}\right]\notag\\
&+&
(-\mu_{0}-\mu)\sum_{i}n_{f,i}
+(V_{\mathrm{eff}}-\mu_{0}-2\mu) \sum_{i}n_{h,i}
\label{eq:H_fermion-boson2}
\end{eqnarray}
where $t_{\mathrm{eff}}=\frac{t'}{2}\langle d\rangle^2$, $g=\frac{t''}{2\sqrt{2}}\langle d\rangle$. In the dilute limit, we can use $\langle d\rangle \sim 1$, and ignore $\mu_0$ corresponding to $n_h+n_d+n_h=1$. Then one can show Eq.(\ref{eq:H_fermion-boson2}) has exactly the same form with Eq.(\ref{eq:H_fermion-boson}). 
}

Our slave boson theory predicts two distinct Fermi liquid phases in the ESD model at finite temperatures, arising from the condensation of either $h$ or $d$ bosons. For $x$ close to $1$, $h$ bosons condense ($\langle h \rangle = \sqrt{n_h} \neq 0$) while $d$ remains gapped, resulting in a conventional Fermi liquid (FL) on top of the product state of $\ket{h}$ at $x=1$. Conversely, for x close to $0$, $d$ bosons condense ($\langle d \rangle = \sqrt{n_d} \neq 0$), and $h$ is gapped, leading to second Fermi liquid (sFL)~\cite{yang2024strong}, which hosts a hole pocket on top of the product state of $\ket{d}$ at $x=0$.  
Within this framework, we demonstrate the BCS'-BEC-BCS crossover in the pairing across from $x = 0$ to $x = 1$ side.

\textbf{Constraint from the particle-hole symmetry} For the special parameter $r=1$, the ESD model exhibits a particle-hole (PH) symmetry, relating the $x<0.5$  regime to the $x>0.5$ regime. Crucially, at $x=0.5$, this PH symmetry maps the sFL phase onto the FL phase. Specifically, for each flavor, the sFL phase at $x=0.5$ features a hole pocket occupying 1/4 of the Brillouin zone (BZ) centered around $\bm{ k}=(\pi,\pi)$ while while the FL phase possesses an electron pocket of the same size centered around $\bm{k}=(0,0)$. These distinct phases demonstrate a discontinuous jump in Fermi surface volume by 1/2 of the BZ, precluding a simple Lifshitz transition. At $x=0.5$, any Fermi liquid phase must necessarily break the PH symmetry, as a symmetric Fermi liquid is incompatible with the Hilbert space of the model. However, the ground state from our DMRG simulation appears to be PH symmetric and thus can not be viewed as a descendant of a symmetric FL. Within our slave boson mean field theory, we obtain a superconducting ground state.  Notably, to ensure adherence to PH symmetry at x = 0.5, the mean-field hopping parameter $t_f$, must vanish. Consequently, the Bogoliubov quasiparticle dispersion becomes dominated by the pairing term, signifying a departure from the conventional BCS mean-field picture built upon a Fermi liquid normal state.

\textbf{Variational wavefunction} We can write down a variational wavefunction based on our slave boson theory:
\begin{eqnarray}
    \ket{\Psi}= P_G\  \langle h \rangle^{N_h} \langle d\rangle^{N_d}\ket{\text{BCS}[f]}
\end{eqnarray}
where $P_G$ is a generalization of the usual Gutzwiller projection \cite{PhysRevLett.10.159}.  For each configuration of $h, d, f$ states in real space, the $\ket{\text{BCS}[f]}$ is the mean field wavefunction of the singlon $f$ described by Eq.~\ref{eq:mean_field}. Importantly, both $\ket{d}$ and $\ket{h}$ states should be identified as the vacuum state $\ket{0}$ within the mean-field theory for $f$. $N_h$ and $N_d$ are the total number of $\ket{h}$ and $\ket{d}$ states, respectively. $\langle h \rangle$ and $\langle d \rangle$ can be obtained from the mean field theory, but can also be viewed as variational parameters. Note that in our case, $N_h+N_d=N-N_f$ is allowed to fluctuate with only $N_d-N_h$ fixed by the density. This wavefunction can be investigated using variational Monte Carlo (VMC) techniques, potentially providing valuable insights into the unusual kinetic superconductivity observed in the ESD model, particularly around $x = 0.5$ where the superconducting state cannot be readily interpreted as a descendant of a conventional Fermi liquid.

\begin{figure}[tb]
    \centering
\includegraphics[width=\linewidth]{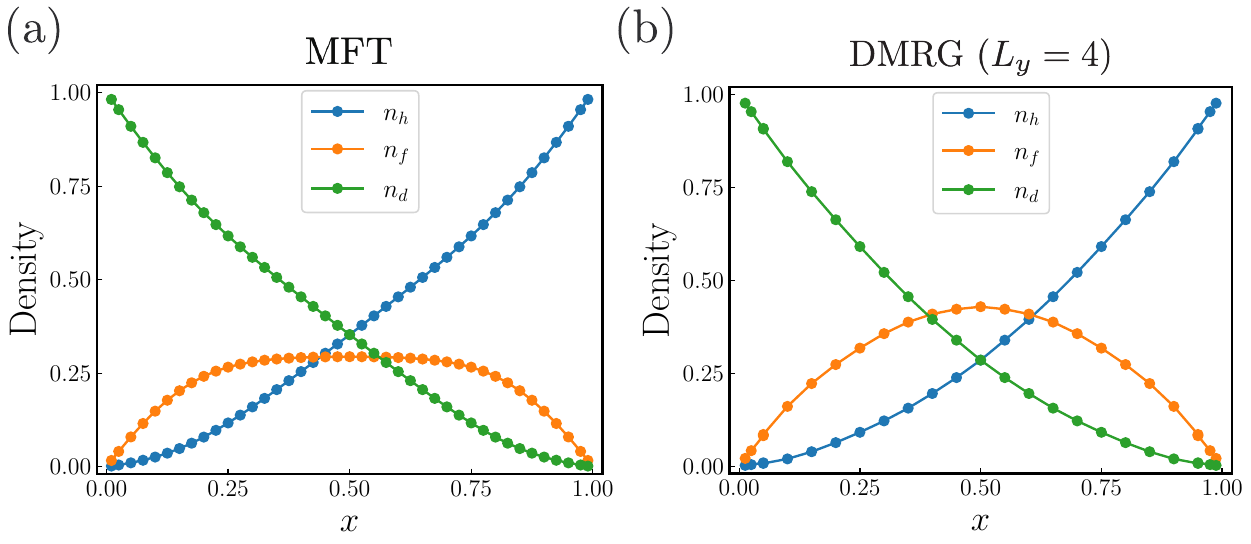}
\caption{\textbf{The hole doping $x$-dependence of particle densities of the ESD  model at $r = 1$ and $\textcolor{black}{V_{\mathrm{eff}}} = 0$.} (a) Mean-field solutions of the ESD model in the two-dimensional limit. Here, we use $L_x=L_y=40$. Here, the density of each particle contents can be carried out by $n_d=\langle d\rangle^2$, $n_h=\langle h\rangle^2$ and $n_f=\sum_{k,l,\sigma}\langle f_{k,l,\sigma}^\dagger f_{k,l,\sigma}\rangle$. (b) The fDMRG results for the ESD model with parameters $L_y = 4$, $L_x = 40$, and $\chi = 3000$. 
}
    \label{fig:8}
\end{figure}

\section{Towards realistic regime of $J_\perp \sim t$}\label{sec:7}
So far, we have demonstrated a superconductor with $T_c/t $ as large as $0.5$ in the infinite $J_\perp$ limit by studying the ideal ESD model. However,the real material might always be in the regime of finite $J_\perp/t$. A crucial question arises: how robust is the superconductor under decreasing $J_\perp/t$? To address this question, we investigate the evolution of the system under variations in $J_\perp/t_0$ within the double-Kondo model described by Eq.~(\ref{eq:Ham_double_kondo}). For simplicity, we still take the $|J_K|\gg t$ limit but keep $J_\perp/t_0$ finite.  We consider the positive and negative $J_K$ sides separately.  $J_K\rightarrow -\infty$ and $J_K\rightarrow +\infty$ correspond to the bilayer type-II t-J model and one-orbital t-J model, respectively.

\subsection{$J_K\rightarrow-\infty$: Bilayer type II t-J model}
\textcolor{black}{Notably, $J_K \to -\infty$ limit is relevant to studying bilayer nickelate 80 K superconductivity under high pressure \cite{sun2023signatures}.
There are various interesting theoretical proposals have been made to account for high critical temperatures of bilayer nickelate. For example, Ref.\cite{Schlömer2024} applied the Feshbach resonance framework to a single-orbital bilayer $t$–$J$ model and reported $T_c$ values reaching up to $\sim 0.5 J_\perp$ near the BEC–BCS crossover under idealized conditions.
To connect the bialyer nickelate to our double Kondo model, the C-layer and S-layer can be identified with the $d_{x^2 - y^2}$ and $d_{z^2}$ orbitals, respectively. The Kondo coupling $J_K = -2J_H < 0$ is viewed as the  Hund’s interaction $J_H$.
The active $d_{x^2 - y^2}$ orbitals primarily have in-plane hopping $t$, resulting in separate charge conservation for each layer of itinerant electrons. In contrast, the $d_{z^2}$ orbital has a significant interlayer exchange interaction $J_\perp$ which introduces strong interlayer correlations in the system. The importance of the interlayer exchange $J_\perp$ arising from the $d_{z^2}$ orbital have been investigated in Refs.~\cite{luo2023bilayer,zhang2023electronic,sakakibara2024possible,tian2024correlation,qin2023high,yang2023minimal,zhan2024cooperation,chen2024non,yang2023possible,gu2023effective,liu2023s,shen2023effective,PhysRevB.109.L201124}. 
More recent works~\cite{oh2023type,yang2024strong,PhysRevLett.132.146002,zhang2023strong,lu2023superconductivity,PhysRevB.110.L081113,qu2024bilayer} have further proposed that an effective $J_\perp$ coupling can be shared with the $d_{x^2 - y^2}$ sector, mediated by strong Hund’s coupling. In this regime, the itinerant electrons in the $d_{x^2 - y^2}$ orbital and the localized spin moments of the $d_{z^2}$ orbital become strongly intertwined together. 
\\
The effective theory in the $J_K \rightarrow -\infty$ limit of the double Kondo model reduces to the bilayer type-II $t$--$J$ model~\cite{zhang2020type,zhang2021fractional}, which captures the spin fluctuations of the $d_{z^2}$ orbital associated with the localized moment in the S-layer. 
In this limit, the conduction electron in the C-layer forms a spin-triplet state with the local spin in the S-layer, to lower the energy penalty from strong ferromagnetic coupling $J_K<0$. Meanwhile, if the conduction electron is unoccupied, the spin itself in the S-layer host spin $1/2$ moment.
This results in five low-energy states at each site of each layer: three $S=1$ states $\ket{T;\alpha}$ and two $S=\frac{1}{2}$ states $\ket{\sigma}$. Here, the subscript $\alpha = \pm 1, 0$ labels $S_{z}=\pm1, 0$.  Then, by projecting onto the five-dimensional Hilbert space, one can reach the bilayer type-II t-J model,
\begin{align}
     \hat{H}=
     -&t_0\sum_{l=t,b}
\sum_{\langle i,j \rangle}
\sum_{\sigma}
\left[Pc^{\dagger}_{i;l;\sigma}
c_{j;l;\sigma}P
+h.c.\right]+V\sum_{i}n_{i,t}n_{i,b}
     \notag\\
+& \sum_{i} \left[J^{ss}_\perp\vec S_{i;t}\cdot \vec S_{i;b}
+J ^{dd}_\perp  \vec T_{i;t}\cdot \vec T_{i;b} \right]\notag\\
+& \sum_{i} J^{sd}_\perp
\left[\vec S_{i;t}\cdot \vec T_{i;b}+\vec T_{i;t}\cdot \vec S_{i;b}\right],
\label{eq:type-II}
\end{align}
with $J^{ss}_{\perp}=2J^{sd}_{\perp}=4J^{dd}_{\perp}=J_{\perp}$. Here, $\vec{S}_{i,l}=\frac{1}{2}\vec{\sigma}_{\alpha\beta}\ket{\alpha}_{i;l}\bra{\beta}_{i;l}$ is the spin operator of the spin $S=\frac{1}{2}$ and  $\vec{T}_{i;l}=\sum_{\alpha,\beta}\vec{T}_{\alpha,\beta}\ket{T;\alpha}_{i;l}\bra{T;\beta}_{i;l}$ is the spin operator of the spin $S=1$.\\
\indent In the context of bilayer nickelates~\cite{yang2024strong}, the spin-one states in the type-II $t-J$ model correspond to a $d^8$ valence configuration with both $e_g$ orbitals occupied, while the spin-half states correspond to the $d^7$ valence with only the $d_{z^2}$ orbital occupied. The bilayer type-II $t-J$ model properly captures the fluctuation of the $d_{z^2}$ orbital~\cite{zhang2020type,zhang2021fractional}, in stark contrast to the simpler bilayer one-orbital model. For instance, an electron in the $d_{x^2 - y^2}$ orbital can bind to an $S = 1$ magnon excitation from the rung singlet formed by the $d_{z^2}$ orbitals, forming a spin polaron.
When $J_\perp$ is finite, the bilayer type II t-J model and the bilayer one-orbital t-J model are quantitatively quite different \cite{yang2024strong}. Especially, focusing on one rung, one can exactly show that the energy splitting of the $S=1$ state and $S=0$ state for the doublon is $\Delta_d=E[n_T=2,S_T=1]-E[n_T=2,S_T=0]=\frac{1}{4} J_\perp$ for the full double Kondo lattice model in the limit of large $J_H$. Meanwhile, this value is $\Delta_d=J_\perp$ for one-orbital t-J model.
This suggests that the effective inter-layer coupling for the $d_{x^2-y^2}$ orbital should be $\tilde J_\perp=\frac{1}{4} J_\perp$ and thus previous works \cite{PhysRevLett.132.146002,zhang2023strong,lu2023superconductivity,qu2024bilayer} using naive one-orbital model with $\tilde J_\perp=J_\perp$ may overestimate the term by a factor of $4$.
Superconducting pairing in the bilayer type-II $t$-$J$
model has been investigated in other studies~\cite{oh2023type,yang2024strong,oh2026Pair-density-wave,oh2026incommensurate}.
}

\begin{figure}[tb]
    \centering
\includegraphics[width=\linewidth]{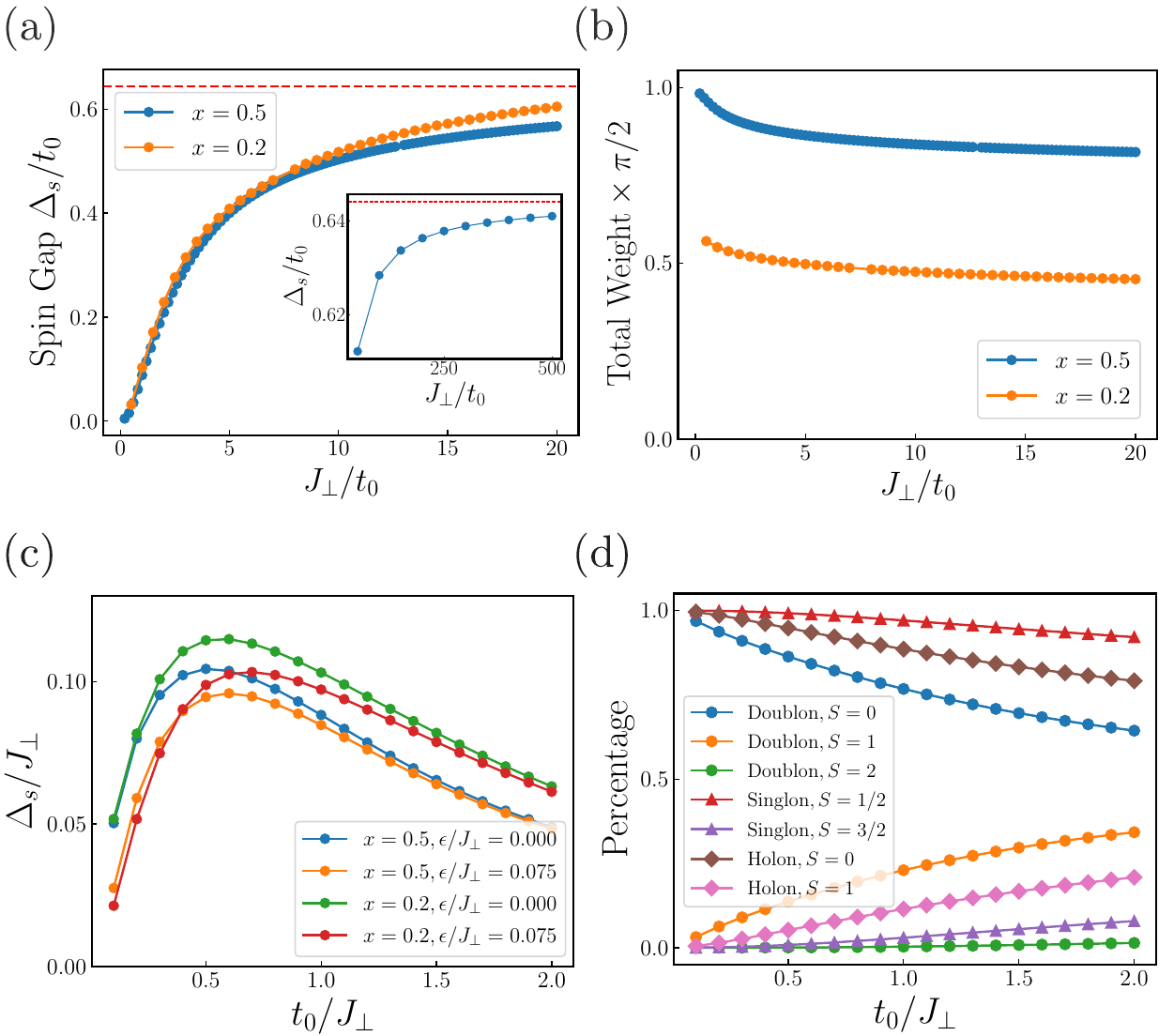}
\caption{\textbf{The DMRG simulation results of the type-II t-J model in Eq.(\ref{eq:type-II}).}  Here, we simulate one dimension with bond dimension $\chi=2000$. The $J_\perp$ dependence of the (a) spin gap and the (b) total weight of type-II t-J model at $x=0.2$ and $x=0.5$. 
    We consider a two-leg ladder configuration 
    ($L_z$ =2, $L_y$ = 1 with $L_x=60$) and set $J_\parallel=0,J_\perp^{ss}=2J_\perp^{sd}=4J_\perp^{dd}=J_\perp$. To explore the kinetic pairing, we set $\textcolor{black}{V_{\mathrm{eff}}}=(V-\frac{1}{4}J_\perp)/2=0$ by choosing $V=J_\perp/4$. 
    The red dashed line of (a) is the spin gap calculated from the corresponding ESD model with $t=3/4 t_0$, $r=\sqrt{2}/\sqrt{3}$, and $\textcolor{black}{V_{\mathrm{eff}}}=0$. 
    In the inset, we zoom in the results for larger $J_\perp$ at $x=0.5$. 
    The result from the type-II t-J model converges to that of the ESD model in the large $J_\perp/t_)$ limit. 
    (c) The  $t_0/J_\perp$ dependence of the spin gap at $x=0.2$ and $x=0.5$. At fixed $J_\perp$, one can see that the pairing gap increases with the hopping $t_0$ when $t_0/J_\perp<0.5$. This proves that the kinetic energy enhances the pairing when $J_\perp>2$. (d) The percentage of various doublon, singlon, and holon states with various total spin $S$. Doublon state with $S=0$, singlon state with $S=1/2$, and holon with $S=0$ dominate in the large $J_\perp/t_0$ limit and justify the ESD model. At a larger $t_0/J_\perp$, the $S=1$ component of the doublon and the $S=1$ component of the holon both grow up. This creates a new channel for single electron to hop and save the kinetic energy, reducing the gain of the Cooper pair in minimizing the kinetic energy.
    }
    \label{fig:9}
\end{figure}

In the limit $J_\perp\gg t_0$, the bilayer type-II t-J model can be further reduced to the ESD model, with $t=\frac{3}{4}t_0$, $r=\sqrt{2}/{\sqrt{3}}$, and $\textcolor{black}{V_{\mathrm{eff}}}=[V-\frac{1}{4}J_\perp]/2$.
We simulate the model Eq.~\ref{eq:type-II} in 1D by DMRG. In Fig.~\ref{fig:9}(a), one can see that the spin gap in the $J_\perp\gg t_0$ limit converges to the value obtained directly from the corresponding ESD model. As $J_\perp/t_0$ decreases, the the spin gap decreases. However, the total weight in Fig.~\ref{fig:9}(b) slightly increases as we decrease $J_\perp/t_0$. As discussed in the previous section, in the $J_\perp\gg t_0$ limit the critical temperature $T_c$ is decided by the phase stiffness. At smaller $J_\perp/t_0$, the pairing gap becomes smaller than the phase stiffness, and $T_c$ is now limited by the pairing gap.

Fig.~\ref{fig:9}(c) illustrates the $t_0/J_\perp$ dependence of spin gap at a fixed $J_\perp=1$. In the region $t_0/J_\perp \lesssim 0.5$, the pairing gap increases with the hopping at a fixed $J_\perp$, consistent with the expected behavior of a kinetic superconductor. Conversely, for $t_0/J_\perp \gtrsim 0.5$, the pairing gap decreases with $t_0/J_\perp$ at a fixed $J_\perp$. Importantly, we note that here we always have $\textcolor{black}{V_{\mathrm{eff}}} \geq 0$, indicating a net repulsive interaction. Consequently, the observed pairing at large $t_0/J_\perp$ cannot be attributed to simple mean-field mechanisms. At larger $t_0/J_\perp$, the low energy degree freedom is not fully restricted to the ESD Hilbert space. As shown in Fig.~\ref{fig:9}(d), the percentage of $S=1$ doublon grows with $t_0/J_\perp$. This provides an alternative channel for single-electron motion, saving the kinetic energy and consequently diminishing the energetic advantage of Cooper pairing. Nevertheless, we believe that kinetic energy remains the dominant driving force for pairing in this regime.

\subsection{$J_K\rightarrow+\infty$: Bilayer one-orbital model}
In the limit $J_K\rightarrow +\infty$, the electron in the C-layer and the localized spin moment from the S layer form a spin-singlet $\ket{s}$ with total spin $S=0$. 
\textcolor{black}{Then, this results in three low-energy states at each site of each layer:} one $S=0$ spin-singlet $\ket{s}$, two spin-$1/2$ states $\ket{\sigma}$ with $\sigma=\uparrow,\downarrow$. As discussed in Ref.~\cite{yang2024strong}, the singlet state can be treated as the new empty state in the one orbital t-J model. We can define the fermionic operators as $\tilde{c}^\dagger_{i;l;\sigma}=\ket{\sigma}_{i;l}\bra{s}_{i;l}$, and the density operators as $\ket{\tilde{n}}_{i;l}=\ket{\sigma}_{i;l}\bra{\sigma}_{i;l}$, and the spin operators $\tilde{\vec{S}}_{i;l}=\vec{\sigma}_{\alpha\beta}\ket{\alpha}_{i;l}\bra{\beta}_{i;l}/2$. Projecting the double-Kondo model onto the $3$-dimensional Hilbert space, we can reach the bilayer one-orbital t-J model,
\begin{align}
   \hat{ H}=-&\tilde{t}_0\sum_{l,\sigma,\langle ij\rangle}P\tilde{c}^\dagger_{i;l;\sigma}\tilde{c}_{i;l;\sigma}P+h.c.\nonumber\\
+&V\tilde{n}_{i;t}\tilde{n}_{i;b}+J_\perp\sum_{i}\tilde{\vec{S}}_{i;t}\cdot\tilde{\vec{S}}_{i;b},
    \label{eq:bilayer_one_orbital_t_J}
\end{align}
where $\tilde{t}_0=-t_0/2$, and $V$ and $J_\perp$ are the same as in the double-Kondo model. In this limit, the electron creation operator should be viewed as the hole creation operator in the original model, and the electron density operator $n_{i}$ changes to $1-\tilde{n}_{i}$.

\begin{figure}[tb]
    \centering
\includegraphics[width=1.0\linewidth]{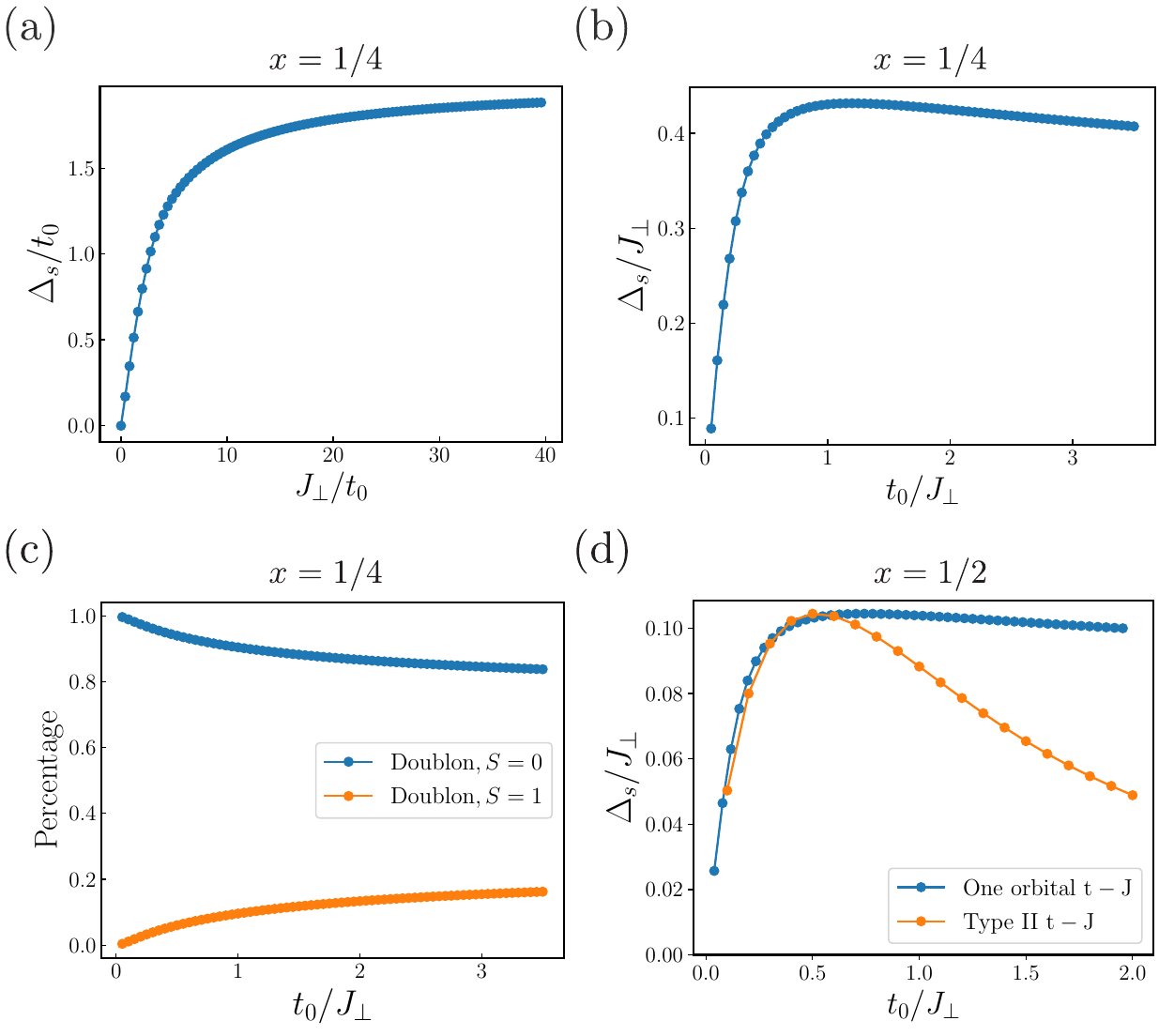}
    \caption{\textbf{The DMRG results of the bilayer one-orbital t-J model Eq.(\ref{eq:bilayer_one_orbital_t_J}).} Here, we simulate one dimension with $L_x=60$ and bond dimension $m=2000$. We set $V=\frac{3}{4}\tilde{J}_\perp$ for imposing $\textcolor{black}{V_{\mathrm{eff}}}=0$ in the ESD model. (a) The $J_\perp/t_0$ dependence of spin gap $\Delta_s$ for $x=0.25$. Similar to that in the bilayer type-II t-J model, the spin gap saturates in the limit $J_\perp\gg t_0$ to the value of the ESD model. (b) The $t_0/J_\perp$ dependence of the spin gap $\Delta_s$ for $x=0.25$. For fixed $J_\perp$, the pairing gap increases with the hopping $t_0$ when $t_0/J_\perp<1.2$. (c) The percentage of the $S=0$ and $S=1$ doublons in the $n_T=2$ doublon state for $x=0.25$. In the large $J_\perp$ limit, corresponding to the ESD model, the $S=0$ doublon dominates. (d) The rescaled spin gap in the one-orbital t-J model for $x=0.5$, here $J_\perp$ is the interlayer coupling in the type-II t-J model. But for the one-orbital model, we use a smaller $J_\perp/a$ with $a=2.56$ for the $J_\perp$ term.  After this rescaling, the two models have similar behaviors for $t_0/J_\perp \lesssim 0.5$. But the two models show qualitatively different trends for larger $t_0/J_\perp$, even after the rescaling of $J_\perp$.} 
    \label{fig:one_orbital}
\end{figure}

\begin{table}[b]
\centering
\begin{tabular}{ c|c c cc}
\hline
\hline
  Spin gap& $J_\perp/t_0=0.4$& $J_\perp/t_0=1$&$J_\perp/t_0=2$&$J_\perp/t_0=5$ \\ 
  \hline
 Type-II& 0.015 $t_0$&0.088 $t_0$ &0.209 $t_0$&0.400 $t_0$\\
One-orbital&0.102 $t_0$&0.267 $t_0$&0.515 $t_0$&0.963 $t_0$\\
 \hline
 \hline
\end{tabular}
\caption{\textbf{Comparison of the spin gap $\Delta_s$ of type-II t-J and one-orbital t-J for various $J_\perp/t_0$ in one dimension at $x=0.5$.}
The results are achieved by DMRG simulation at $L_x=60$ with $\chi=2000$. 
}
\label{spin_gap_table}
\end{table}
In the limit $J_\perp\gg \tilde t_0$,  the bilayer one-orbital model is also reduced to the ESD model, with $t=\tilde{t}_0$, $r={1}/{\sqrt{2}}$, and $\textcolor{black}{V_{\mathrm{eff}}}=\frac{1}{2}[V-\frac{3}{4}J_\perp]$. 
In Fig.~\ref{fig:one_orbital}, we simulate the Eq.~(\ref{eq:bilayer_one_orbital_t_J}) with $V=3J_\perp/4$, corresponding to $\textcolor{black}{V_{\mathrm{eff}}}=0$ in the ESD model.
In Figs.~\ref{fig:one_orbital}(a-c) reveal that the spin gap has a qualitatively similar, but quantitatively stronger trend compared to that in the type-II t-J model. Furthermore, the effect of $J_\perp$ in the bilayer one-orbital t-J model is apparently stronger than in the type-II t-J model. The energy splitting of the $S=1$ and $S=0$ doublons in the bilayer one-orbital model is $4$ times the splitting in the bilayer type-II t-J model. In Table.~\ref{spin_gap_table}, we provide the spin gap $\Delta_s$ at different values of $J_\perp/t_0$ for the bilayer type II t-J model in the $J_K\rightarrow -\infty$ and the bilayer one-orbital model in the $J_K \rightarrow +\infty$ limit. It is clear that the one-orbital model has a relatively larger pairing gap, indicating stronger pairing in the $J_K>0$ side of the double Kondo model. However, for both signs of $J_K$, we note that the pairing gap reaches the order of $10\% t_0$ for moderate $J_\perp/t_0 \sim 1$. This is already an unprecedentedly huge pairing energy scale for a model without net attraction in a realistic regime.

In the context of the bilayer nickelate, some works~\cite{PhysRevLett.132.146002,zhang2023strong,lu2023superconductivity,qu2024bilayer} propose to utilizing the bilayer one-orbital model with the interlayer coupling $J_\perp$ term simply equal to that of the $J_\perp$ from the $d_{z^2}$ orbital. Our above analysis clearly shows that this is not appropriate and would overestimate the pairing strength a lot. A more accurate description of bilayer nickelates necessitates the use of the double-Kondo model with negative $J_K$ or, equivalently the bilayer type II t-J model in the large $|J_K|$ limit. It is not equivalent to the simple bilayer one-orbital model. To match the type II t-J model, we need to rescale the $J_\perp$ parameter in the one-orbital model. In Fig.~\ref{fig:one_orbital}(d), for fixed $t_0$ and $J_\perp$ in the double Kondo model, we use a rescaled parameter $\tilde{t}_0=t_0$ and $J_\perp/a$ with $a=2.56$ in the one orbital t-J model.  Empirically, after rescaling, the one-orbital model has a roughly similar behavior as the type II t-J model for  $t_0/J_\perp<0.5$ region, suggesting that the effect of $J_\perp$ in the one-orbital model is overestimated by a factor of $2.56$. For $t_0/J_\perp>0.5$, the two models show qualitatively very different behaviors, suggesting that the two models are essentially different and not simply equivalent under a change of parameter.

\begin{figure}[tb]
    \centering
\includegraphics[width=\linewidth]{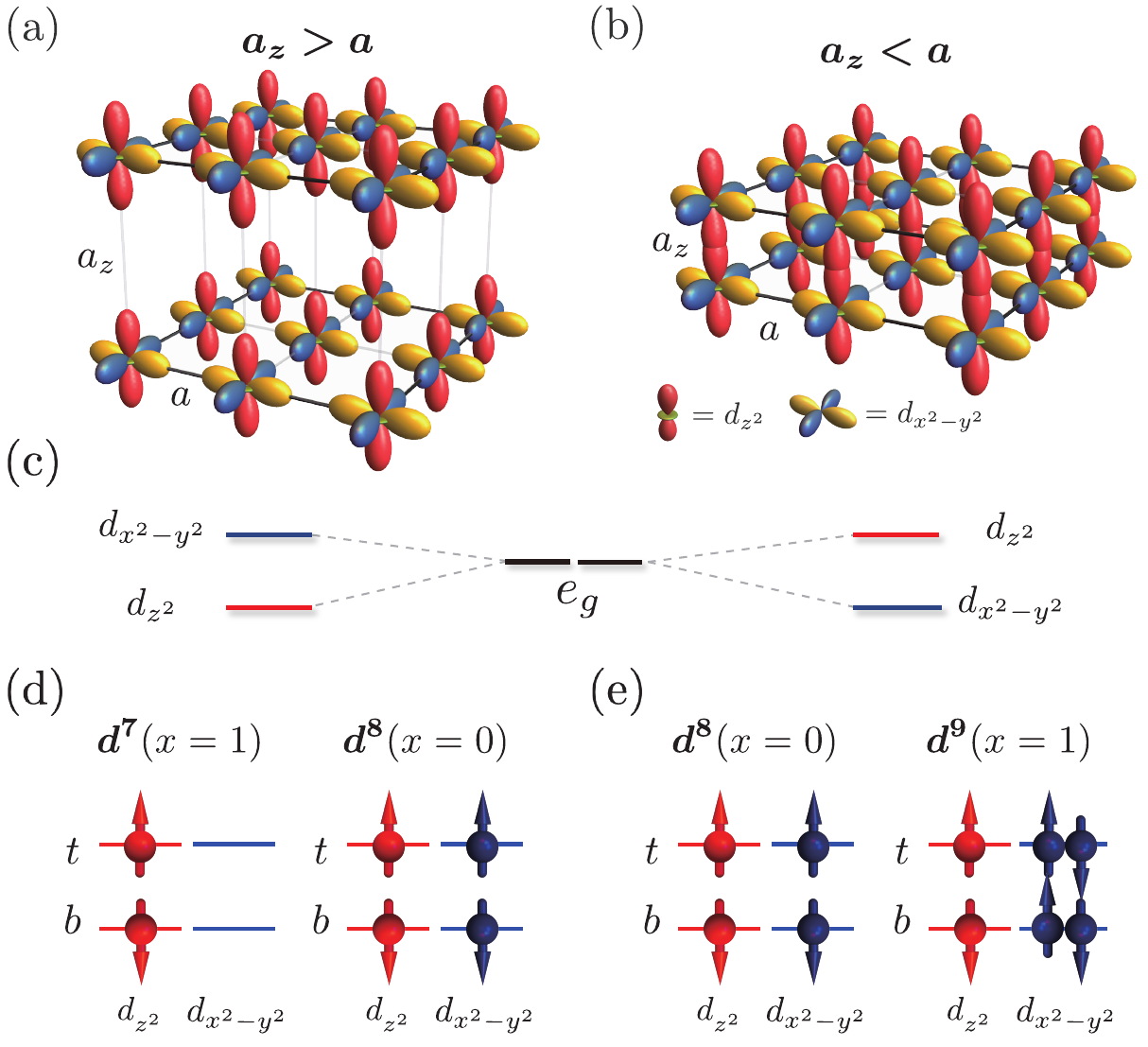}
    \caption{\textbf{Material realization of the high $T_c$ superconductivity.} (a,b) One way to realize the large $J_K,J_\perp$, is to reduce the lattice constant $a_z$ in $z$ direction compared to $a$, the inplane lattice constant in a bilayer nickelate. Conventionally we have $a_z>a$. We propose to search for bilayer materials with $a_z<a$. (c) Comparison of the crystal field energy splitting of the $e_g$ orbitals for $a_{z}>a$ (left) and $a_z<a$ (right). (d) In the conventional $a_{z}>a$ case, one should target the valence $d^{8-x}$ with holes doped into the $d_{x^2-y^2}$ orbital relative to the $d^8$ parent state. (e) In the $a_{z}<a$ case, the crystal field splitting is reversed. Now we should target the valence $d^{8+x}$ to dope electrons into the $d_{x^2-y^2}$ orbital relative to the $d^8$ parent state.  Both (d)(e) should be described by the bilayer type II t-J model, but with a stronger $J_\perp$ (and $V$) in the $a_z<a$ case, leading to a stronger $T_c$. We expect the optimal doping to be slightly smaller than $x<0.5$. }
    \label{fig:10}
\end{figure}

\section{Discussion on pairing mechanism and  material realization}\label{sec:8}

We have demonstrated a quite strong superconductor in the limit $J_\perp\gg t$. Counterintuitively, we show that a net attractive interaction is not necessary and the superconductivity is robust against adding a repulsion $V$ term. Note that the $J_\perp$ term has two separate effects: (1) It provides a net attractive interaction to lower the energy of the doublon state. This is equivalent to the $V n_{i;t}n_{i;b}$ term with $V<0$.  (2) It splits the energy of the $S=1$ doublon relative to the energy of the $S=0$ doublon: $\Delta_d=E(n_T=2,S_T=1)-E(n_T=2,S_T=0)\propto J_\perp$. Conventional mean-field theories rely on the effect (1) to get a pairing from decoupling. This is justified only if we ignore the inter-layer Coulomb repulsion, which may be comparable to the attraction from $J_\perp$. In this work, we add the repulsion $V$ to exactly offset the attraction caused by $J_\perp$. In the end there is no net attraction and mean field theory would predict the absence of a superconductor. Contrary to the mean field expectation, we have shown that the effect (2) of $J_\perp$ term is enough to drive pairing, but in an indirect way. The $J_\perp$ term penalizes the spin-triplet doublon and blockades the free moving of a single electron. In the large $J_\perp$ regime (even with no net attraction or even with a net repulsion), the low energy physics is restricted to a reduced Hilbert space. It turns out that now a Fermi liquid is energetically penalized and the ground state is a strong superconductor driven by purely kinetic energy.

We estimated that the paring gap can be larger than $1.5 t$ in the large $J_\perp$ limit. We expect $T_c$ to be as large as $0.5 t$, limited by the phase stiffness. When decreasing $J_\perp/t$, we expect that the pairing gap decreases, while the phase stiffness actually increases.  This suggests that the $T_c$ first increases and then decreases, with the maximal $T_c$ at a large but finite $J_\perp$. Given that the hopping $t$ is usually several thousand Kelvin in typical solid state systems, the maximal $T_c$ may reach even one thousand Kelvin if one can freely tune $J_\perp$ and $J_K$ in the double Kondo model.

One promising way to realize the double Kondo model with large $J_K,J_\perp$ is to optimize a bilayer nickelate system, analogous to the experimentally investigated La$_3$Ni$_2$O$_7$. In this system, $J_K=-2J_H$ arises from the on-site Hund's coupling between the $d_{z^2}$ and $d_{x^2-y^2}$ orbital. As Hund's coupling is a part of the Coulomb interaction, it is natural to expect $|J_K|\gg t$. However, the primary challenge lies in enhancing the interlayer coupling $J_\perp$, which mediates super-exchange interactions between the $d_{z^2}$ orbital. To enhance $J_\perp/t$, it is desirable to reduce the lattice constant $a_z$ in the $z$ direction compared to the intra-layer lattice constant $a$. The current La$_3$Ni$_2$O$_7$ is still in the regime with $a_z>a$.  We propose to search for bilayer materials with $a_z<a$. In this case, the energy levels of the $d_{z^2}$ and $d_{x^2-y^2}$ orbital would be reverted. But we can look for a superconductor for filling corresponding to the valence $d^{8+x}$ instead of $d^{8-x}$ (see Fig.~\ref{fig:10}).  We estimate the optimal $x$ to be smaller than $x=0.5$ ($x\approx 0.3$ according to Fig.~\ref{fig:3}(b)).

\section{Conclusion}\label{sec:9}
In this work, we demonstrated a kinetic-energy-driven superconductor in a simple model with only nearest-neighbor hopping projected to a restricted Hilbert space. The ESD model hosts a robust superconducting phase with $T_c/t$ potentially at order of 0.5, suggesting a very high $T_c$  if realized in solid-state systems. While real materials can not fully match this ideal model, it suggests a pathway to high $T_c$ superconductors by closely approaching the strong coupling limit, potentially through bilayer nickelates with a reduced z-axis lattice constant. The recently discovered La$_3$Ni$_2$O$_7$ system under pressure may already align with our bilayer model, albeit with a finite $J_\perp/t$. Future work will investigate the kinetic pairing mechanism in the regime of finite $J_\perp/t$.  Theoretically, at the doping level of $x=0.5$, a particle-hole symmetry at a specific parameter precludes a symmetric Fermi liquid, providing a unique opportunity to explore superconductivity in the absence of a traditional Fermi liquid normal state.

\textit{Acknowledgement.} ---
This work was supported by a startup fund from Johns Hopkins University and and the Alfred P. Sloan Foundation through a Sloan Research Fellowship (YHZ).

\textit{Data Availability Statement.} --- The numerical data and custom code supporting the findings
of this article are available from the authors upon reasonable
request. The DMRG calculations were performed using the
publicly available TeNPy package~\cite{tenpy}.

%

\onecolumngrid
\newpage
\clearpage
\setcounter{equation}{0}
\setcounter{table}{0}
\setcounter{page}{1}
\setcounter{section}{0}

\maketitle 
\makeatletter
\renewcommand{\theequation}{S\arabic{equation}}
\renewcommand{\thefigure}{S\arabic{figure}}
\renewcommand{\thetable}{S\arabic{table}}
\renewcommand{\thesection}{S\arabic{section}}

\appendix
\onecolumngrid

\begin{center}
\vspace{10pt}
\textbf{\large Appendix of 
``High-temperature superconductivity from kinetic energy''}
\end{center} 
\begin{center} 
{Hanbit Oh$^{\ \textcolor{red}{*}}$, Hui Yang$^{\ \textcolor{red}{*}}$, and Ya-Hui Zhang$^{\ \textcolor{red}{\dagger}}$
}\\
\emph{William H. Miller III Department of Physics and Astronomy, \\
Johns Hopkins University, Baltimore, Maryland, 21218, USA}
\vspace{5pt}
\end{center}

\section{Mapping from the double-Kondo model to ESD model}\label{sec:s1}
We start with the double-Kondo model Hamiltonian (see Fig.\ref{fig:1}(a)),
\begin{align}
 \hat{H}_{\mathrm{DK}}=&
 -t_{0}\sum_{l=t,b} \sum_{\langle i,j \rangle}
Pc^\dagger_{i;l;\sigma} c_{j;a;\sigma}P
+J_{K}
\sum_{i}\sum_{l=t,b} \vec s_{c;i;a} \cdot \vec S_{i;l}+J_{\perp}\sum_{i}
\vec S_{i;t}\cdot \vec S_{i;b}
+V\sum_{i}n_{i;t}n_{i;b}, 
\end{align}
where $c_{i;l;\sigma}$ is the creation operator of the conduction electron, and 
its spin operator of the is defined as $\vec s_{c;i;l}=\frac{1}{2} \sum_{\sigma,\sigma'}c^\dagger_{i;l\sigma} \vec \sigma_{\sigma \sigma'}c_{i;l\sigma'}$, with a spin quantum number $\sigma=\uparrow,\downarrow$. 
$\vec S_{i;l}$ is  the localized moment of the electron at site $i$ and layer $l=t,b$. 
Starting from the above double Kondo model, we consider the limit of $t\rightarrow 0$ where we can safely ignore the and just consider local spin interaction defined at every site $i$,
\begin{align}
 H_{local}=&
J_{K}
\sum_{a=t,b} \vec s_{c;a} \cdot \vec S_{a}+J_{\perp}
\vec S_{t}\cdot \vec S_{b}
+V n_{t}n_{b},
\end{align}
where we omit the site index $i$ here. 
Depending on the particle number of C-layer, $n_{T}=\sum_{a,\sigma}c^{\dagger}_{a;\sigma}c_{a;\sigma}$ at each site, we can categorizes the states into three cases $(n_C=0,1,2)$. 
Along with $n_C$ and $S_T$, 
we can classify the six states  
\begin{eqnarray*}
 |h\rangle &=&  \frac{1}{\sqrt{2}}\left[
   |\uparrow\downarrow\rangle-|\downarrow\uparrow\rangle
   \right]_{2,3},
   \end{eqnarray*}
\begin{eqnarray*}
|t,\sigma\rangle&
  \sim&
  \left[
|\sigma\sigma\overline{\sigma}\rangle
-
\left(y+\alpha(y)\right)
|\sigma\overline{\sigma}\sigma\rangle
+
\left(y+\alpha(y)-1\right)
|\overline{\sigma}\sigma\sigma\rangle
\right]_{1,2,3}
   \\
|b,\sigma\rangle&
  \sim&
  \left[
|\sigma\sigma\overline{\sigma}\rangle
-
\left(y+\alpha(y)\right)
|\sigma\overline{\sigma}\sigma\rangle
+
\left(y+\alpha(y)-1\right)
|\overline{\sigma}\sigma\sigma\rangle
\right]_{4,3,2}
\end{eqnarray*}
\begin{eqnarray*}
|d\rangle &
  \sim&
  \left[
 |\uparrow \uparrow \downarrow\downarrow \rangle
 + |\downarrow \downarrow\uparrow \uparrow \rangle
-(2y+\beta(y))  
( |\uparrow \downarrow \uparrow\downarrow \rangle
 + |\downarrow \uparrow\downarrow \uparrow \rangle)
+(2r+\beta(y)-1)  
( |\uparrow \downarrow \downarrow\uparrow \rangle
 + |\downarrow \uparrow\uparrow \downarrow \rangle)
\right]_{1,2,3,4}.
\end{eqnarray*}
The subscript $1,4$ denotes the top/bottom layer of conduction electron layer, and $2,3$ denotes top/bottom layer of localized moments, as illustrated in Fig.\ref{fig:1}. We have defined $ \alpha(y) =\sqrt{1 -y+ y^2} , 
   \beta(y) =\sqrt{1 -2y+ 4y^2}$
with $y=J_{K}/J_{\perp}$. 
Also, $\sim$ denotes the normalization factor.
The associated energies of each state are 
\begin{eqnarray}
\epsilon_h&=&-\frac{3}{4}J_\perp,
\quad
   \epsilon_f=\frac{J_\perp}{4}\left[-y
   -1-2\alpha(y)\right],  \quad 
   \epsilon_d=\frac{J_\perp}{4}\left[-2y-1-2\beta(y)\right]+V. 
     \label{eq:onsite}
   \nonumber
\end{eqnarray}
Then, the resultant ESD model Hamiltonian within the six states is 
\begin{eqnarray}
    H&=& -t \sum_{\langle i,j \rangle}
Pc^\dagger_{i;l;\sigma} c_{j;l;\sigma}P
+\sum_{i}\textcolor{black}{V_{\mathrm{eff}}} (n_{h,i}+n_{d,i}),
\label{esd_ham}
\end{eqnarray}
with 
\begin{eqnarray}
    \textcolor{black}{V_{\mathrm{eff}}}[y]&=&\epsilon_{0}[y]+\frac{V}{2}, \quad 
  \epsilon_{0}[y]= \frac{J_{\perp}}{4} [-1 - 
   \beta(r)
   +2 \alpha(r)],
\end{eqnarray}
and $y$ dependence of $\epsilon_{0}$ is illustrated in Fig.\ref{fig:1} in the main-text. Moreover, the electron annihilation operator of the conduction electrons within the six states is written as
\begin{eqnarray}
    c_{i,l,\sigma}
    &=& 
     a_{11}(y) |h \rangle_{i}
    \langle l,\sigma|_{i}
   + \textcolor{black}\epsilon_{\sigma,\overline{\sigma}}
    a_{12}(y)
    |\overline{l},\overline{\sigma}\rangle_{i}
    \langle d|_{i},
    \label{parton}
\end{eqnarray}
with
\begin{eqnarray*}
  a_{11}(r)&=&-\frac{y - 2  - 2 \alpha(y)}{2 \sqrt{
 2y^2 -y \alpha(y) + 2  (-y + 1 + \alpha(y))}},
 \end{eqnarray*}
 \begin{eqnarray*}
a_{12}(y)&=&\frac{
    2y^2 +y(2\alpha(y)+\beta(y))
   +(1+\alpha(y))(1+\beta(y))  
    }{2
    \sqrt{2(2y^2-y\alpha(y)
    +2(-y+1+\alpha(y))
    }
\sqrt{2+4(2y^2+y\beta(y))-(4y+\beta(y))}
    }\label{parton}
\end{eqnarray*}
In the main text, we control $(t/t_{0},r)$ instead of $a_{11},a_{12}$, using the definition  $a_{11}^2=t/t_{0}$ and $a_{12}/a_{11}=r$. 
\begin{figure}
    \centering
    \includegraphics[width=0.35\linewidth]{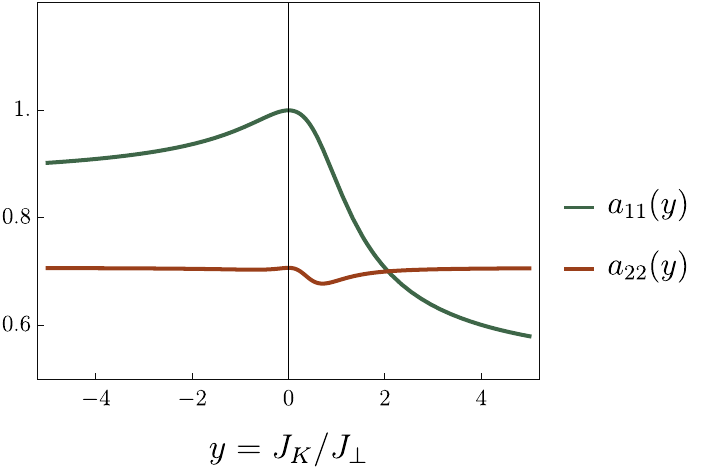}
    \caption{The $J_{K}/J_{\perp}=y$ dependence of $a_{11}$ and $a_{12}$ of the parton construction. In $J_K/J_\perp\rightarrow -\infty$ limit, we have $a_{11}\rightarrow\sqrt{3}/2$ and $a_{12}\rightarrow1/\sqrt{2}$. In $J_K/J_\perp\rightarrow +\infty$ limit, we have $a_{11}\rightarrow 1/2$ and $a_{12}\rightarrow1/\sqrt{2}$. In the main-text, we illustrate $a_{11}^2=t/t_{0}$ and $a_{12}/a_{11}=r$ dependence. }
    \label{fig:s0}
\end{figure}

\section{Details on the numerical simulation of ESD model}\label{sec:s2}
\subsection{fDMRG simulation}
\begin{figure}[h]
    \centering
    \includegraphics[width=0.6\linewidth]{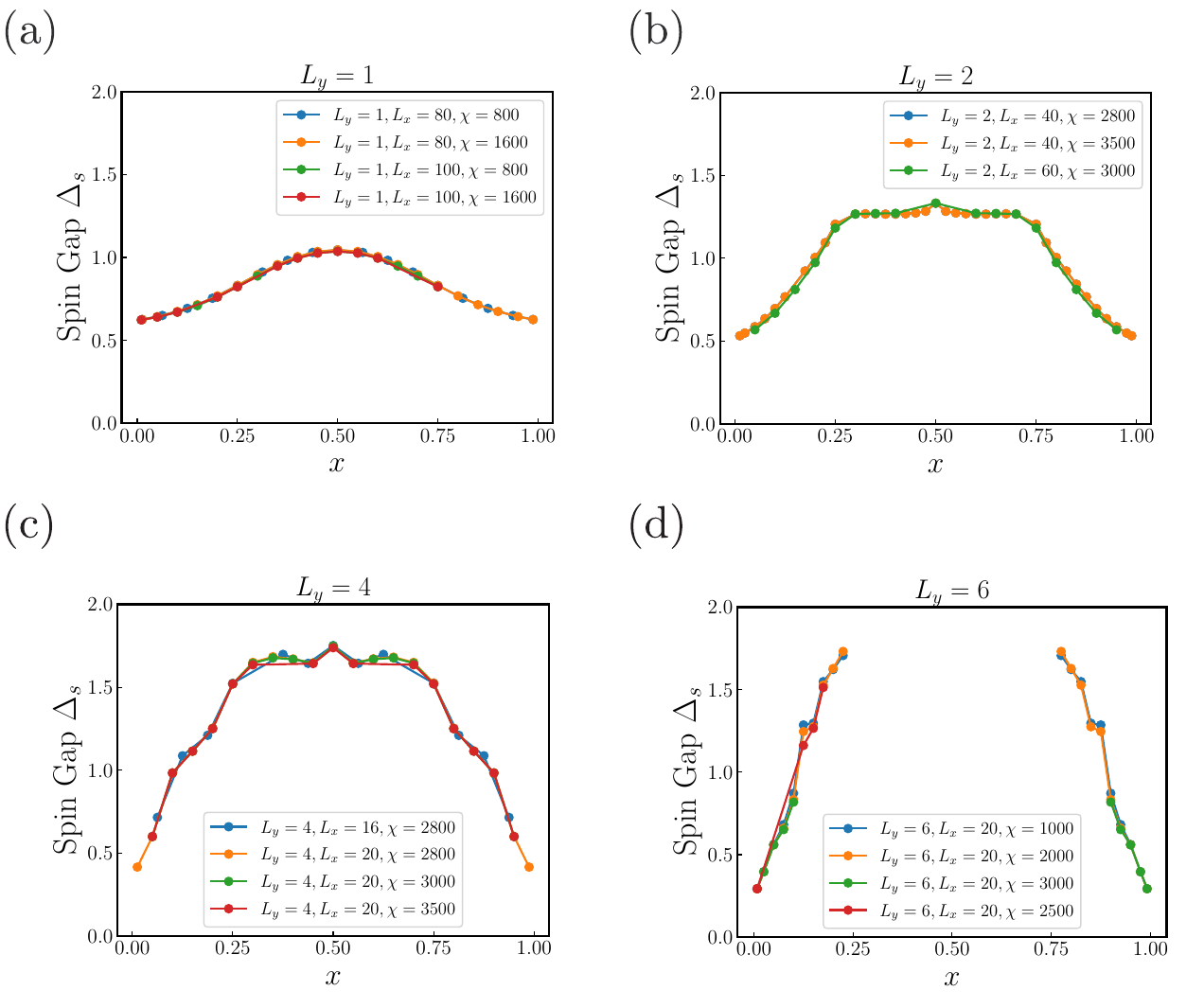}
    \caption{The system size and bond dimension dependence of the spin gap at $L_y=1,2,4,6$. For $L_y=1,2,4$, we achieved the convergent results of the spin gap for all ranges of $x$. However, for $L_y=6$, we only obtain the convergent results in $0<x<0.1$ and $0.9<x<1$, as presented in the main text. At least, based on the not fully convergent data, the scale of the spin gap almost reaches $1.7t$ at $x=0.25$.  }
    \label{fig:s1}
\end{figure}

\begin{figure}[h]
    \centering
\includegraphics[width=0.28\linewidth]{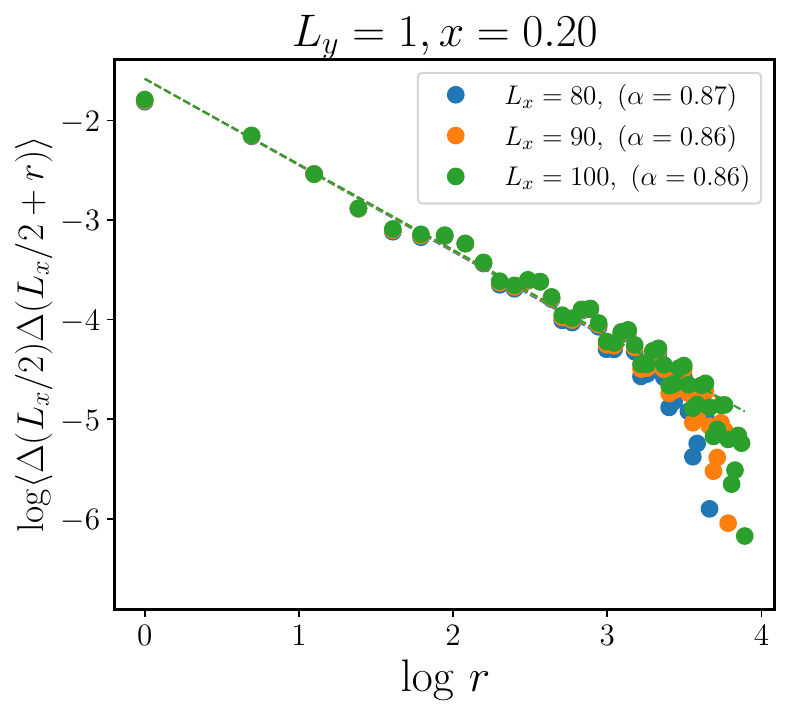}
\includegraphics[width=0.28\linewidth]{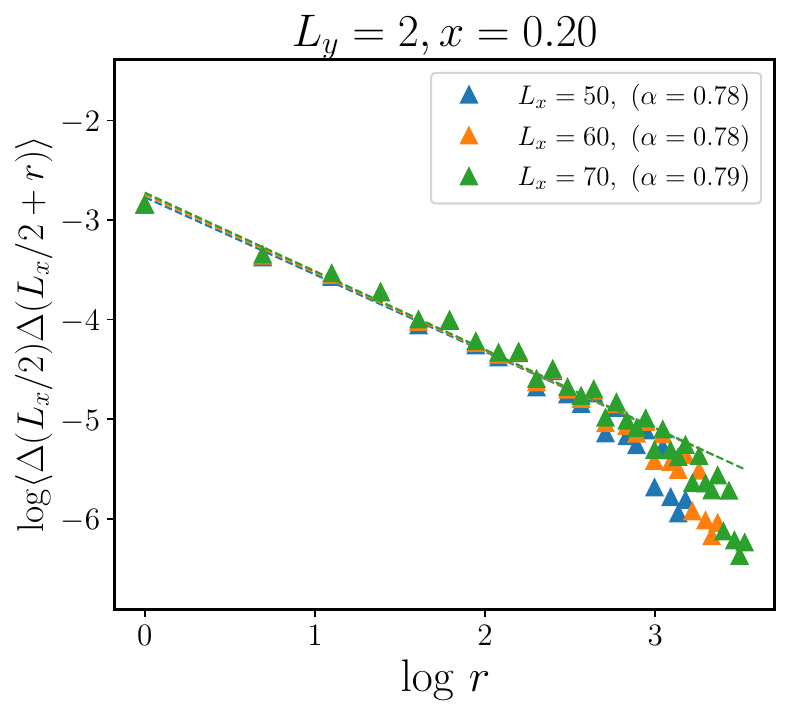}
\includegraphics[width=0.28\linewidth]{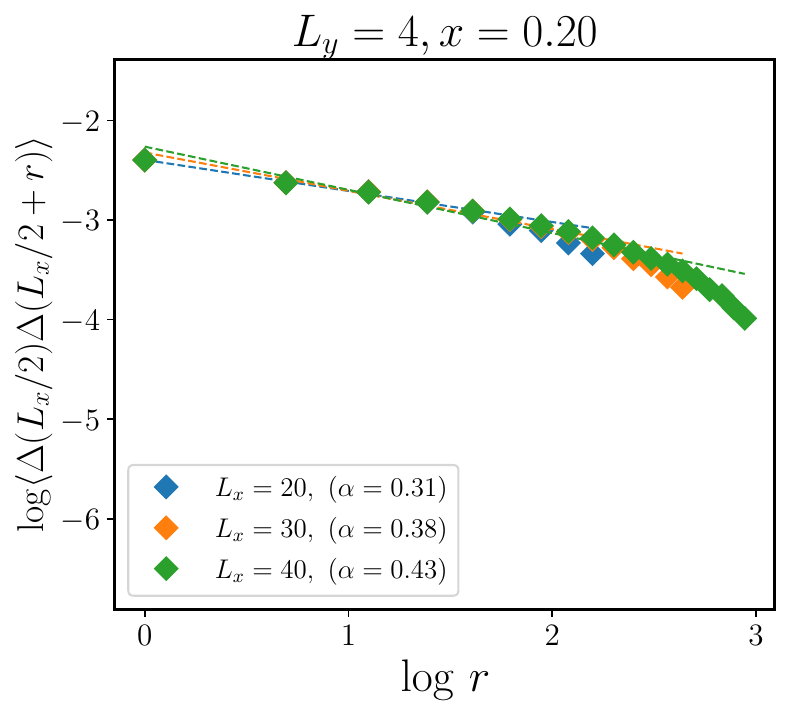}
    \caption{System size $L_x$ dependence of the pair-correlation function with fDMRG at fixed $L_y=1,2,4$ (Compare with Fig.\ref{fig:3}(d) in the main text). 
    Here, we use $\chi=3000$ for all cases. 
    }
    \label{fig:R1}
\end{figure}

In Fig.\ref{fig:s1}, we support the convergence of the spin gap calculation for various system size $L_y=1,2,4,6$ and various bond dimensions. 
\textcolor{black}{In Fig.\ref{fig:R1}, we plot the system size $L_x$ dependence of the pair-pair correlation of $L_y=2$. }

\begin{figure}[h]
    \centering
\includegraphics[width=0.62\linewidth]{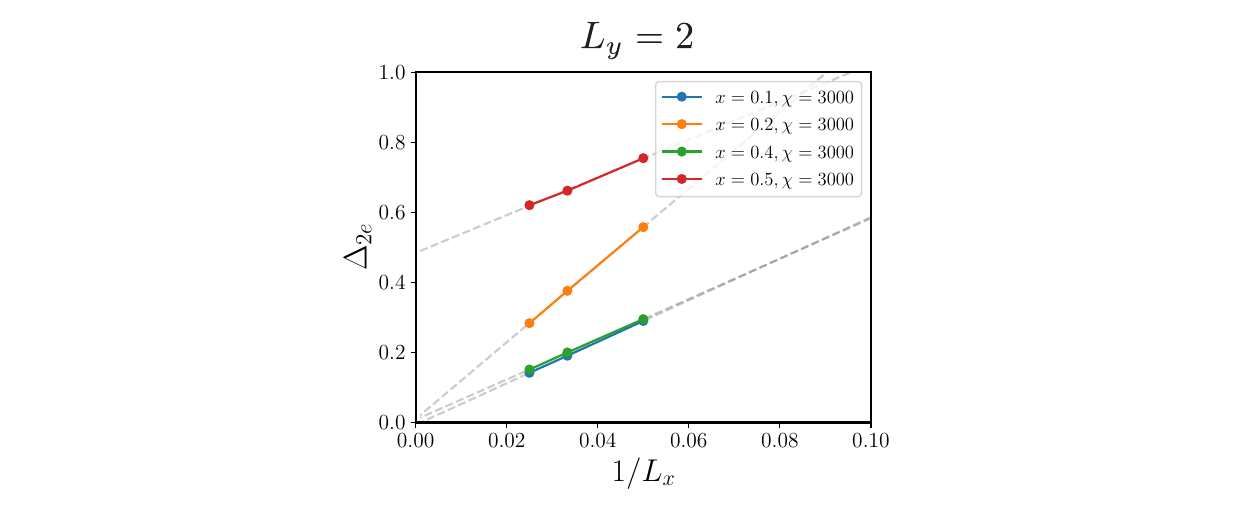}
    \caption{The system size dependence of the charge 2e gap, $\Delta_{2e}$, at $L_y=2$. The charge 2e gap is calculated at bond dimension $\chi=3000$. For $x=0.5$, the charge 2e gap does not vanish even in the large $L_x$ limit. 
  Even in the Drude weight calculation, we have already seen it vanish at $x=0.5$ of $L_y=2$. 
We presume that this special point turns out to be an insulator.  At $x=0.5$, the Cooper pair is at filling $1/2$ per site. For $L_y=2$, we have one Cooper pair per unit cell, so an insulator is allowed.
    }
    \label{fig:s2}
\end{figure}

\begin{figure}[h]
    \centering
\includegraphics[width=0.65\linewidth]{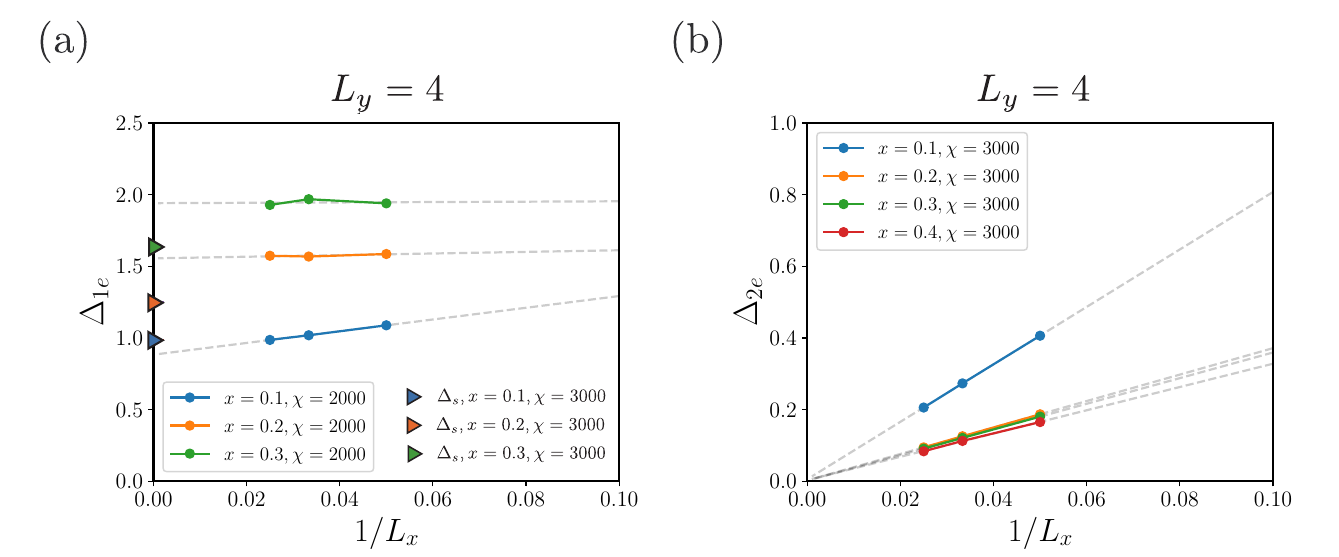}
    \caption{  The system size dependence of the (a) charge 1e gap, $\Delta_e$, and (b) charge 2e gap, $\Delta_{2e}$, at $L_y=4$ and $x = 0.1, 0.2, 0.3$. (a) For the charge 1e gap, we choose bond dimension $\chi=2000$. In the large $L_x$ limit, the  $\Delta_e$ converges to the finite value. The different colored triangles denote the spin gap $\Delta_s$ calculated at $L_y=4$ and $\chi=3000$, which are comparable size with the $\Delta_{1e}$ (b) The charge 2e gap, is calculated at bond dimension $\chi=3000$. For all $x$, the charge 2e gap vanishes in the large $L_x$ limit, indicating that Cooper pairs are gapless in this regime. 
    }
    \label{fig:s3}
\end{figure}

In Fig.\ref{fig:s2} and Fig.\ref{fig:s3}, we provide the data for the charge-1e and the charge-2e gap at  $L_y=2$ and $L_y=4$, respectively. Here, the charge-1e gap is defined as $\Delta_{1e}=E[N+1] +E[N-1]-2 E[N]$, and the charge-2e gap is defined as $\Delta_{2e}=E[N+2] +E[N-2] - 2E[N]$. From Fig.~\ref{fig:s2}, the charge-2e gap is zero for $x=0.1,0.2,0.3,0.4$, while it is finite for $x=0.5$, indicating an insulator at $x=0.5$ with one Cooper pair per unit cell. Fig.~\ref{fig:s3}, we can see, for $L_x=4$, the charge-1e gap is finite, while the charge-2e gap is zero in the large $L_x$ limit. A gapped electron and gapless Cooper pair indicate a superconductor phase.  

\begin{figure}[h]
    \centering
    \includegraphics[width=0.65\linewidth]{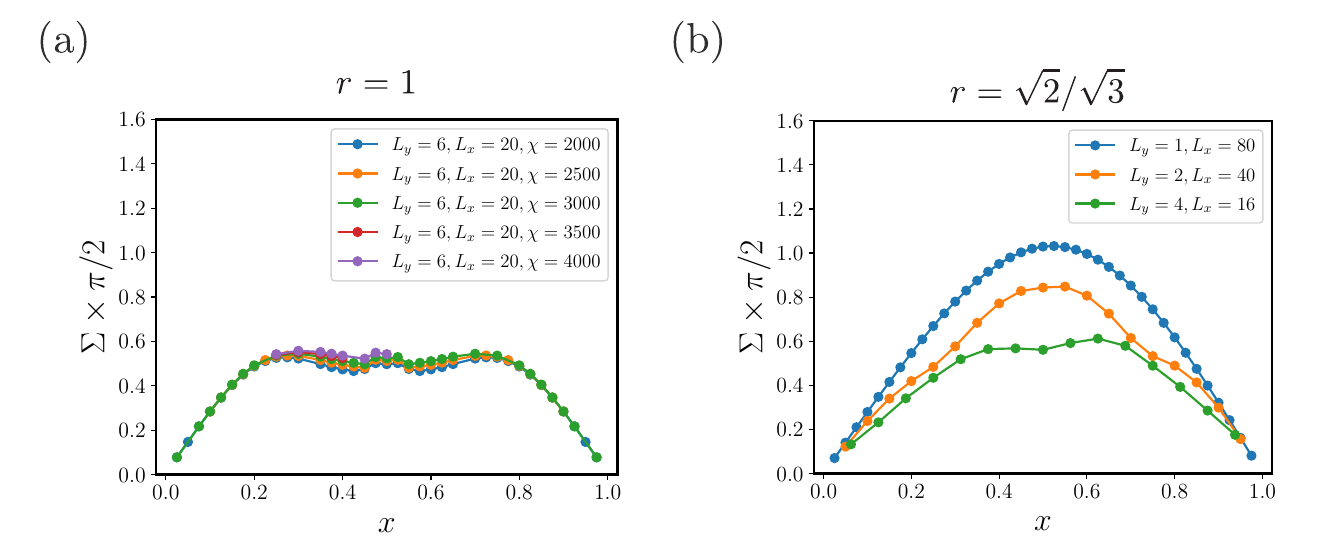}\vspace{-10pt}
    \caption{The $x$ dependence of the total weight at (a) $r=1$ and (b) $r=\sqrt{2}/\sqrt{3}$. (a) For $r=1$, we here show all $x$ ranges of total weight for $L_y=6$, which is not fully convergent. Based on fully convergent results,  the optimal value is almost $\Sigma\times \pi/2\simeq 0.7 t$ at $x\simeq 0.3$. (b) For $r=\sqrt{2}/\sqrt{3}$, we show the fully convergent results of $L_y=1,2,4$ with $\chi=3000$. The overall dome of the total weight for$r=\sqrt{2}/\sqrt{3}$ is shifted from the $r=1$ case. 
    }
    \label{fig:total_weight}
    
\end{figure}

\begin{figure}
    \centering
\includegraphics[width=0.65\linewidth]{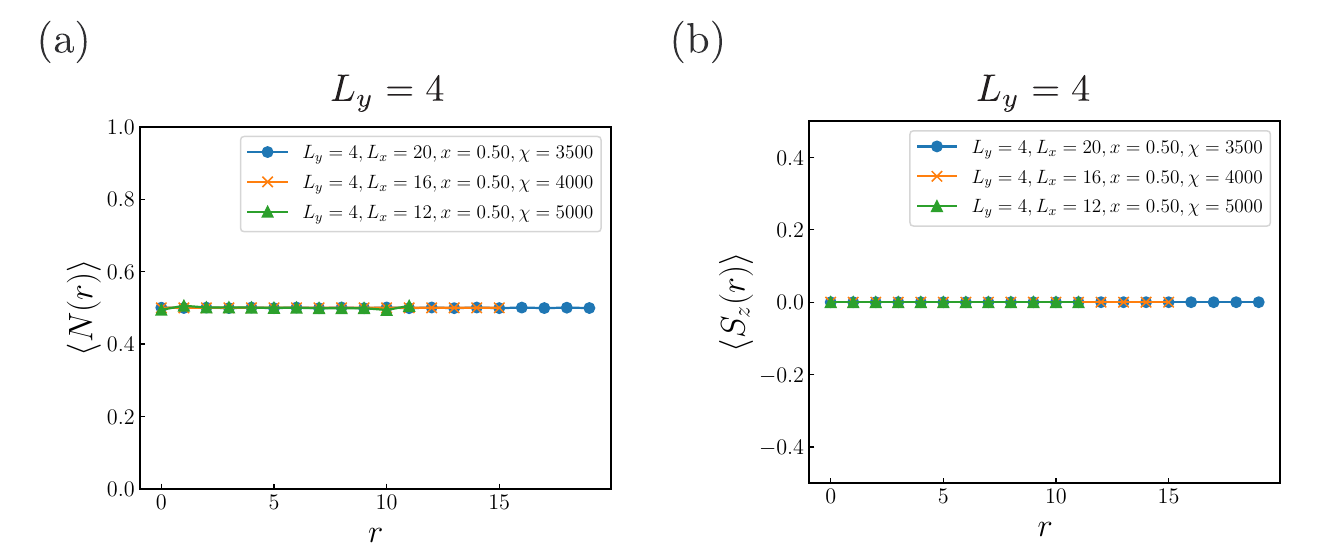}
 \vspace{-5pt}
    \caption{Distribution of $\langle N(r)\rangle$ and $\langle S_{z}(r)\rangle$ at $L_y=4$ and $x=0.5$. There is no translational symmetry-breaking signature. We tend to believe $x=0.5$ is still a superconductor instead of an insulator of Cooper pair in the two-dimensional limit. }
    \label{fig:s4}
\end{figure}

In Fig. \ref{fig:total_weight}, we show the total weight results of $L_y=6$ for all ranges of $x$. Note that in the main text, we only showed the fully convergent region $0<x<0.25$ and $0.75<x<1$. 
In Fig.\ref{fig:s4}, we provide the evidence of a uniform $\langle N(r)\rangle$ and $\langle S_z(r)\rangle=0$ with no translation symmetry breaking for $x=0.5$. 

\subsection{iDMRG simulation}
In Fig.~\ref{fig:s5}, we show the correlation length of various operators for different bond dimensions. In our DMRG simulation, we consider two symmetries, two-particle conservations in top and bottom layer $U_t(1)$, $U_b(1)$ and spin $S^z$ conservation $U(1)$. The correlation length can be calculated from the transfer matrix method in different symmetry sectors. The density operator, spin operator, electron operator, and interlayer pairing operator correspond to different symmetry sectors with $(\delta N_t,\delta N_b, \delta S^z)=$ $(0,0,0)$, $(0,0,1)$, $(1,0,\frac{1}{2})$, $(1,1,0)$. From the plot, we find that the correlation length $\xi_{\Delta}$ seems to go to infinity as we increase the bond dimension, while the correlation lengths of the spin operator and the electron operator remain finite as the bond dimension increases.
\begin{figure}
    \centering
\includegraphics[width=0.4\linewidth]{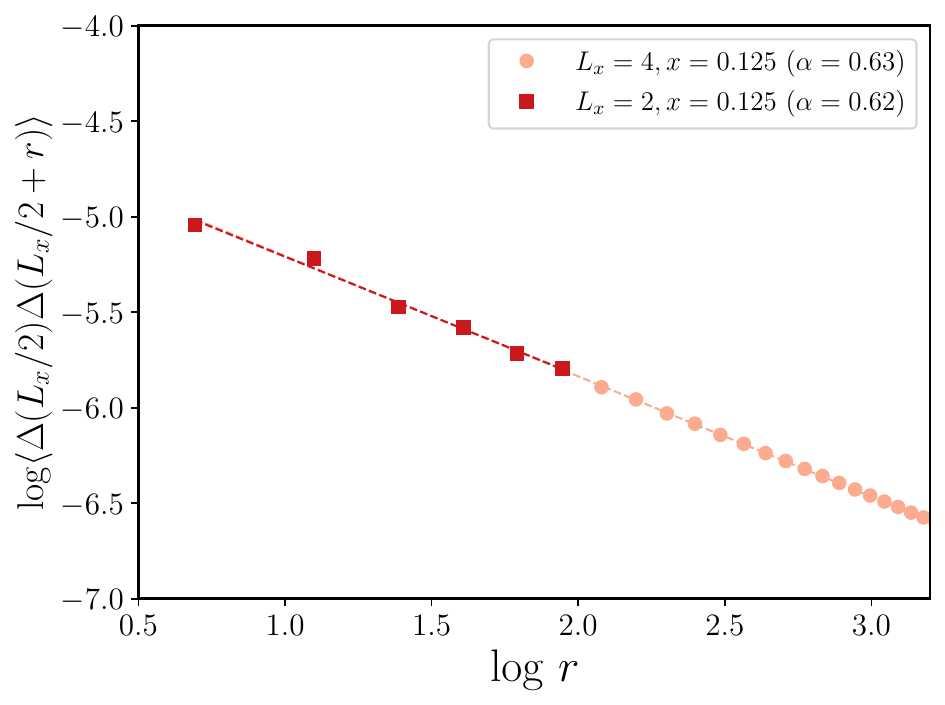}
\caption{\textcolor{black}{\textbf{The pair-correlation function with iDMRG at fixed $L_y=4$ and $L_x=2,4$ (Compare with Fig.6(b) in the main-text).} Here, we use $\chi=12000$ for $L_x=4$ and we use $\chi=10000$ for $L_x=2$.} }
    \label{fig:R1_2}
\end{figure}

\begin{figure}[h]
    \centering
\includegraphics[width=.65\linewidth]{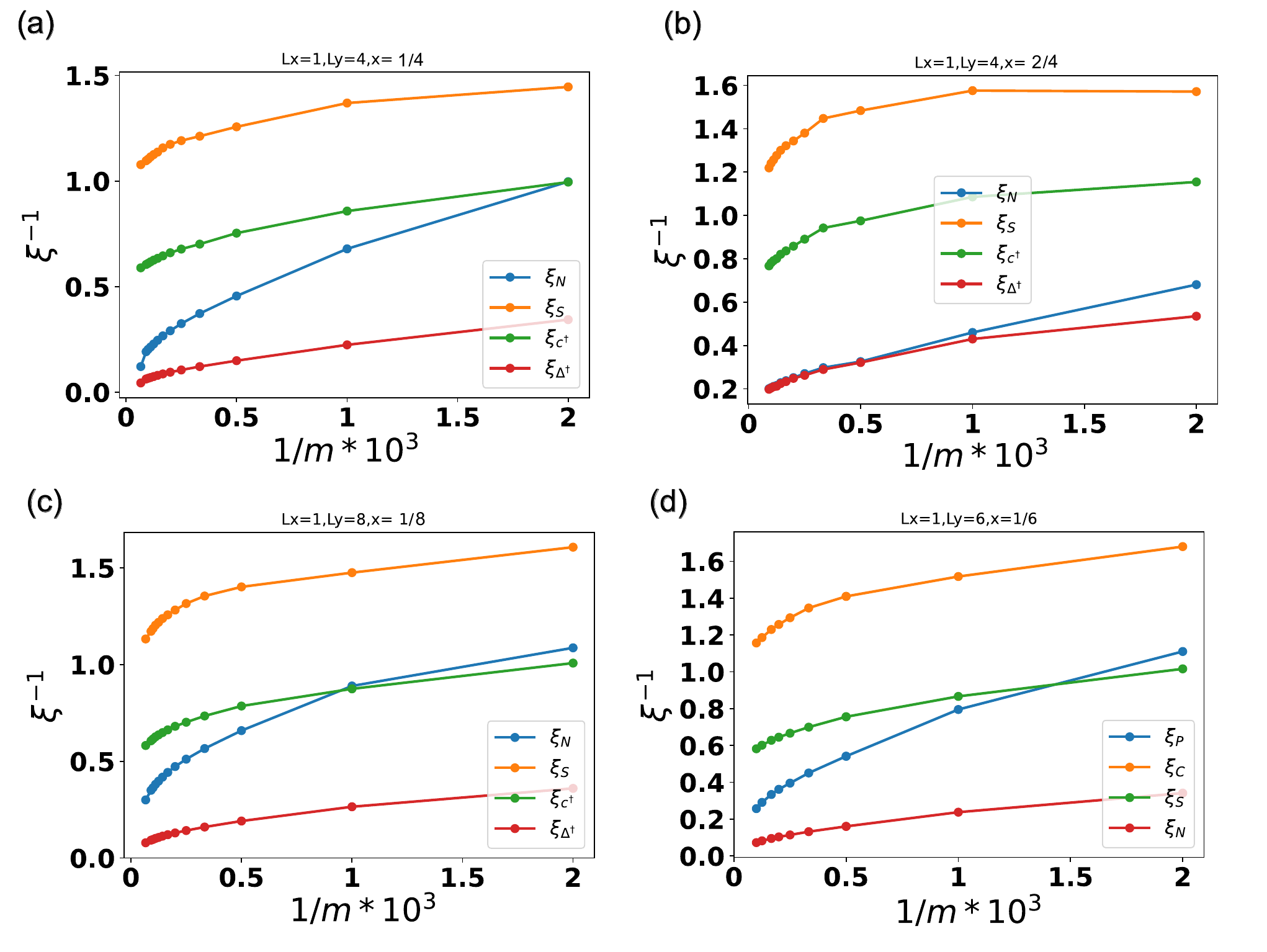}\vspace{-10pt}
    \caption{The correlation length of the ESD model from the iDMRG simulation. (a) $L_x=1,L_y=4,x=1/4$, (b) $L_x=1,L_y=4,x=2/4$, (c)$L_x=1,L_y=8,x=1/8$, (d) $L_x=1,L_y=6,x=1/6$. Here $\xi_N$, $\xi_S$, $\xi_{c^\dagger}$, $\xi_{\Delta^\dag}$ correspond to the correlation length of the density operator, spin operator, electron operator, and the interlayer pairing operator, respectively.} 
    \label{fig:s5}
\end{figure}

\clearpage
\section{Details on the \textbf{exact solution of two-particle problems} calculation}\label{sec:s3}
In this section, we present a detailed derivation showing how the two-electron or two-hole problem of the ESD model can be mapped onto a tight-binding Hamiltonian with impurities located adjacent to the origin (see Fig. \ref{fig:5}).
\subsection{$N_e=2$ case}
Let us consider electron relative to the fully empty vacuum near $x = 1$, which corresponds to a product state of $|h\rangle$, $\ket{G} = \prod_i \ket{h}_i$. Relative to this ground state, we define a single-electron state as $\tilde{c}_{i,l,\sigma}^\dagger \ket{G}$, where the singlon creation operator is given by $\tilde{c}_{i,l,\sigma}^\dagger = \ket{l,\sigma}_i \bra{h}_i$. We can also define an creation operator of Cooper pair of electron, $\Delta_i^\dagger = \ket{d}_i \bra{h}_i  $.
We can construct the two electron states separated by a distance of $\bm{R}$, forming a spin singlet, 
\begin{eqnarray}
\ket{\Psi^{(2e)};\bm{R}}&=&\begin{cases}
\frac{1}{\sqrt{N_s}} \sum_{\bm{i}} 
\Delta^{\dagger}_{i}
\ket{G}, \quad \mathrm{if\ }\bm{R}=(0,0)\\
\frac{1}{\sqrt{2N_s}} \sum_{\bm{i}} \left[ 
\tilde{c}_{\bm{i},t,\uparrow}^\dagger
\tilde{c}_{\bm{i}+\bm{R},b,\downarrow}^\dagger
-\tilde{c}_{\bm{i},t,\downarrow}^\dagger
\tilde{c}_{\bm{i}+\bm{R},b,\uparrow}^\dagger
\right]\ket{G},\quad \mathrm{Otherwise}. \\
\end{cases}
\end{eqnarray}
where $N_s$ is the total number of sites. 
Taking the ESD Hamiltonian onto the $\ket{\Psi^{(2e)};\bm{R}}$, one can find that 
 \begin{eqnarray}
H_{\mathrm{ESD}}\ket{\Psi^{(2e)};\bm{R}\neq \bm{0}}
=  \sum_{\bm{\xi}}
\left[
\left[ -2t|a_{11}|^2 \left(1-\delta_{\bm{R}+\bm{\xi},0}\right)
-2\sqrt{2}ta_{11}a_{12}^{*} \delta_{\bm{R}+\bm{\xi},0}\
\right]\ket{\Psi^{(2e)}_{-};\bm{R}+\bm{\xi}} \right]+(N_s-2)\textcolor{black}{V_{\mathrm{eff}}}\ket{\Psi^{(2e)};\bm{R}}, 
\label{eq:hop1}
    \end{eqnarray}
and 
\begin{eqnarray}
    H_{\mathrm{ESD}} \ket{\Psi^{(2e)};\bm{R}=\bm{0}}   
&=& 
 \sum_{\bm{\xi}} \left[-2\sqrt{2} t a_{11}^{*}a_{12}
\ket{\Psi^{(2)};\bm{\delta} }
\right]+N_s \textcolor{black}{V_{\mathrm{eff}}} \ket{\Psi^{(2e)};\bm{0} }, 
\label{eq:hop2}
    \end{eqnarray}  
where $\bm{\xi}$ denotes the nearest neighbor vector. 
Then, we can write $H^{(2e)}_{\bm{R},\bm{R}'}=\langle\Psi^{(2e)};\bm{R}|H|\Psi^{(2e)};\bm{R}'\rangle$ in a compact form, 
\begin{eqnarray}
    H^{(2e)}_{\bm{R},\bm{R}'}
    &=&-t'\sum_{\bm{\xi}}\delta_{\bm{R}',\bm{R}+\bm{\xi}}
    (1-\delta_{\bm{R},\bm{0}})(1-\delta_{\bm{R}',\bm{0}})
    -t''\sum_{\bm{\xi}}\delta_{\bm{R}',\bm{R}+\bm{\xi}}
[\delta_{\bm{R},\bm{0}}+\delta_{\bm{R}',\bm{0}}]
+w\delta_{\bm{R},\bm{0}}\delta_{\bm{R}',\bm{0}}.
\end{eqnarray}
Here, we have $t' = 2t|a_{11}|^2 $, $t'' = 2\sqrt{2}a_{11}^* a_{12}t$ and $w=2\textcolor{black}{V_{\mathrm{eff}}}$. Note that we redefine the zero energy shifting by $(N_s-2)\textcolor{black}{V_{\mathrm{eff}}}$. If we only control $a_{12}=r$ with fixing $a_{11}=1$, we have $t' = 2t$, $t'' = 2\sqrt{2}rt$ and $w=2\textcolor{black}{V_{\mathrm{eff}}}$. 
In the main-text, we provide the system size $(L_x, L_y)$ dependence of the binding energy at $r=1,\textcolor{black}{V_{\mathrm{eff}}}=0$. 
Notably, the binding energy remains as a considerable size $E_{b}= 0.127t$, even in the true two-dimensional limit as $L_x, L_y \rightarrow \infty$.

\subsection{$N_h=2$ case}
Let us consider two holes relative to the fully occupied vacuum near $x = 0$, which corresponds to a product state of $|d\rangle$, $\ket{G} = \prod_i \ket{d}_i$. Relative to this ground state, we define a single-hole state as $\tilde{c}_{i,l,\sigma}^\dagger \ket{G}$, where the hole creation operator is given by $\tilde{c}_{i,l,\sigma}^\dagger = \ket{\overline{l},\overline{\sigma}}_i \bra{d}_i$. We can also define an creation operator of Cooper pair of holes, $\Delta_i^\dagger = \ket{h}_i \bra{d}_i $.
We can construct the two hole states separated by a distance of $\bm{R}$, forming a spin singlet, 
\begin{eqnarray}
\ket{\Psi^{(2h)};\bm{R}}&=&\begin{cases}
\frac{1}{\sqrt{N_s}} \sum_{\bm{i}} 
\Delta^{\dagger}_{i}
\ket{G}, \quad \mathrm{if\ }\bm{R}=(0,0)\\
\frac{1}{\sqrt{2N_s}} \sum_{\bm{i}} \left[ 
\tilde{c}_{\bm{i},t,\uparrow}^\dagger
\tilde{c}_{\bm{i}+\bm{R},b,\downarrow}^\dagger
-\tilde{c}_{\bm{i},t,\downarrow}^\dagger
\tilde{c}_{\bm{i}+\bm{R},b,\uparrow}^\dagger
\right]\ket{G},\quad \mathrm{Otherwise}. \\
\end{cases}
\end{eqnarray}
Then, one can find the two-body Hamiltonian  $H^{(2h)}_{\bm{R},\bm{R}'}=\langle\Psi^{(2h)};\bm{R}|H|\Psi^{(2h)};\bm{R}'\rangle$ in the two-hole case is written as
\begin{eqnarray}
    H^{(2h)}_{\bm{R},\bm{R}'}
    &=&-t'\sum_{\bm{\xi}}\delta_{\bm{R}',\bm{R}+\bm{\xi}}
    (1-\delta_{\bm{R},\bm{0}})(1-\delta_{\bm{R}',\bm{0}})
    -t''\sum_{\bm{\xi}}\delta_{\bm{R}',\bm{R}+\bm{\xi}}
[\delta_{\bm{R},\bm{0}}+\delta_{\bm{R}',\bm{0}}]
+w\delta_{\bm{R},\bm{0}}\delta_{\bm{R}',\bm{0}}. 
\end{eqnarray}
Here, we find $t' = 2t|a_{12}|^2 $, $v = 2\sqrt{2}a_{12}^* a_{11}t$ and $w=2\textcolor{black}{V_{\mathrm{eff}}}$. 
Therefore, with $a_{11},a_{12}$, we have $E_{b}[a_{11},a_{12};N_h=2]=E_{b}[a_{12},a_{11};N_e=2]$. 
With $a_{11}=1$ and $a_{12}=r$, we have $t' = 2r^2t$, $t'' = 2\sqrt{2}rt$ and $w=2\textcolor{black}{V_{\mathrm{eff}}}$. In this case, we can find  $E_{b}[r;N_h=2]=E_{b}[1/r;N_e=2]r^2$ at $\textcolor{black}{V_{\mathrm{eff}}}=0$. 

\begin{figure}[tb]
    \centering
\includegraphics[width=.8\linewidth]{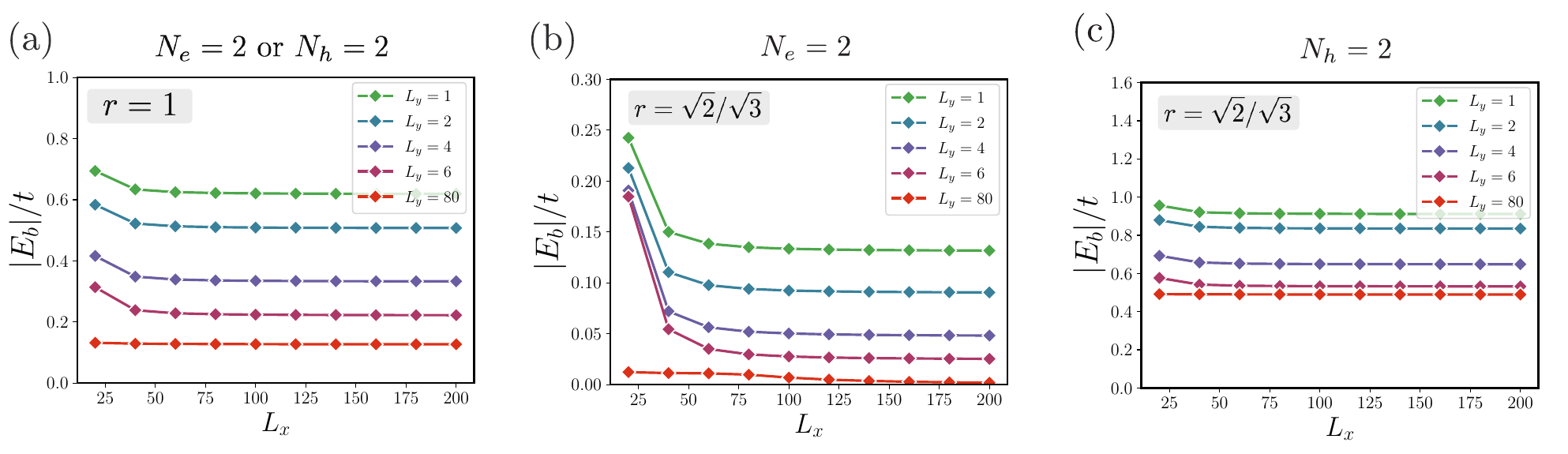}
\vspace{-10pt}
\caption{The system size dependence of the binding energy via ED calculations with various $r$ and $L_y$. (a) For $r=1$, due to the particle-hole symmetry, $E_b[N_e=2]=E_b[N_h=2]$ for all $L_y$. (b,c) For $r={\sqrt{2}}/{\sqrt{3}}$, the binding energy is only finite at $N_h=2$ case in the two-dimension limit $L_y\rightarrow \infty$. 
}
    \label{fig:enter-label}
\end{figure}

\subsection{{Analytic solution in the two-dimensional limit}}

{

The two-particle problem in Eq.~(13) can be solved exactly in the
two-dimensional thermodynamic limit. The relative-coordinate
Hamiltonian is a uniform square lattice with hopping $t'$, except for
the four bonds connected to $\bm R=0$, where the hopping is $t''$,
and the on-site potential $w$ at $\bm R=0$.

To solve this defect problem, we introduce the Green's function of the
uniform square lattice,
\begin{equation}
G(E;\bm R)
=
\int_{\mathrm{BZ}}
\frac{d^2k}{(2\pi)^2}
\frac{e^{i\bm k\cdot\bm R}}
{E+2t'(\cos k_x+\cos k_y)}.
\label{eq:G_square}
\end{equation}
In particular, $G(E;\bm 0)$ is the local Green's function at the
origin. Integrating out the states with $\bm R\neq0$ reduces the
bound-state problem to the scalar equation
\begin{equation}
\left[
\left(
\frac{t''^{\,2}}{t'^{\,2}}-1
\right)E+w
\right]
G(E;\bm 0)
=
\frac{t''^{\,2}}{t'^{\,2}}.
\label{eq:Eb_secular}
\end{equation}
For the two-dimensional square lattice, the local Green's function
below the continuum can be evaluated exactly as
\begin{equation}
G(E;\bm 0)
=
\frac{2}{\pi E}
K\!\left[
\left(\frac{4t'}{E}\right)^2
\right],
\qquad
E<-4t',
\label{eq:G_square_exact}
\end{equation}
where $K$ is the complete elliptic integral of the first kind.

The bottom of the continuum is $-4t'$. Therefore, the energy of a
bound state can be written as
\begin{equation}
E=-4t'-E_b,
\qquad
E_b>0.
\end{equation}
Substituting Eq.~\eqref{eq:G_square_exact} into
Eq.~\eqref{eq:Eb_secular}, we obtain
\begin{equation}
\left[
\frac{t''^{\,2}}{t'^{\,2}}-1
-\frac{w}{4t'+E_b}
\right]
\frac{2}{\pi}
K\!\left[
\left(\frac{4t'}{4t'+E_b}\right)^2
\right]
=
\frac{t''^{\,2}}{t'^{\,2}}.
\label{eq:Eb_exact}
\end{equation}
Thus, for given $t'$, $t''$, and $w$, the positive root of
Eq.~\eqref{eq:Eb_exact} gives the exact binding energy in the
two-dimensional thermodynamic limit. This equation is obtained
directly from the thermodynamic-limit Green's function and does not
require a finite-size extrapolation.

The critical interaction $V_{\mathrm{eff},c}$ can also be obtained
analytically. As $E_b\rightarrow0^{+}$, the elliptic integral in
Eq.~\eqref{eq:Eb_exact} diverges. Therefore, the prefactor multiplying
the elliptic integral must vanish at the binding threshold. Using
$w=2V_{\mathrm{eff}}$, we obtain
\begin{equation}
V_{\mathrm{eff},c}
=
2t'
\left(
\frac{t''^{\,2}}{t'^{\,2}}-1
\right).
\label{eq:Veffc_exact}
\end{equation}
Using the values of $t'$ and $t''$ defined above for the two-electron
and two-hole problems gives
\begin{equation}
V_{\mathrm{eff},c}^{(2e)}
=
4t(2r^2-1),
\qquad
V_{\mathrm{eff},c}^{(2h)}
=
4t(2-r^2).
\label{eq:Veffc_2e2h}
\end{equation}
A normalizable bound state therefore exists for
$V_{\mathrm{eff}}<V_{\mathrm{eff},c}$, while
$E_b\rightarrow0$ as
$V_{\mathrm{eff}}\rightarrow V_{\mathrm{eff},c}^{-}$.
Exactly at $V_{\mathrm{eff}}=V_{\mathrm{eff},c}$, the state reaches
the bottom of the continuum and is no longer a normalizable isolated
bound state.

The corresponding bound-state wave function can also be obtained from
the same Green's function. Writing
\begin{equation}
|\Psi\rangle_{\mathrm{bound}}
=
\sum_{\bm R}
f(\bm R)
|\Psi^{(2h)};\bm R\rangle ,
\end{equation}
the amplitudes for $\bm R\neq0$ satisfy
\begin{equation}
\frac{f(\bm R)}{f(\bm 0)}
=
\frac{t''}{t'}
\frac{G(E;\bm R)}
{G(E;\bm 0)},
\qquad
\bm R\neq0,
\label{eq:fR_exact}
\end{equation}
where $E=-4t'-E_b$, and $f(\bm 0)$ is fixed by the normalization
condition
\begin{equation}
\sum_{\bm R}|f(\bm R)|^2=1.
\end{equation}
Therefore, once $E_b$ is obtained from Eq.~\eqref{eq:Eb_exact},
the full bound-state wave function is determined without any
finite-size diagonalization.

For a shallow bound state, $E_b\ll t'$, the long-distance behavior of
Eq.~\eqref{eq:fR_exact} is
\begin{equation}
f(\bm R)
\propto
K_0\!\left(
\sqrt{\frac{E_b}{t'}}\,|\bm R|
\right),
\end{equation}
where $K_0$ is the modified Bessel function. Thus, the characteristic
decay length grows as
\begin{equation}
\xi
\simeq
\sqrt{\frac{t'}{E_b}},
\label{eq:xi_exact}
\end{equation}
and diverges rapidly as the binding threshold is approached.

To verify the analytical solution, Fig.~\ref{fig:Eb_finite_size}
compares the thermodynamic-limit value obtained from
Eq.~\eqref{eq:Eb_exact} with finite-size exact diagonalization.

\begin{figure}
    \centering
    \includegraphics[width=0.8\linewidth]{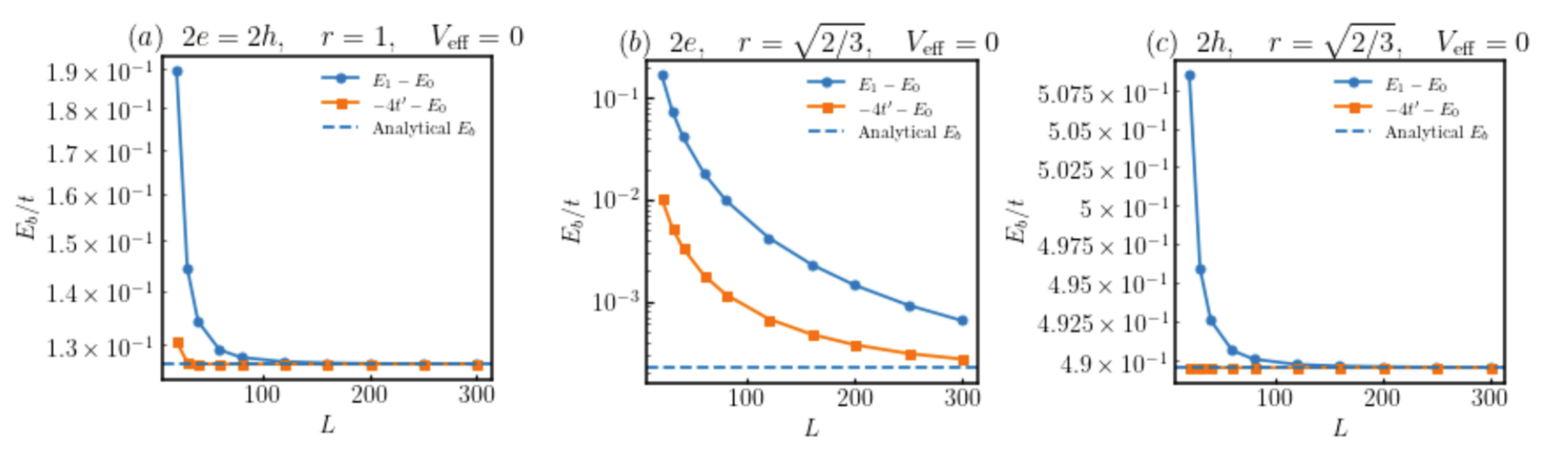}
    \caption{{
    \textbf{Comparison between the thermodynamic-limit value obtained from
Eq.~\eqref{eq:Eb_exact} with finite-size exact diagonalization.}
    We plot both $E_1-E_0$ and $-4t'-E_0$ as functions of
$L=L_x=L_y$. Both quantities approach the analytical binding energy
with increasing $L$. The convergence is especially slow for the
$N_e=2$, $r=\sqrt{2}/\sqrt{3}$, $V_{\mathrm{eff}}=0$ case because
the binding energy is very small,
$E_b/t=0.000223232$, corresponding to a spatially extended bound
state.}}
    \label{fig:Eb_finite_size}
\end{figure}
}

\subsection{Analytic solution in the one-dimensional limit}
\begin{figure}[h]
    \centering
\includegraphics[width=0.7\linewidth]{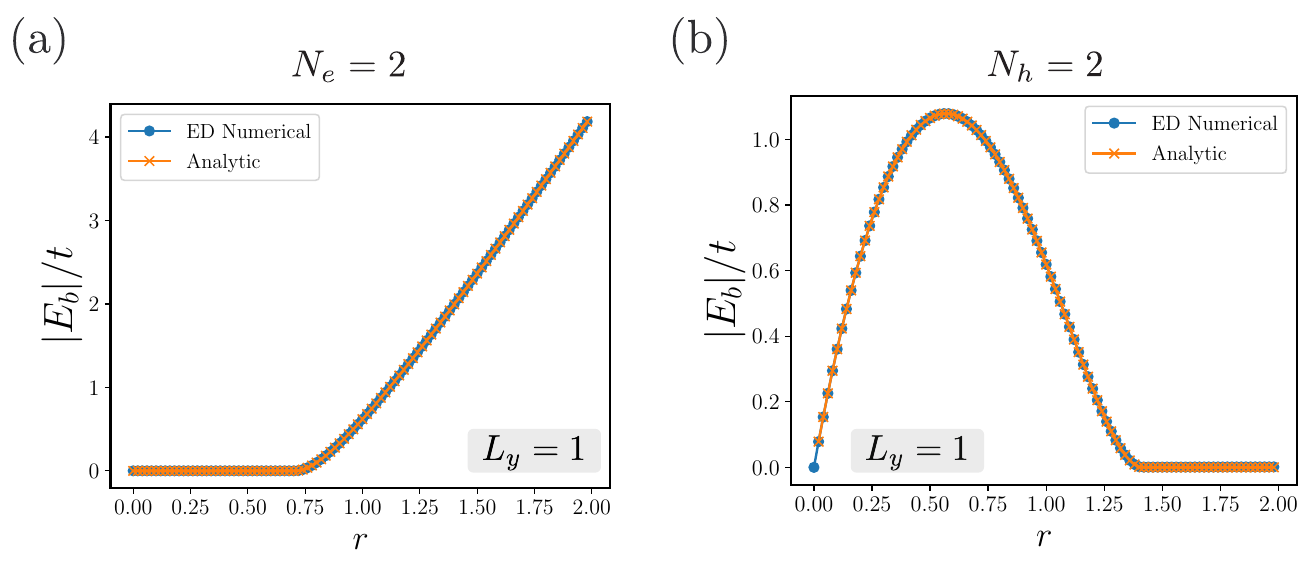}
\vspace{-10pt}
    \caption{The analytic solution of binding energy for (a) $N_e=2$ and (b) $N_h=2$. The analytical solution Eq.\ref{eq:analy} and numerical diagonalization of Eq.\ref{e2} offer consistent results for both $N_e=2$ and $N_h=2$ cases. }
    \label{fig:ed_compare}
\end{figure}

For $L_y=1$, we can show the analytic solution of $E_b$.
We consider the tight-binding Hamiltonian in the presence of a single impurity potential and a single bond disorder, in Eq.(\ref{eq:H_dis}),
\begin{eqnarray}
    H^{(2e)}
    &=& -t' \sum_{i=-L/2+1}^{L/2}  c_{i+1}^{\dagger}c_{i}
    +h.c.+
 wc_{0}^{\dagger}c_{0}
 +(t'-t'')[c_{0}^{\dagger}(c_{1}+c_{-1})+ h.c.]. 
 \label{e2}
 \end{eqnarray}
The $t''$, $v$ term only changes the matrix component near $a_{0},a_{1},a_{-1}$, so we can take the ansatz, 
\begin{eqnarray}
    a_{n}=\alpha e^{-\lambda|n-1|}a_{0}, 
    \quad 
E=-2t'\cosh[\lambda].
\end{eqnarray}
The remaining variables $\lambda$,$\alpha$ are determined by two independent equations, 
\begin{eqnarray}
    -t''[a_{-1}+a_{1}]+w a_{0}= 
    E a_{0}, 
    \quad
    -t'' a_{0}-t'a_{2} =E a_{1}, 
\end{eqnarray}
or equivalently, 
\begin{eqnarray}
    2 t''\alpha-w=t(e^{\lambda}+e^{-\lambda}), 
    \quad
    t'' +t'\alpha e^{-\lambda}=t'(e^{\lambda}+e^{-\lambda}). 
\end{eqnarray}
Equating the two equation leads $e^{\lambda}=\sqrt{\frac{2(t'')^2}{t'^2}-1+\left(\frac{w}{2t'}\right)^2}-\frac{w}{2t'}$ and the boundedness condition restricts $e^{\lambda}>1$. Hence, the binding energy is given by 
\begin{eqnarray}
    E_{b}(t',t'',w)&=&-
   \left[ \left(\sqrt{\frac{2(t'')^2}{t'^2}-1+\left(\frac{w}{2t'}\right)^2}-\frac{w}{2t'}\right)
   +\left(\sqrt{\frac{2(t'')^2}{t'^2}-1+\left(\frac{w}{2t'}\right)^2}-\frac{w}{2t'}\right)^{-1}
   -2t'
   \right]\nonumber \\
   &&\times 
   \theta
\left(\sqrt{
   2(t'')^2-t'^2+\frac{w}{2}}-\frac{w}{2}-1
   \right).\label{eq:analy}
\end{eqnarray}
Putting  $t'=2t,t''=2\sqrt{2}rt,w=2\textcolor{black}{V_{\mathrm{eff}}}$ for $N_e=2$ and $t'=2r^2t,t''=2\sqrt{2}rt,w=2\textcolor{black}{V_{\mathrm{eff}}}$ for $N_h=2$ 
leads to the expression of the binding energy. In Fig.\ref{fig:ed_compare} we compare the obtained analytic results and the numerical matrix diagonalization results to check the consistency.

\section{Layer dependent Zeeman field and mapping to the free fermion model}\label{sec:5}
\begin{figure}[b]
    \centering
\includegraphics[width=.7\linewidth]{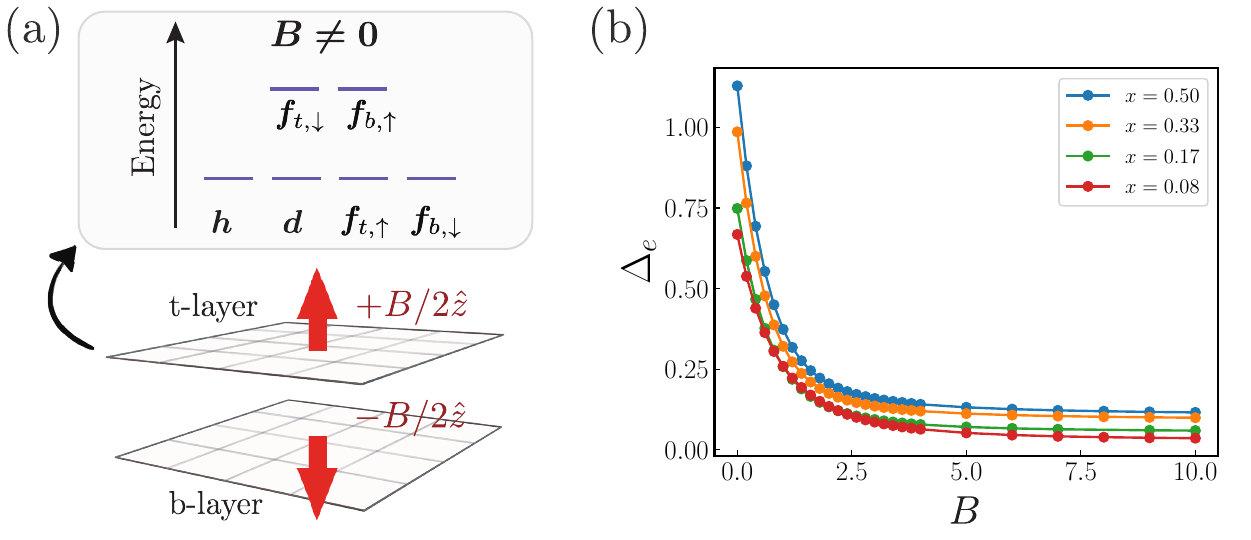}
\caption{
\textbf{The ESD model coupled to a layer-dependent Zeeman field $B$.}
(a) Setup for $\hat{H}_B$ (Eq.~\ref{eq:esd+zeeman}). Applying Zeeman fields in opposite directions to the two layers, along with setting $\textcolor{black}{V_{\mathrm{eff}}} = -B/2$, lifts the ground-state degeneracy. In the limit of large $B$, only four states ($|h\rangle$, $|d\rangle$, $|t, \uparrow\rangle$, and $|b, \downarrow\rangle$) are relevant, enabling a mapping of the model to a free-electron model with additional interactions ($\hat{H}_{\mathrm{eff}}$).
(b) Dependence of the charge-1e gap $\Delta_{1e}$ (in units of $t$) on $B$ at $r = 1$ for various doping levels $x$. The results were obtained using fDMRG simulations with $L_y = 1$, $L_x = 60$, and bond dimension $\chi = 3000$. A finite gap is observed for all $x$ in the $B \gg t$ regime, and it is suppressed as $B \to 0$.
} 
    \label{fig:7}
\end{figure}

In this section, we introduce another limit in which we can gain intuitive insight into the pairing mechanism using analytical techniques. We consider the ESD model with an additional layer-opposite Zeeman field $B$, as illustrated in Fig.~\ref{fig:7}(a). The Hamiltonian is then
\begin{equation}
\hat{H}_B = \hat{H}_{\mathrm{ESD}} - B \sum_{i} (S^z_{i,t} - S^z_{i,b}),
\label{eq:esd+zeeman}
\end{equation}
where $S^z_{i,l} = \sum_{\sigma,\sigma'} |l,\sigma\rangle_i \sigma^z_{\sigma,\sigma'} \langle l,\sigma'|_i / 2$.

Crucially, we also include an on-site energy term $\textcolor{black}{V_{\mathrm{eff}}} = -B/2$ in the ESD model to compensate for the repulsive interaction induced by the $B$ term. This splits the six states into two groups: $|h\rangle$, $|d\rangle$, $|t, \uparrow\rangle$, and $|b, \downarrow\rangle$ at lower energy, separated by an energy gap $B > 0$ from $|t, \downarrow\rangle$ and $|b, \uparrow\rangle$ (see Fig.~\ref{fig:7}(a)).

We again focus on $r=1$. We consider the $B\gg t$ limit and demonstrate that the model in this limit is mapped to the free-electron model. In $B\rightarrow \infty $ limit, we can relabel $\ket{h}$ as $\ket{0}$ and $\ket{d}$ as the doubly occupied state in the familiar spin-1/2 Hubbard model.
Then, the electron operators within the four ground states follow a simple free electron algebra, 
\begin{eqnarray}
    c_{t,\uparrow}^\dagger&=& 
    \ket{t,\uparrow}
    \bra{h}
    +
    \ket{d}
    \bra{b,\downarrow}
    \equiv \eta_{\uparrow}^\dagger
    \\
c_{b,\downarrow}^\dagger&=& 
    \ket{b,\downarrow}
    \bra{h}
    -
    \ket{d}
    \bra{t,\uparrow}   
    \equiv 
    \eta_{\downarrow}^\dagger.
\end{eqnarray}
Note that $\eta_{\sigma}^{\dagger}$ is a free electron creation operator and does not have any layer index. 
The four ground states can be generated by the $\eta_{\sigma}^{\dagger}$ operator: $\ket{h}=\ket{0}$, $\ket{t,\uparrow}=\eta^\dagger_{\uparrow}\ket{0}$, $\ket{b,\downarrow}=\eta^\dagger_{\downarrow}\ket{0}$, and $\ket{d}=\eta ^\dagger_{\uparrow}\eta^\dagger_{\downarrow}\ket{0}$ in the $B\rightarrow + \infty$ limit. 

At leading order, we have a spin-$1/2$ free fermion, and the ground state is known to be a Fermi liquid. We now reduce $B$ and derive the second-order terms in $t/B$. The effective Hamiltonian becomes
\begin{align}
\hat{H}_{\mathrm{eff}} &= -t \sum_{\sigma} \sum_{\langle i,j \rangle} (\eta^\dagger_{i,\sigma} \eta_{j,\sigma} + \eta^\dagger_{j,\sigma} \eta_{i,\sigma}) \nonumber \\
&\quad - \frac{2t^2}{B} \sum_{\langle i,j \rangle} n_{i,\uparrow} n_{i,\downarrow} (1 - n_{j,\uparrow})(1 - n_{j,\downarrow}) \nonumber \\
&\quad - \frac{2t^2}{B} \sum_{\langle i,j \rangle} (\eta^\dagger_{i,\uparrow} \eta^\dagger_{i,\downarrow}) (\eta_{j,\downarrow} \eta_{j,\uparrow}) + \text{h.c.} + \cdots,
\label{eq:Heff}
\end{align}
where $n_{i,\sigma} = \eta^\dagger_{i,\sigma} \eta_{i,\sigma}$ is the density operator.
The first term represents the kinetic energy of free electrons, while the last two terms describe interactions generated by hopping processes involving the higher-energy states. The second term represents an attractive density-density interaction, which energetically favors configurations with a doublon ($n_{i,\uparrow} n_{i,\downarrow} = 1$) and a holon ($(1 - n_{j,\uparrow})(1 - n_{j,\downarrow}) = 1$) on neighboring sites. The third term is a pair-hopping term. The $\cdots$ indicates omitted three-site processes for simplicity.
Thus, the ESD model with a layer-dependent Zeeman field can be interpreted as a free-electron model with an effective attractive interaction. Simple mean-field theory then predicts an $s$-wave superconducting state with a pairing strength that increases with $t^2/B$.

To support this picture, we performed finite DMRG simulations of $\hat{H}_B$. As shown in Fig.~\ref{fig:7}(b), the pairing gap, characterized by the charge-1e gap, increases as $B$ decreases, reaching a maximum at $B = 0$. In the limit $B \gg t$, the model can be treated with conventional mean-field theory based on the attractive interaction, allowing us to confidently assert the existence of a superconducting phase in the two-dimensional limit. As $B$ decreases, the pairing strength is enhanced because the effective attractive interaction becomes stronger. Although this analysis breaks down for $B \lesssim t$, it is reasonable to conjecture that the pairing strength continues to increase for $B < t$ until $B = 0$, consistent with our finding that the ESD model at $B = 0$ is in a superconducting phase at any doping in the two-dimensional limit.

\clearpage
\section{Details on the Mean-field theory}\label{sec:s4}
We start with ESD t-J model Hamiltonian ($r=1,\textcolor{black}{V_{\mathrm{eff}}}=0$),
\begin{align}
 H&=-t\sum_{l,\sigma} \left[\sum_{i} Pc^\dagger_{i;l;\sigma} c_{i+x;l;\sigma} P+h.c.
 +\sum_{i} Pc^\dagger_{i;l;\sigma} c_{i+y;l;\sigma}P+h.c.
 \right]
\end{align}

\begin{figure}[h]
    \centering
    \includegraphics[width=0.5\linewidth]{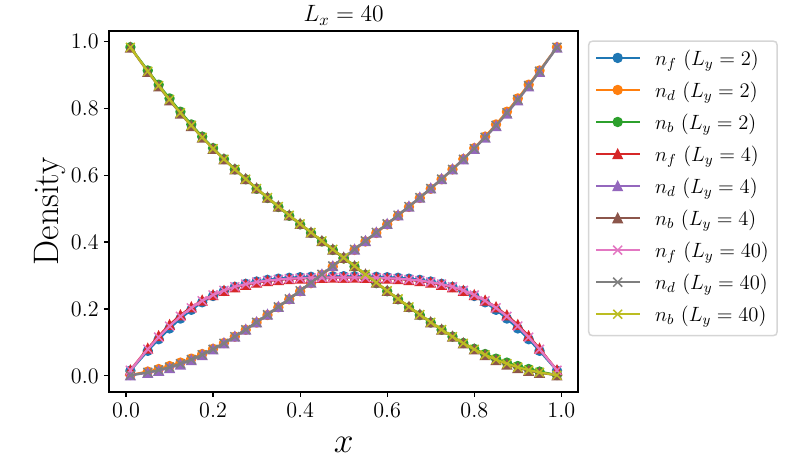}
    \caption{The $x$ dependence of mean-field theory of the ESD model at $L_x=40$ with various $L_y=2,4,40$. 
   There are almost no changes on $L_y$. 
   Here, we can predict the two distinct Fermi-Liquid near $x=0$ and $x=1$ limit. The enhancement of the pairing toward $x\rightarrow 0.5$ starting from $x=0$ or $x=1$ can be understood by the dominance of the virtual Cooper pair resulting in the Feshbach resonance mechanism. 
    }
    \label{fig:mft}
\end{figure}

Upon the conventional mean-field decoupling the the ESD Hamiltonian is decouped as, 
\begin{eqnarray}
    H_{MF} 
    &=&H_{f}+H_{hd},
\end{eqnarray}
\begin{eqnarray}
  H_{f} &=&-t_{f;x} \sum_{i,l,\sigma} [f^{\dagger}_{i;l;\sigma}f_{i+x;l;\sigma}
+f^{\dagger}_{i+x;l;\sigma}f_{i;l;\sigma}]-t_{f;y} \sum_{i,l,\sigma} [f^{\dagger}_{i;l;\sigma}f_{i+y;l;\sigma}
+f^{\dagger}_{i+y;l;\sigma}f_{i;l;\sigma}]
  \nonumber \\
    &&+\Delta_{f;x} \sum_{i}
\left[
f^{\dagger}_{i;t;\uparrow}f^{\dagger}_{i+x;b;\downarrow}
-f^{\dagger}_{i;t;\downarrow}f^{\dagger}_{i+x;b;\uparrow}
+(t\leftrightarrow b)
\right] +h.c. \notag\\
    &&+\Delta_{f;y} \sum_{i}
\left[
f^{\dagger}_{i;t;\uparrow}f^{\dagger}_{i+y;b;\downarrow}
-f^{\dagger}_{i;t;\downarrow}f^{\dagger}_{i+y;b;\uparrow}
+(t\leftrightarrow b)
\right] +h.c.  +\delta_{f}\sum_{i}n_{f;i}\nonumber\\
H_{hd}&=& 
\sum_{i}
[\lambda_{h} h_{i} +h.c.]
+
\sum_{i}
[\lambda_{d} d_{i} +h.c.]
+\delta_{h}\sum_{i}n_{h;i}
+\delta_{d}\sum_{i}n_{d;i} 
\end{eqnarray}
with 
\begin{eqnarray*}
    t_{f;x} &=&
    t_{f;y}=t\left[  \langle h \rangle^2 - \langle d \rangle^2\right], \\
    \Delta_{f;x} &=& 
    \Delta_{f;y} = -2 t \langle h \rangle \langle d\rangle,
\end{eqnarray*}
\begin{eqnarray*}
    \lambda_{h} & = &-2t\big[
    \langle h \rangle
    \left[
    \chi_{x}
    +\chi_{y}
    \right]
    -\langle d \rangle
 \left[\Delta_{x}
 + \Delta_{y}\right]
 \big]\times 2,\notag\\
    \lambda_{d} &=& 2t\big[  
    \langle d\rangle 
\left[
\chi_{x}
+\chi_{y}
\right]
+ \langle d\rangle
\left[
\Delta_{x}
+\Delta_{y}
\right]\times 2,
\end{eqnarray*}
\begin{eqnarray*}
    \delta_{f} &= &-\mu_{0}-\mu, 
    \quad \delta_{d}=-\mu_{0}-2\mu,
\quad \delta_{b}=-\mu_{0}
\end{eqnarray*}
where two chemical potentials, $\mu_{0},\mu$ are introduced for the two constraints: $n_{d}+n_{f}+n_{b}=1$, and $n_{f}+2n_{d}=2x$. The order parameters are defined as 
\begin{eqnarray}
    \chi_{x(y)}&=& \sum_{\sigma} \langle f_{i;l;\sigma}^{\dagger}f_{i+x(y);l;\sigma}
\rangle, \\
    \Delta_{x(y)} & = &
\big\langle
f_{i;t;\uparrow}f_{i+x(y);b;\downarrow} -
f_{i;t;\downarrow}f_{i+x(y);b;\uparrow} 
     \big \rangle, \notag\\
    \langle d \rangle &=&\frac{\lambda_d}{-\delta_{d}}, \ \mathrm{for}\ \delta_{d}>0,
    \quad 
    \langle b \rangle =\frac{\lambda_b}{-\delta_{b}}, \ \mathrm{for}\ \delta_{b}>0,\notag
\end{eqnarray}
Here, we split all order parameters $x$ and $y$ separately to see the $(L_x,L_y)$ dependence. In $L_x=L_y$ case, we can show $\chi_x=\chi_y$ and $\Delta_x=\Delta_y$. In Fig.\ref{fig:mft}, we present the mean-field results at $r=1,\textcolor{black}{V_{\mathrm{eff}}}=0$ and show the system size dependence. Here, the density of each particle content can be carried out with $n_d=\langle d\rangle^2$, $n_h=\langle h\rangle^2$ and $n_f=\sum_{k,l,\sigma}\langle f_{k,l,\sigma}^\dagger f_{k,l,\sigma}\rangle$. 

\newpage
\section{Details on the realistic bilayer models}\label{sec:s5}
\subsection{Two-site problem}
In this section, we consider the two-site problem of type-II t-J model with the following Hamiltonian
\begin{align}
    H=J_\perp\vec{S}_{i;t}\cdot \vec{S}_{i;b}+\frac{J_\perp}{2}(\vec{S}_{i;t}\cdot \vec{T}_{i;b}+\vec{T}_{i;t}\cdot \vec{S}_{i;b})+\frac{J_\perp}{4}\vec{T}_{i;t}\cdot\vec{T}_{i;b}+\frac{J_\perp}{4}n_{i;t}n_{i;b},
\end{align}
here we consider the case $V=\frac{1}{4}J_\perp$ corresponding to $\textcolor{black}{V_{\mathrm{eff}}}=0$.
In the $n_T=2$ sector, two spin-one moments coupling, there are one eigenstate with energy $E=-\frac{1}{2}J_\perp+V$,
\begin{align}
    \ket{S=0,S^z=0}=\frac{1}{\sqrt{3}}(\ket{1,1}\ket{1,-1}-\ket{1,0}\ket{1,0}+\ket{1,-1}\ket{1,1})
\end{align}
three eigenstates with energy $E=-\frac{1}{4}J_\perp+V$,
\begin{align}
    \ket{S=1,S^z=1}=&\frac{1}{\sqrt{2}}(\ket{1,1}\ket{1,0}-\ket{1,0}\ket{1,1}),\\
    \ket{S=1,S^z=0}=&\frac{1}{\sqrt{2}}(\ket{1,1}\ket{1,-1}-\ket{1,-1}\ket{1,1}),\\
    \ket{S=1,S^z=-1}=&\frac{1}{\sqrt{2}}(\ket{1,0}\ket{1,-1}-\ket{1,-1}\ket{1,0}),
\end{align}
five eigenstates with energy $E=\frac{1}{4}J_\perp+V$,
\begin{align}
    \ket{S=2,S^z=2}=&\ket{1,1}\ket{1,1},\\
    \ket{S=2,S^z=1}=&\frac{1}{\sqrt{2}}(\ket{1,1}\ket{1,0}+\ket{1,0}\ket{1,1}),\\
    \ket{S=2,S^z=0}=&\frac{1}{\sqrt{6}}(\ket{1,1}\ket{1,-1}+2\ket{1,0}\ket{1,0}+\ket{1,-1}\ket{1,1}),\\
    \ket{S=2,S^z=-1}=&\frac{1}{\sqrt{2}}(\ket{1,0}\ket{1,-1}+\ket{1,-1},\ket{1,0}),\\
    \ket{S=2,S^z=-2}=&\ket{1,-1}\ket{1,-1},
\end{align}
in the $n_T=1$ sector, one spin-one moment couples with spin-$1/2$ moment, there are two states with energy $E=-\frac{1}{2}J_\perp$,
\begin{align}
    \ket{S=\frac{1}{2},S^z=\frac{1}{2}}=&\frac{\sqrt{2}}{\sqrt{3}}\ket{1,1}\ket{\frac{1}{2},-\frac{1}{2}}-\frac{1}{\sqrt{3}}\ket{1,0}\ket{\frac{1}{2},\frac{1}{2}},\\
    \ket{S=\frac{1}{2},S^z=-\frac{1}{2}}=&\frac{1}{\sqrt{3}}\ket{1,0}\ket{\frac{1}{2},-\frac{1}{2}}-\frac{\sqrt{2}}{\sqrt{3}}\ket{1,-1}\ket{\frac{1}{2},\frac{1}{2}},
\end{align}
four states with $E=\frac{1}{2}J_\perp$
\begin{align}
    \ket{S=\frac{3}{2},S^z=\frac{3}{2}}=&\ket{1,1}\ket{\frac{1}{2},\frac{1}{2}},\\
    \ket{S=\frac{3}{2},S^z=\frac{1}{2}}=&\frac{1}{\sqrt{3}}\ket{1,1}\ket{\frac{1}{2},-\frac{1}{2}}+\frac{\sqrt{2}}{\sqrt{3}}\ket{1,0}\ket{\frac{1}{2},\frac{1}{2}},\\
    \ket{S=\frac{3}{2},S^z=-\frac{1}{2}}=&\frac{\sqrt{2}}{\sqrt{3}}\ket{1,0}\ket{\frac{1}{2},-\frac{1}{2}}+\frac{1}{\sqrt{3}}\ket{1,-1}\ket{\frac{1}{2},\frac{1}{2}},\\
    \ket{S=\frac{3}{2},S^z=-\frac{3}{2}}=&\ket{1,-1}\ket{\frac{1}{2},-\frac{1}{2}}.
\end{align}
There are another $2+4$ states with the exchange of the top layer and bottom layer.
In the $n_T=0$ sector, two spin-$1/2$ coupling, there is one state with $E=-\frac{3}{4}J_\perp$,
\begin{align}
    \ket{S=0,S=0}=\frac{1}{\sqrt{2}}(\ket{\frac{1}{2},\frac{1}{2}}\ket{\frac{1}{2},-\frac{1}{2}}-\ket{\frac{1}{2},-\frac{1}{2}}\ket{\frac{1}{2},\frac{1}{2}}),
\end{align}
there are three states with $E=\frac{1}{4}J_\perp$,
\begin{align}
    \ket{S=1,S^z=1}=&\ket{\frac{1}{2},\frac{1}{2}}\ket{\frac{1}{2},\frac{1}{2}},\\
    \ket{S=1,S=0}=&\frac{1}{\sqrt{2}}(\ket{\frac{1}{2},\frac{1}{2}}\ket{\frac{1}{2},+\frac{1}{2}}-\ket{\frac{1}{2},-\frac{1}{2}}\ket{\frac{1}{2},\frac{1}{2}}),\\
    \ket{S=1,S^z=-1}=&\ket{\frac{1}{2},-\frac{1}{2}}\ket{\frac{1}{2},-\frac{1}{2}}
\end{align}
\subsection{Details on the calculation of percentages}
The density of holon, singlon and doublon are $x_0$, $x_1$ and $x_2$, respectively. They should satisfy $x_0+x_1+x_2=1$. The total number of holes are $2x_0+x_1=2(1-n)$, where $n$ is the density of electrons $\langle N\rangle$. The average $n_tn_b=\langle N_tN_b\rangle$ measures the density of doublons $x_2$. Finally, we can get
\begin{align}
    x_0=&1-2n+n_tn_b,\nonumber\\
    x_1=&2(n-n_tn_b),\nonumber\\
    x_2=&n_tn_b,
\end{align}
In the doublon sector, we have $\vec{T}_t\cdot\vec{T}_b=-2,-1,1$ for the total spin $S=0,1,2$, respectively. We can define the following projecting operators to project into the subspace with total spin $S$,
\begin{align}
    P_{S=0}=&(-1-\vec{T}_t\cdot\vec{T}_b)(1-\vec{T}_t\cdot\vec{T}_b),\\
    P_{S=1}=&(-2-\vec{T}_t\cdot\vec{T}_b)(1-\vec{T}_t\cdot\vec{T}_b),\\
    P_{S=2}=&(-1-\vec{T}_t\cdot\vec{T}_b)(-2-\vec{T}_t\cdot\vec{T}_b),
\end{align}
the density of $S=0,1,2$ doublons are $n_{d,S=0}=\frac{1}{3}\langle P_{S=0}N_tN_b\rangle$, $n_{d,S=1}=\frac{1}{-2}\langle P_{S=1}N_tN_b\rangle$, $n_{d,S=2}=\frac{1}{6}\langle P_{S=2}N_tN_b\rangle$, respectively, where the factor in front of the expectation value is the normalization factor. In calculating the density of different spin sectors, we first project the wave function into the sector with $n_T=2$ and then project into the subspace with total spin $S$. Similarly, we can calculate the density of different singlons and holons, with
\begin{align}
    n_{s,S=1/2}=&\frac{2}{3}\langle P^t_{S=\frac{1}{2}}N_t(1-N_b)+P^b_{S=\frac{1}{2}}(1-N_t)N_b\rangle,\\
    n_{s,S=3/2}=&-\frac{2}{3}\langle P^t_{S=\frac{3}{2}}N_t(1-N_b)+P^b_{S=\frac{3}{2}}(1-N_t)N_b\rangle,
\end{align}
where $P^t_{S=1/2}=(\frac{1}{2}-\vec{T}_t\cdot\vec{S}_b)$, $P^t_{S=3/2}=(-1-\vec{T}_t\cdot\vec{S}_b)$, $P^b_{S=1/2}=(\frac{1}{2}-\vec{S}_t\cdot\vec{T}_b)$ and $P^t_{S=3/2}=(-1-\vec{S}_t\cdot\vec{T}_b)$.
\begin{align}
    n_{h,S=0}=&\langle P_{h,S=0}(1-N_t)(1-N_b)\rangle,\\
    n_{h,S=0}=&-\langle P_{h,S=1}(1-N_t)(1-N_b)\rangle,
\end{align}
with $P_{h,S=0}=\frac{1}{4}-\vec{S}_t\cdot\vec{S}_b$ and $P_{h,S=1}=-\frac{3}{4}-\vec{S}_t\cdot\vec{S}_b$.
Then we can calculate the percentage of different spin sectors.
\end{document}